\documentclass[12pt]{report}

\usepackage[utf8]{inputenc}
\usepackage{charter}
\usepackage{fullpage}
\usepackage{hyperref}
\usepackage{graphicx}
\usepackage{lipsum}
\usepackage[sort&compress,numbers]{natbib}
\usepackage{hyperref}
\hypersetup{hidelinks}
\usepackage[total={6.5in,9in},top=1in,headsep=0.1in,headheight=1in]{geometry}
\usepackage{siunitx}
\usepackage{units}

\usepackage{tiffany}
\usepackage{color}

\usepackage{ulem}
\usepackage[nottoc,notlot,notlof]{tocbibind}

\newcommand\snowmass{
\begin{center}
  \rule[-0.2in]{\hsize}{0.01in}\\
  \rule{\hsize}{0.01in}\\
  Submitted to the Proceedings of the US Community Study\\ 
  on the Future of Particle Physics (Snowmass 2021)\\
  \vskip 0.1in
  \rule{\hsize}{0.01in}\\
  \rule[+0.2in]{\hsize}{0.01in}\\[-2em]
\end{center}
}

\definecolor{hookgreen}{rgb}{0.0,0.44,0.0}

\newcommand{\msun}{M_\odot}

\newcommand{\nocontentsline}[3]{}

\usepackage{acro}
\DeclareAcronym{LGRB}{
    short = {LGRB},
    long = {long gamma-ray burst},
    long-plural-form = {long gamma-ray bursts}
}

\def\apjl{Astrophys.~J.~Lett.}

\usepackage[firstpage=true]{background}
\backgroundsetup{contents={\parbox{6.5in}{\snowmass}}, scale=1,placement=top,opacity=1,color=black,position={3.25in,1.2in}}

\usepackage{fancyhdr}
\fancypagestyle{plain}{%
  \fancyhf{}%
  \fancyhead[C]{}
  \fancyfoot[C]{\thepage}
}

\fancypagestyle{empty}{%
  \fancyhf{}%
  \fancyhead[C]{{\it Snowmass2021 CF07 Multimessenger Facilities \& Experiments}}
  \fancyfoot[C]{\thepage}
}
\date{}

\begin{document}


\clearpage
\pagenumbering{roman}



\vspace*{\fill}
\begin{center}{\textbf{\LARGE Advancing the Landscape of Multimessenger Science in the Next Decade}}
\end{center}

\vspace*{0.2cm}
\begin{center}{{\large \textsc{Edited by}}\\
\vspace*{0.2cm}

{Kristi~Engel \orcidlink{0000-0001-5737-1820}}$^{1,2}$, 
{Tiffany~Lewis}\orcidlink{0000-0002-9854-1432}$^{3}$, 
{Marco~Stein~Muzio}\orcidlink{0000-0003-4615-5529}$^{4}$\\[1mm]
{Tonia~M.~Venters \orcidlink{0000-0002-4188-627X}}$^{5}$ 
}\end{center}
\vspace*{\fill}

\clearpage

\thispagestyle{plain}	
\renewcommand{\thefootnote}{\fnsymbol{footnote}}


\vspace*{-0.2cm}
\begin{center}{{\Large \textsc{Contributors}}\\
\vspace*{0.2cm}
{Markus~Ahlers}$^{6}$,
{Andrea~Albert}$^{1}$,
{Alice~Allen}$^{2}$,
{Hugo~Alberto~Ayala~Solares}$^{7}$,
{Samalka~Anandagoda}$^{8}$,
{Thomas~Andersen}$^{62}$,
{Sarah~Antier}$^{9}$,
{David~Alvarez-Castillo}$^{10}$,
{Olaf~Bar}$^{59}$,
{Dmitri~Beznosko}$^{57}$,
{\L{}ukasz~Bibrzyck}$^{59}$,
{Adam~Brazier}$^{11}$,
{Chad~Brisbois}$^{2}$,
{Robert~Brose}$^{12}$,
{Duncan~A.~Brown}$^{13}$,
{Mattia~Bulla}$^{14}$,
{J.~Michael~Burgess}$^{15}$,
{Eric~Burns}$^{16}$,
{Cecilia~Chirenti}$^{2,5,17,18}$,
{Stefano~Ciprini}$^{19,20}$,
{Roger~Clay$^{58}$}
{Michael~W.~Coughlin}$^{21}$,
{Austin~Cummings}$^{7}$,
{Valerio~D'Elia}$^{19}$,
{Shi~Dai}$^{22}$,
{Tim~Dietrich}$^{23,24}$,
{Niccol\`{o}~Di~Lalla}$^{25}$,
{Brenda~Dingus$^{2,1}$},
{Mora Durocher$^{1}$},
{Johannes~Eser}$^{26}$,
{Miroslav~D.~Filipovi\'{c}$^{22}$},
{Henrike~Fleischhack$^{27,5,17}$},
{Francois~Foucart}$^{28}$,
{Michał~Frontczak}$^{59}$,
{Christopher~L.~Fryer}$^{1}$,
{Ronald~S.~Gamble}$^{29,2}$,
{Dario~Gasparrini}$^{19,20}$,
{Marco~Giardino}$^{19}$,
{Jordan~Goodman}$^{2}$,
{J.~Patrick~Harding}$^{1,30}$,
{Jeremy~Hare}$^{3}$,
{Kelly~Holley-Bockelmann}$^{31}$,
{Piotr~Homola}$^{56}$,
{Kaeli~A.~Hughes}$^{32}$,
{Brian~Humensky}$^{2}$,
{Yoshiyuki~Inoue}$^{33,34,35}$,
{Tess~Jaffe}$^{36}$,
{Oleg~Kargaltsev}$^{37}$,
{Carolyn~Kierans}$^{5}$
{James~P.~Kneller}$^{38}$,
{Cristina~Leto}$^{19}$,
{Fabrizio~Lucarelli}$^{19,39}$,
{Humberto~Mart\'{i}nez-Huerta}$^{40}$,
{Alessandro~Maselli}$^{19,39}$,
{Athina~Meli}$^{41,42}$,
{Patrick~Meyers}$^{43}$,
{Guido~Mueller}$^{44}$,
{Zachary~Nasipak}$^{3}$,
{Michela~Negro}$^{45,5,17}$,
{Michał~Nied\'{z}wiecki}$^{81}$,
{Scott~C.~Noble}$^{46}$,
{Nicola~Omodei}$^{25}$,
{Stefan~Oslowski}$^{47}$,
{Matteo~Perri}$^{19,39}$,
{Marcin~Piekarczyk}$^{59}$,
{Carlotta~Pittori}$^{19,39}$,
{Gianluca~Polenta}$^{19}$,
{Remy~L.~Prechelt}$^{48}$,
{Giacomo~Principe}$^{48}$,
{Judith~Racusin}$^{5}$,
{Krzysztof~Rzecki}$^{60}$,
{Rita~M.~Sambruna}$^{29}$,
{Joshua~E.~Schlieder}$^{49}$,
{David~Shoemaker}$^{50}$,
{Alan~Smale}$^{36}$,
{Tomasz~Sośnicki}$^{60}$,
{Robert~Stein}$^{51}$,
{Sławomir~Stuglik}$^{56}$,
{Peter~Teuben}$^{2}$,
{James~Ira~Thorpe}$^{46}$,
{Joris~P.~Verbiest}$^{52,53}$,
{Franceso~Verrecchia}$^{19,39}$,
{Salvatore~Vitale}$^{54}$,
{Zorawar~Wadiasingh}$^{2,46,17}$,
{Tadeusz~Wibig}$^{63}$,
{Elijah~Willox}$^{2}$,
{Colleen~A.~Wilson-Hodge}$^{55}$,
{Joshua~Wood}$^{55}$,
{Hui~Yang}$^{37}$,
{Haocheng~Zhang}$^{3}$
}\end{center}

\vspace*{-0.2cm}
\begin{center}{\it\footnotesize
$^{1}${Physics Division, Los Alamos National Laboratory, Los Alamos, NM, 87545, USA}\\
$^{2}${University of Maryland, College Park, College Park, MD 20742, USA}\\
$^{3}${NASA Postdoctoral Program Fellow, NASA Goddard Space Flight Center, Greenbelt, MD 20771, USA}\\
$^{4}${NSF MPS-Ascend Fellow, Department of Physics, Pennsylvania State University, State College, PA 16801, USA}\\
$^{5}${Astroparticle Physics Laboratory, NASA Goddard Space Flight Center, Greenbelt, MD 20771, USA}\\
$^{6}${Niels Bohr International Academy, Blegdamsvej 17, 2100 Copenhagen, Denmark}\\
$^{7}${Department of Physics, Pennsylvania State University, State College, PA 16801, USA}\\
$^{8}${Clemson University, Department of Physics \& Astronomy, Clemson, SC 29634, USA}\\
$^{9}${ARTEMIS UMR 7250 UCA CNRS OCA, Boulevard de l’Observatoire, CS 34229, 06304 Nice
CEDEX 04, France}\\
$^{10}${Institute of Nuclear Physics PAN, Cracow 31-342, Poland}\\
$^{11}${Cornell Center for Astrophysics and Planetary Science and Department of Astronomy, Cornell University, Ithaca, NY 14853, USA}\\
$^{12}${Dublin Institute for Advanced Studies, Astronomy \& Astrophysics Section, 31 Fitzwilliam
Place, D02 XF86 Dublin 2, Ireland}\\
$^{13}${Department of Physics, Syracuse University, Syracuse, NY 13244, USA}\\
$^{14}${The Oskar Klein Centre, Department of Astronomy, Stockholm University, AlbaNova, SE-106
91 Stockholm, Sweden}\\
$^{15}${Max-Planck Institut fur Extraterrestrische Physik, Giessenbachstrasse 1, 85740 Garching, Germany}\\
$^{16}${Louisiana State University, Baton Rouge, LA, 70803, USA}\\
$^{17}${Center for Research and Exploration in Space Science and Technology, NASA Goddard Space
Flight Center, Greenbelt, Maryland 20771, USA}\\
$^{18}${Center for Mathematics, Computation and Cognition, UFABC, Santo Andr`e, SP 09210-580,
Brazil}\\
$^{19}${Space Science Data Center, Italian Space Agency, via del Politecnico snc, 00133, Roma, Italy}\\
$^{20}${INFN-Sezione di Roma Tor Vergata\^{\i}, 00133, Roma, Italy}\\
$^{21}${University of Minnesota, Minneapolis, MN, 55455, USA}\\
$^{22}${1Western Sydney University, Locked Bag 1797, Penrith, NSW 2751, Australia}\\
$^{23}${Institute of Physics and Astronomy, University of Potsdam, Karl-Liebknecht-Str. 24/25, 14476, Potsdam, Germany}\\
$^{24}${Max Planck Institute for Gravitational Physics (Albert Einstein Institute), Am Muhlenberg 1, D-14476 Potsdam, Germany}\\
$^{25}${W. W. Hansen Experimental Physics Laboratory, Kavli Institute for Particle Astrophysics and Cosmology, Department of Physics and SLAC National Accelerator Laboratory, Stanford University, Stanford, CA 94305, USA}\\
$^{26}${University of Chicago, Chicago, IL 60637, USA}\\
$^{27}${Catholic University of America, Washington DC 20064, USA}\\
$^{28}${Department of Physics and Astronomy, University of New Hampshire, 9 Library Way, Durham New Hampshire 03824, USA}\\
$^{29}${Astrophysics Science Division, NASA Goddard Space Flight Center, Greenbelt, MD 20771, USA}\\
$^{30}${Michigan State University, East Lansing, MI, 48824, USA}\\
$^{31}${Department of Physics and Astronomy, Vanderbilt University, Nashville, Tennessee 37235,
USA}\\
$^{32}${Department of Physics, Enrico Fermi Institute, Kavli Institute for Cosmological Physics, University of Chicago, Chicago, IL 60637}\\
$^{33}${Osaka University, Toyonaka, Osaka 560-0043, Japan}\\
$^{34}${Interdisciplinary Theoretical \& Mathematical Science Program (iTHEMS), RIKEN, 2-1 Hirosawa, Saitama 351-0198, Japan}\\
$^{35}${Kavli Institute for the Physics and Mathematics of the Universe (WPI), UTIAS, The University of Tokyo, Kashiwa, Chiba 277-8583, Japan}\\
$^{36}${HEASARC Office, NASA Goddard Space Flight Center, Greenbelt, MD 20771, USA}\\
$^{37}${Department of Physics, The George Washington University, 725 21st St. NW, Washington,
DC 20052, USA}\\
$^{38}${North Carolina State University, Raleigh NC 27695, USA}\\
$^{39}${INAF-OAR, via Frascati 33, 00078 Monte Porzio Catone (RM), Italy}\\
$^{40}${Universidad de Monterrey, San Pedro Garza Garc\'ia NL, 66238, Mexico}\\
$^{41}${North Carolina Agricultural and Technical State University, Greensboro, NC 27411, USA}\\
$^{42}${Universite de Liege, 4000 Liege, Belgium}\\
$^{43}${Theoretical Astrophysics Group, California Institute of Technology, Pasadena, CA 91125, USA}\\
$^{44}${Department of Physics, University of Florida, Gainesville, Florida 32611, USA}\\
$^{45}${University of Maryland, Baltimore County, Baltimore, MD 21250, USA}\\
$^{46}${Gravitational Astrophysics Laboratory, NASA Goddard Space Flight Center, Greenbelt, Maryland 20771, USA}\\
$^{47}${Manly Astrophysics, 15/41-42 East Esplanade, Manly, NSW 2095, Australia}\\
$^{48}${Department of Physics \& Astronomy, University of Hawai'i M\=anoa, Honolulu, HI 96822}\\
$^{49}${Exoplanets and Stellar Astrophysics Laboratory, NASA Goddard Space Flight Center, Greenbelt, Maryland 20771, USA}\\
$^{50}${LIGO Laboratory, Massachusetts Institute of Technology, Cambridge, Massachusetts 02139, USA}\\
$^{51}${Department of Astronomy, California Institute of Technology, Pasadena, CA 91125, USA}\\
$^{52}${Fakult{\"a}t f{\"u}r Physik, Universit{\"a}t Bielefeld, Postfach 100131, 33501 Bielefeld, Germany}\\
$^{53}${Max-Planck-Institut für Radioastronomie, Auf dem H{\"u}gel 69, 53121 Bonn, Germany}\\
$^{54}${Kavli Institute for Astrophysics and Space Research and Department of Physics, Massachusetts Institute of Technology, Cambridge, Massachusetts 02139, USA}\\
$^{55}${NASA Marshall Space Flight Center, Huntsville, AL 35805}
$^{56}${Institute of Nuclear Physics Polish Academy of Sciences, Radzikowskiego, Krak\'{o}w, Poland}
$^{57}${Clayton State University, Morrow, Georgia, USA}
$^{58}${University of Adelaide, Adelaide, S.A., Australia}
$^{59}${Pedagogical University of Kraków: Krakow, Małopolska, PL}
$^{60}${AGH University of Science and Technology in Krakow, Poland}
$^{61}${Department of Computer Science, Cracow University of Technology, Warszawska, kraków, Poland}
$^{62}${NSCIR - 046516 Meaford, Ontario N4L 1W7, Canada}
$^{63}${Faculty of Physics and Applied Informatics, University of Lodz, 90-236 Łódź, Pomorska 149/153, Poland}
}\end{center}

\begin{center}
  \rule[-0.2in]{\hsize}{0.01in}\\[.15mm]
  \rule{\hsize}{0.01in}\\
\end{center}

\vspace*{-0.2cm}
\begin{center}{{\Large \textsc{Endorsers}}\\
\vspace*{0.2cm}
{},
}\end{center}





\clearpage
\pagestyle{plain}
\chapter*{Executive Summary}\addcontentsline{toc}{chapter}{\protect\textbf{Executive Summary}}

The last decade has brought about a profound transformation in multimessenger science. Ten years ago, facilities had been built or were under construction that would eventually discover the nature of objects in our universe could be detected through multiple messengers. Nonetheless, multimessenger science was hardly more than a dream. The rewards for our foresight were finally realized through IceCube's discovery of the diffuse astrophysical neutrino flux, the first observation of gravitational waves by LIGO, and the first joint detections in gravitational waves and photons and in neutrinos and photons. Today we live in the dawn of the multimessenger era.

The successes of the multimessenger campaigns of the last decade have pushed multimessenger science to the forefront of priority science areas in both the particle physics and the astrophysics communities. Multimessenger science provides new methods of testing fundamental theories about the nature of matter and energy, particularly in conditions that are not reproducible on Earth. This white paper will present the science and facilities that will provide opportunities for the particle physics community renew its commitment and maintain its leadership in multimessenger science.

\setcounter{secnumdepth}{5}
\setcounter{tocdepth}{3}
\tableofcontents

\pagenumbering{arabic}



\chapter{Introduction}
\label{sec-introduction}


The last decade of physics and astrophysics have brought us to the dawn of an exciting new era -- the emergence of multimessenger science of which we could only previously dream. Prior to the last decade, multimessenger science scarcely even existed. Only two of the four messengers, cosmic rays and photons, were being observed, and coordination between the two was inconceivable. Nonetheless, the promise of new insight into the workings of our universe inspired us to take the bold step of building facilities to observe the last two messengers, undeterred by the challenges of their detection. The rewards for our courage were finally realized over the course of this last decade with IceCube's discovery of the diffuse astrophysical neutrino flux~\cite{Aartsen:2013jdh}, the first observation of gravitational waves by LIGO~\cite{LIGOScientific:2016aoc}, and the first joint detections in gravitational waves and photons~\cite{2017ApJ...848L..12A} and in neutrinos and photons~\cite{IceCube:2018dnn}. Now multimessenger science is no longer a dream, but a reality.  


The guiding philosophy of multimessenger science is grounded in the recognition of the particular strengths of each messenger. Gravitational waves are sensitive to sites of extreme gravity. Cosmic rays encompass the most energetic particles observed, reaching energies of up to ten million times the energies achieved by the Large Hadron Collider. Photons are the universal messenger of the transfer of energy between phenomena with gamma rays signifying particle acceleration to and collisions at high energies. Neutrinos portend hadrons engaging in these processes. 

Multimessenger science fully leverages the unique qualities of each messenger by combining them to provide new methods of testing fundamental theories about the nature of matter and energy. 
Photons and gravitational waves together test General Relativity, constrain models of quantum gravity, and measure the expansion of the universe. Neutrinos, gamma rays, and cosmic rays unite to reveal the most powerful accelerators in the universe. All four messengers work in partnership to search for clues to the nature of dark matter and relics from early universe processes. All four messengers are connected to the history of structure formation in the universe, each messenger highlighting a particular facet of the cultivation of cosmic environments on all scales.

The enormous potential of multimessenger science for fundamental physics is without question~\cite{NAP26141} (also cross reference Snowmass WPs discussing multimessenger science). At the same time, we must acknowledge that \textit{the promise of fundamental physics with multimessenger is brought about by astrophysics.} Multimessenger facilities observe astrophysical sources in order to provide crucial fundamental physics measurements; as such, the fundamental physics terrain that is unique to multimessenger science is necessarily \textit{entwined} with astrophysics. Nonetheless, the astrophysical sources that are the targets of multimessenger observations feature the most extreme environments in existence, and thus, they provide an unparalleled opportunity to study matter in conditions that are not reproducible on Earth. In this sense and for this reason, \textit{astrophysics is fundamental physics}.

\section{Current Landscape of Multimessenger Science}
\label{sec-sample}


The first multimessenger co-detection was announced on 16 October 2017 -- on 17 August 2017, a binary neutron star merger had been detected by the LIGO/Virgo Collaboration and 1.7 seconds later, from the same area of sky, {\it Fermi}-GBM detected a short gamma-ray burst~\cite{LIGOVirgoGW170817press,NASAGW170817press}. This was the first observation that explicitly linked a short gamma-ray burst to a binary neutron star merger, a concept that had long been theorized~\cite[e.g.,][]{Paczynski:1986px,Eichler:1989ve,Narayan:1992iy} and had gained traction through \textit{Swift} observations~\cite{Fox:2005kv,Barthelmy:2005bx}, but hithereto had eluded direct evidence. The joint detection in gravitational waves and gamma rays spurred the largest follow-up campaign ever conducted with searches for counterparts across the electromagnetic spectrum~\cite{LIGOScientific:2017ync} and in neutrinos~\cite{ANTARES:2017bia}. The follow-up campaign not only succeeded in localizing the merger to the host galaxy, NGC4993, it also provide the first unambiguous detection of a kilonova, the broadband signature of $r$-process nucleosynthesis in the merger ejecta~\cite[e.g.,][]{Li:1998bw,Kulkarni:2005jw,Metzger:2010sy}.


The first extragalactic gamma-ray--neutrino co-detection was announced on 13 July 2018~\cite{IC170922Apress,NASAIC170922Apress}. The IceCube, {\it Fermi}-LAT, MAGIC, {\it AGILE}, HAWC, H.E.S.S., {\it INTEGRAL}, and KANATA collaborations jointly announced that the blazar TXS 0506+056 produced neutrinos simultaneous with a gamma-ray flare on 22 September 2017. Further examination of archival data additionally suggests that the same blazar experienced an orphan neutrino flare at the end of 2014~\cite{IceCube:2018cha}. One of the longstanding questions about blazar jets is whether they include hadronic particles and whether protons can be accelerated to high energies in that environment. Though tentative, the detection of neutrinos from a blazar suggests that they can. This finding has major implications for our understanding of particle energetics near supermassive black holes, as well as the origin(s) of cosmic rays and astrophysical neutrinos~\cite{Aartsen:2013jdh}. 

The successes of these early multimessenger campaigns have pushed multimessenger science to the forefront of priority science areas in both the particle physics and the astrophysics communities. By our count, there $N$ Snowmass white papers that address multimessenger topics (REFS to Snowmass WPs). Multimessenger science was also a key theme of the recent Decadal Survey of Astronomy and Astrophysics (Astro2020), which called for the expansion of facilities operating across the electromagnetic spectrum both on the ground and in space in order to fully exploit the potential of this area of science~\cite{NAP26141}. The Astro2020 also highlighted the need for replacing the crucial capabilities currently being provided by aging facilites, such as the {\it Fermi Gamma-ray Space Telescope} and the {\it Neil Gehrels Swift Observatory}) for which no obvious successors have been identified (see Figure~\ref{fig:gantt_chart}). It is worth acknowledging the major roles the facilities play in following up multimessenger events, including the two highlighted here. As such, the lack of obvious successors is especially concerning and could leave the fate of multimessenger science in a precarious place if not addressed over the next decade. Multimessenger science is not possible without the ongoing support of gamma-ray facilities.

The report of Astro2020 Panel on Particle Astrophysics and Gravitation recognized the outsize role of wider physics community in bringing about the dawn of the multimessenger era. Facilities in gravitational waves, neutrinos, gamma rays, and cosmic rays have largely been developed, funded, and carried out as part of physics programs~\cite{NAP26141}. This white paper will present several opportunities for the particle physics community to renew its commitment to these programs and maintain its leadership in this crucial area of science.





\section{Multimessenger Science in the Next Two Decades} 


Over the next two decades, some of the biggest physics questions in astronomy will be related to cosmology and dark matter. We will make more precise measurements of the Hubble constant, search for primordial black holes, dark matter and Lorentz invariance violation (Chapter \ref{sec-astroBSM}).  We will delve into the discrepancies between different observational methods to reveal new ideas and test established theories. The astrophysical background is provided in Chapter \ref{sec-astrophysics} for objects that theorists predict can be observed with more than one messenger.  Section \ref{sec-facilities} discusses the current facilities that will continue to make these measurements and the landscape of future facilities that will carry multimessenger science into its golden age. Chapter \ref{sec-Infrastructure} describes the collaborative infrastructure through which the work will be accomplished, and points to key opportunities to support scientific excellence. 

\begin{figure}[ht]
    \includegraphics[width=\textwidth]{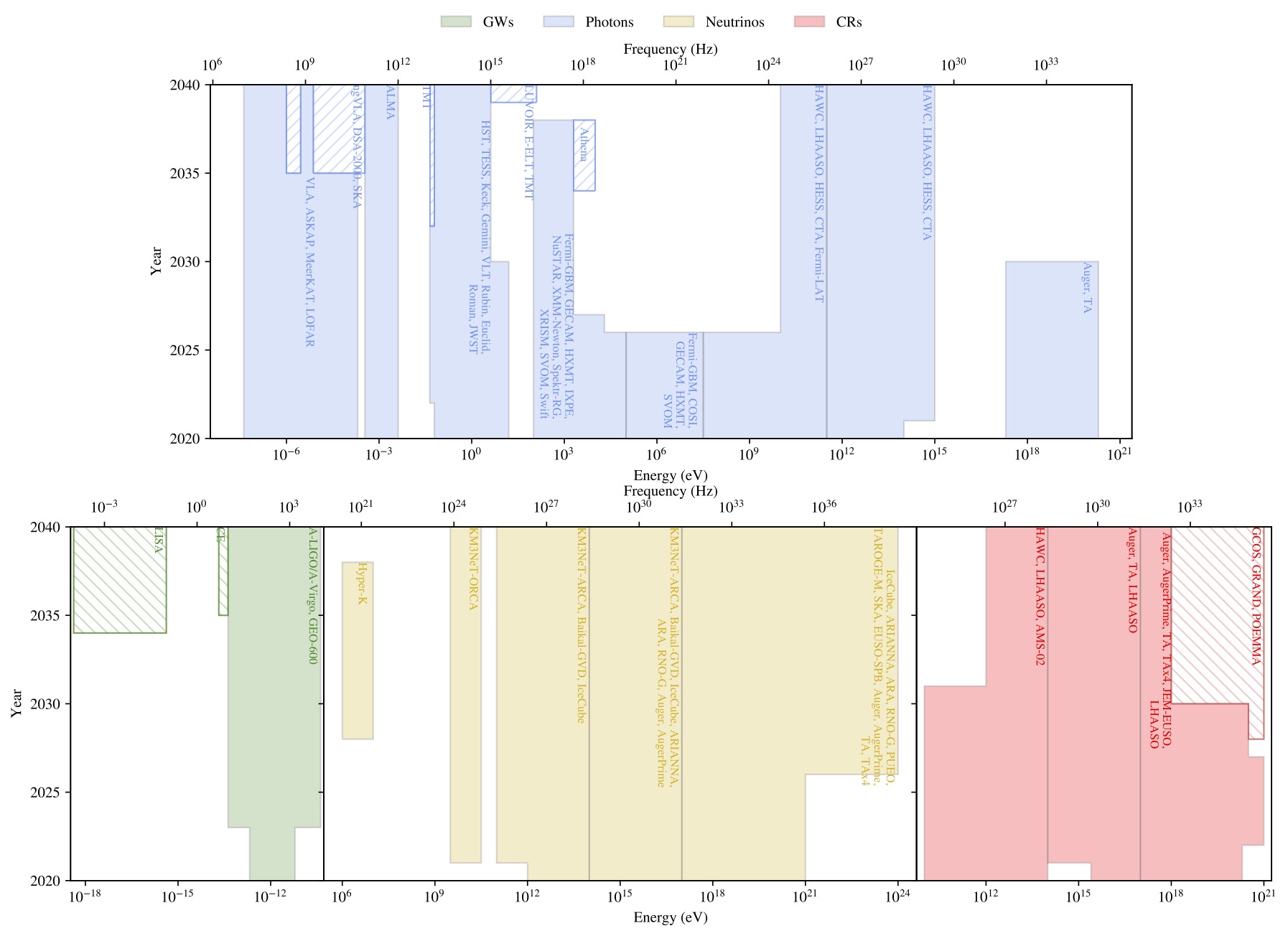}
    \caption{Timeline of current and proposed photon, gravitational wave (GW), neutrino, and cosmic-ray (CR) facilities. Hatched regions indicate energies which proposed experiments would observe that would not be simultaneously observed by any current facilities. Over time, most messengers plan to increase their spectral coverage. The the photon frame in blue illustrates continuous multi-wavelength coverage for the next two decades, with the glaring exception of MeV, GeV, and ultra-high-energy gamma rays. This impending gamma-ray gap is concerning to the broader multimessenger community.}
    \label{fig:gantt_chart}
\end{figure}

The individual instruments involved in multimessenger efforts are some of the most finely tuned human hands have developed. While redundancies can and should be built into facilities, agency program managers rarely have that luxury. In multiwavelength astronomy, NASA has a stated priority for completeness in spectral coverage because we cannot see what we are not looking at.  Similarly, the programmatic management of multimessenger science will be key to optimal scientific output over the next several decades. Gravitational wave, cosmic ray, and neutrino facilities plan generally to increase their spectral coverage over the next two decades, with some currently unfunded future facilities picking up where current ones sunset. Notably, current MeV and GeV gamma-ray facilities are presently expected to end before 2030 with no long term plan to fill that gap in coverage that will impact intrinsically MeV and GeV science as well as make it impossible to collaborate with other wavelengths and messengers, effectively ending multimessenger science as we currently conceive of it (Figure \ref{fig:gantt_chart}). 

\begin{figure}[ht]
    \includegraphics[width=0.5\textwidth, trim = 7cm 0 7cm 0, clip]{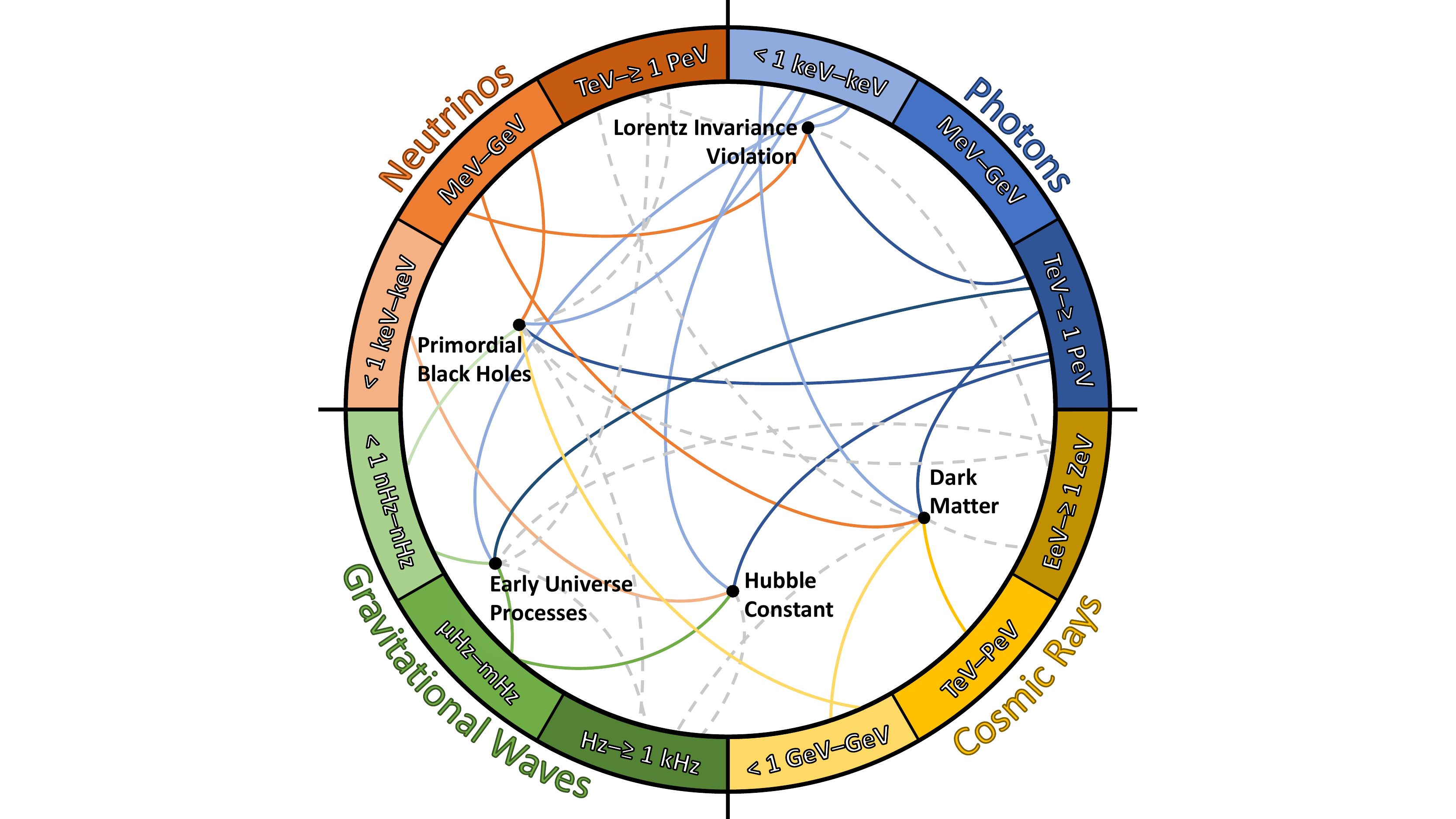}\includegraphics[width=0.5\textwidth, trim = 7cm 0 7cm 0, clip]{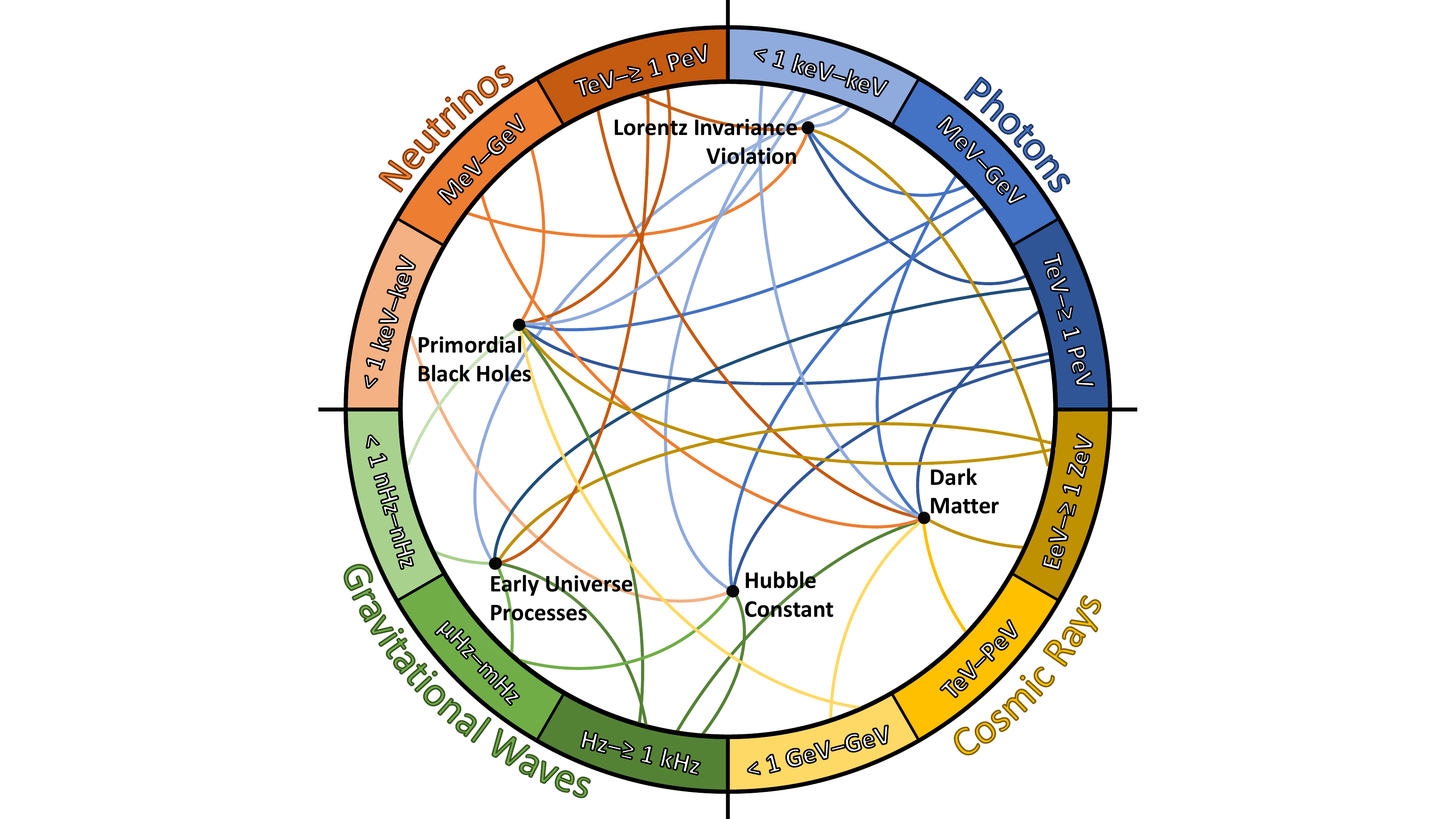}
    \includegraphics[width=0.5\textwidth, trim = 7cm 0 7cm 0, clip]{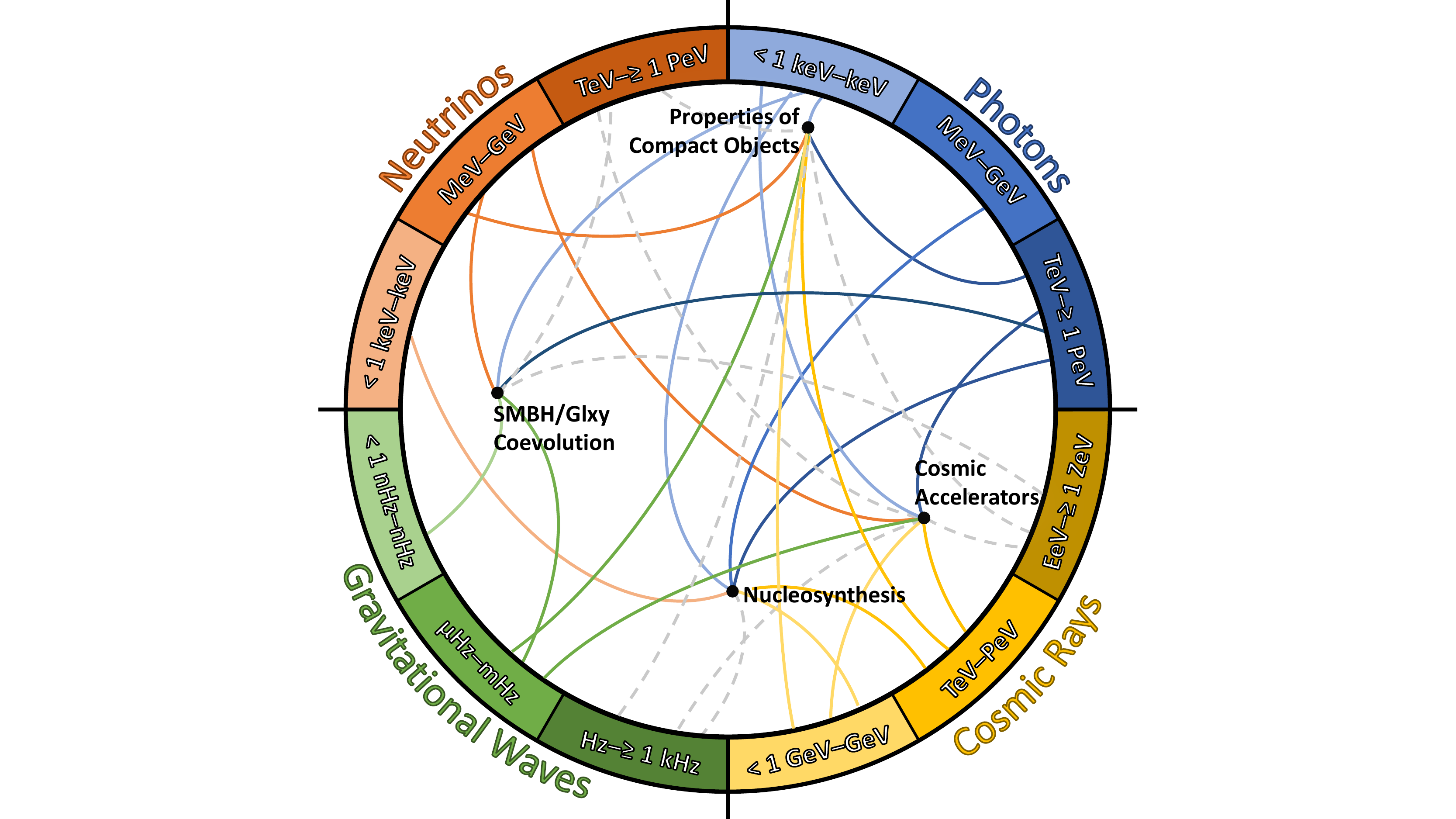}\includegraphics[width=0.5\textwidth, trim = 7cm 0 7cm 0, clip]{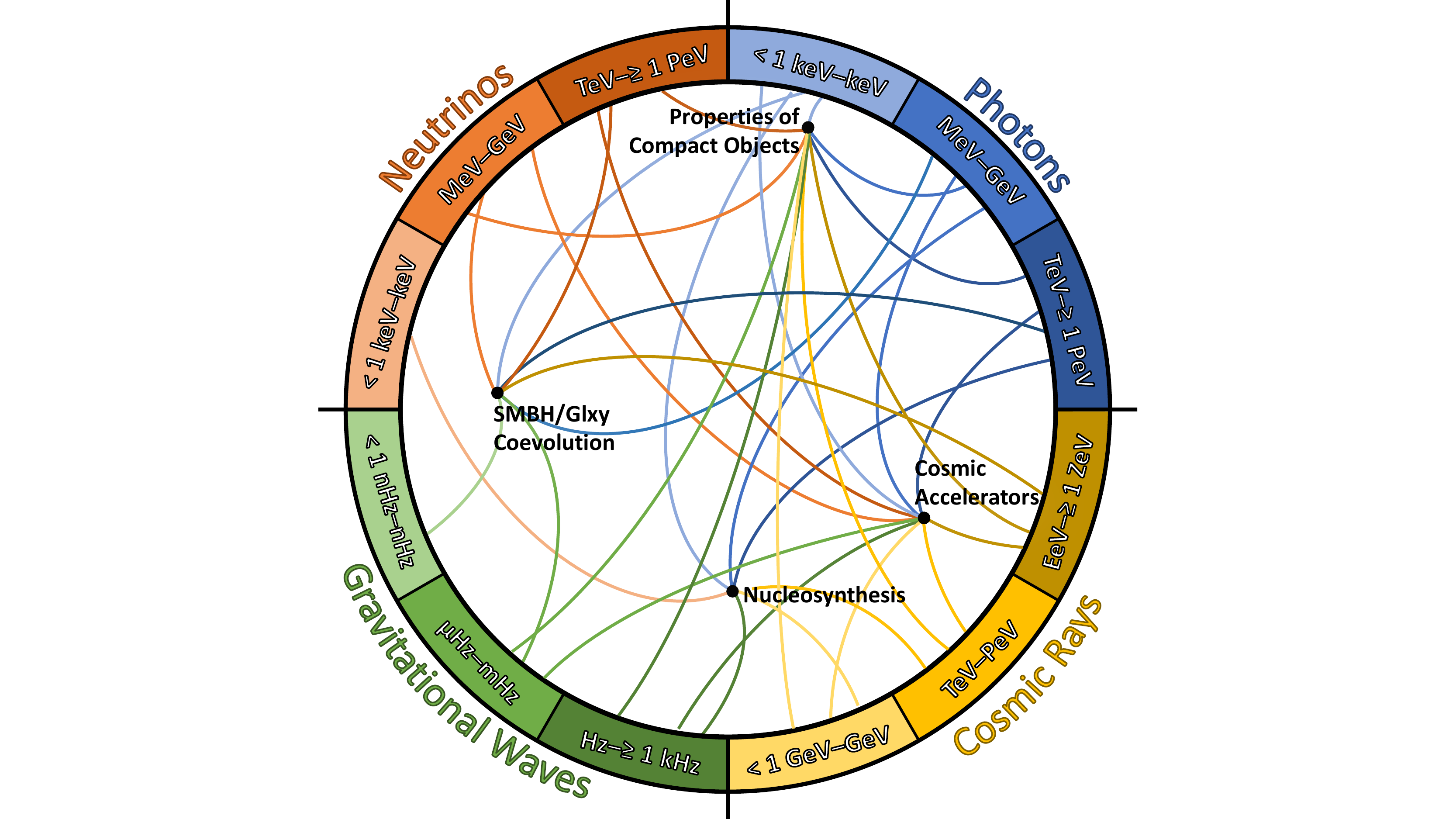}
    \caption{{\it Top panels:} Connections between messengers and fundamental physics topics. {\it Bottom panels:} Connections between messengers and particle astrophysics topics. {\it Left panels:} Future multimessenger landscape with current facilities that are planned to continue operating and future facilities that are already funded. {\it Right panels:} Future multimessenger landscape with enhanced capabilities provided by proposed facilities.}
    \label{fig:chord_plots}
\end{figure}

The loss of instrumental coverage in the MeV-GeV gap has broad implications for the goals of fundamental physics through the study of astronomical objects. Gamma-rays are pivotal in the study of every major physics question in the coming decade. The lack of planned funding for this photon band, in addition to ultra high-energy neutrinos, cosmic-rays and low frequency gravitational waves, which are probed through pulsar timing arrays, should be truly alarming to those who have borne witness to the magnitude of recent multimessenger discoveries. The possible connections between fundamental physics questions, the astronomical objects through which they are studies and observations that probe them by messenger and energy are shown in Figure \ref{fig:chord_plots}, alongside the potential loss of scientific excellence if key instrument classes are not prioritized over the next decade. 


\chapter{Searches for Beyond-Standard-Model and Tests of Fundamental Physics}
\label{sec-astroBSM}

\chapterauthor[ ]{ }
\vspace{-0.2in}
 \addtocontents{toc}{
     \leftskip3cm
    \scshape\small
    \parbox{5in}{\raggedleft James P. Kneller $\quad \ $}
    \upshape\normalsize
    \string\par
    \raggedright
    \vskip -0.19in
    }

 \noindent
 \nocontentsline\chapterauthor[]{James P. Kneller }\orcidlink{0000-0002-3502-3830}
 \\
 \begin{affils}
   \chapteraffil[]{Department of Physics, North Carolina State University, Raleigh NC 27695, USA}
 \end{affils}

The development of new ways to observe the Universe using gravitational waves, neutrinos, and cosmic rays, allows us to significantly advance our understanding of some of the most extreme environments found in the cosmos. Sources that are bright in two or more of these messengers include supernovae, magnetars, black holes, and active galactic nuclei (AGN). These sources are environments where our theories are pushed to their limits due to the incredible density, temperatures and magnetic fields found within them, and where Beyond the Standard Model (BSM) physics can influence the emission. In addition to being a particle beyond the standard model all by itself, one may also envision the future addition of the axion to the list of messenger particles~\cite{2017PhRvD..95f3017O}. While the information we can extract from a signal of any one of the cosmic messengers is valuable, when they are combined we are able to construct a much more vivid and complete view of each source. One need look no further than our own Sun to see how multi-messenger astronomy has changed our understanding of the most prominent object in our sky and revealed BSM physics. After all, the original goal of the Davis experiment was to determine the central temperature of the Sun which the models showed to be dependent upon the chemical composition and the opacity~\cite{1983SSRv...36..426B}, not to learn anything about neutrinos. The discrepancy between theory and Davis's experiment immediately led to proposals for non-standard solar models which were filtered through the Scientific Method to eventually arrive at the present day understanding that neutrinos oscillate. 

There are now several other examples of how the complimentarity of multiple astrophysical messengers reveals more insight than we could obtain from one messenger alone. The gravitational waves and electromagnetic radiation detected from GW170817 clearly established the long-suspected, but previously unproven, conjecture that short duration gamma ray bursts were the merger of neutron stars. The difference of the arrival times of the gravitational waves and the gamma ray flash permit a test of the Equivalence Principle~\cite{Wei_2017}. Similarly, the neutrinos from SN~1987A by themselves confirmed the basic paradigm of core-collapse supernovae and the time difference between the arrival of the neutrinos and electromagnetic radiation allows us to place upper limits on the neutrinos mass~\cite{1989NYASA.571..601L}.  
Finally, the detection of neutrinos from the blazar TXS~0506+056 in coincidence with a flair seen in gamma rays provides information about the baryon content of the relativistic material in the jet---which photons by themselves cannot constrain---upending the long-held belief that electrons were the dominant source of the gamma rays~\cite{2021arXiv211001687H}.

These examples of the added value from multiple astrophysical messengers are a mouthwatering hors d'oeuvre of the future astronomical observations and the search for BSM physics within the most extreme environments nature can concoct. In the sections which follow we describe several new fields where the complementarity of the information that only multi-messenger astronomy can provide will furnish new probes of BSM physics, namely measurments of the Hubble constant (Section \ref{sec-HubConst}), primordial black holes (Section \ref{sec-PBHs}), dark matter (Section \ref{sec-DM}), and Lorentz invariance violation (Section \ref{sec-LIV}).


\section{Hubble Constant Measurements}\label{sec-HubConst}

\noindent
 \chapterauthor[1]{Michael W. Coughlin}\orcidlink{0000-0002-8262-2924}
  \chapterauthor[2]{Sarah Antier}\orcidlink{0000-0002-7686-3334}
   \chapterauthor[3]{Mattia Bulla}\orcidlink{0000-0002-8255-5127}
    \chapterauthor[4,5]{Tim Dietrich}\orcidlink{0000-0003-2374-307X}
 \\
 \begin{affils}
    \chapteraffil[1]{School of Physics and Astronomy, University of Minnesota, Minneapolis, MN, 55455, USA}
    \chapteraffil[2]{ARTEMIS UMR 7250 UCA CNRS OCA, Boulevard de l'Observatoire, CS 34229, 06304 Nice CEDEX 04, France}
    \chapteraffil[3]{The Oskar Klein Centre, Department of Astronomy, Stockholm University, AlbaNova, SE-106 91 Stockholm, Sweden}
    \chapteraffil[4]{Institute of Physics and Astronomy, University of Potsdam, Karl-Liebknecht-Str. 24/25, 14476, Potsdam, Germany}
    \chapteraffil[5]{Max Planck Institute for Gravitational Physics (Albert Einstein Institute), Am M\"uhlenberg 1, D-14476 Potsdam, Germany}
 \end{affils}
 

 It has been known for many decades that the multi-messenger analysis of compact binary systems provides an additional pathway to measuring $H_0$~\cite{Sch1986} beyond cosmic microwave background~\cite{2016A&A...594A..13P} and type Ia supernovae~\cite{RiCa2019} measurements.
Using gravitational waves emitted from compact binary mergers, to measure the expansion rate of the universe is particularly appealing since, unlike the other analyses, this measurement does not rely on a cosmic distance ladder or assumes any cosmological model as a prior; except for assuming that general relativity is correct. The combination of the distance measurement via gravitational waves and the redshift from the electromagnetic counterpart makes tight constraints on $H_0$ possible.

This approach was vitalized by GW170817 and its electromagnetic counterpart AT2017gfo, with an $H_0$ measurement provided by many teams with various levels of assumptions. In the following, we include the variety of ``flavors'' possible with kilonova-based $H_0$ measurements.

\paragraph*{Gravitational waves as standard sirens}

One direct measurement with the fewest modeling assumptions entails the use of gravitational waves to measure the distance and the host galaxy of the electromagnetic counterpart to measure the redshift. 
GW170817 led to a $H_0$ measurement of $H_0=70^{+12}_{-8}\,$km/s/Mpc in the case of GW170817~\cite{AbEA2019b}. This measurement is predominantly limited by the uncertainty on the distance measurement due to a large degeneracy between the luminosity distance and inclination angle of the gravitational waves signal. Based on this, it has been estimated that $\sim\,$50--100~GW events with identified optical counterparts would be required to have a $H_0$ precision measurement of $\sim\,$2\%~\cite{ChFa2017}. 

\paragraph*{Constraining the inclination angle using an associated gamma-ray burst}

In addition to the observation of AT2017gfo, GW170817 was associated with sGRB170817A, which proved that at least some of the observed sGRBs originate from the merger of compact binaries (Section \ref{sec-NS-NS}). 
In fact, GRBs are known to be produced by internal shocks during the propagation of a highly relativistic jet powered by a compact central engine, which emits gamma rays and hard X-rays~\cite{WiRe1997,MeRe1998}. The GRB is then followed by an afterglow visible in X-rays, optical, and radio for hours to months after the initial prompt gamma-ray emission created by the interaction of the jet with the external medium~\cite{SaPiNa1998}.
The observation and modelling of the GRB afterglow provide constraints on the inclination angle of the system and help to break the distance-inclination degeneracy of the gravitational-wave signal.  This technique has been applied to the sGRB170817A afterglow and obtained $H_0=75.5^{+14.0}_{-7.3}$ \cite{Guidorzi:2017ogy}. 

\paragraph*{Constraining the inclination angle using the superluminal motion}

The resulting $H_0$ measurements can be further improved with, for example, high angular resolution imaging of the radio counterpart. 
The measurements of the observing angle depend on the fact that both, the measured superluminal motion and the observed light curve depend on the jet opening angle as well as the angle between the observer and the jet.
Hence, measurements of the superluminal motion of gravitational-wave counterparts are a potential channel for improving the inclination angle constraints~\cite{MoDe2018}. This technique has been applied for GW170817 and obtained $H_0=68.9^{+4.7}_{-4.6}$ km/s/Mpc \cite{HoNa2018}.

\paragraph*{Constraining the inclination angle using the kilonova}

Kilonovae, which are the byproduct of r-process nucleosynthesis in binary neutron star
mergers (Section \ref{sec-NS-NS}), produce light curves which depend on the viewing angle, which implies the possibility to constrain the inclination further. Significant theoretical modeling prior to and after GW170817 has made it possible to study AT2017gfo in great detail, including measurements of the masses, velocities, and compositions of the different ejecta types.
These measurements rely on models employing both simplified semi-analytical descriptions of the observational signatures (e.g., \cite{Metzger:2016pju}) and modeling using full-radiative transfer simulations (e.g., \cite{KaMe2017,Kaw2018}). This technique has been applied for GW170817 using full-radiative transfer simulations~\cite{Bulla:2019muo} and obtained $H_0=72.4^{+7.9}_{-7.3}$~km/s/Mpc \cite{HoNa2018}.

\paragraph*{Using kilonovae as standardizable candles}

Given that the underlying physical processes triggering the kilonova are universal, it is possible to make an $H_0$ measurement using only kilonovae~\cite{CoDi2019c}.
This approach uses techniques borrowed from the type-Ia supernova community to measure distance moduli based on kilonova light curves using known dependencies of the modeled light curves on the ejecta mass, ejecta velocity, and lanthanide fraction.
With this technique, models for the intrinsic luminosity of kilonovae based on observables, such that the light curves become ``standardizable'', such as standard candles. 
This technique was used to constrain $H_0 = 85^{+22}_{-17}\,$km $\mathrm{s}^{-1}$ $\mathrm{Mpc}^{-1}$ and $H_0 = 79^{+23}_{-15}\,$km $\mathrm{s}^{-1}$ $\mathrm{Mpc}^{-1}$ employing two different kilonova models \cite{CoDi2019c}.

\paragraph*{Joint multi-messenger analyses}

The statistical framework for performing joint standard candle-standard siren measurements using gravitational waves, electromagnetic follow-up data, and simulations of electromagnetic counterparts is summarized in Reference \citenum{Doc2019}.
Bayesian analyses joining the above measurements of GW170817, AT2017gfo, and GRB170817A improve on what is possible analyzing the objects independently. 
Using this technique, a Hubble constant measurement of {$66.2^{+4.4}_{-4.2}\ \rm km \,Mpc^{-1}\, s^{-1}$} is reported at $1\sigma$ uncertainty \cite{DiCo2020}.






\section{Primordial Black Holes}
\label{sec-PBHs}
\noindent
 \chapterauthor[1,2]{Kristi Engel}\orcidlink{0000-0001-5737-1820}
 \chapterauthor[2,3]{J. Patrick Harding}\orcidlink{0000-0001-9844-2648}
 \chapterauthor[3]{Alison Peisker}
 \\
 \begin{affils}
    \chapteraffil[1]{University of Maryland, College Park, College Park, MD 20742, USA}
    \chapteraffil[2]{Physics Division, Los Alamos National Laboratory, Los Alamos, NM, 87545, USA}
    \chapteraffil[3]{Michigan State University, East Lansing, MI, 48824, USA}
 \end{affils}
 
\begin{figure}[htb]
\centering
	\includegraphics[width=0.8\textwidth]{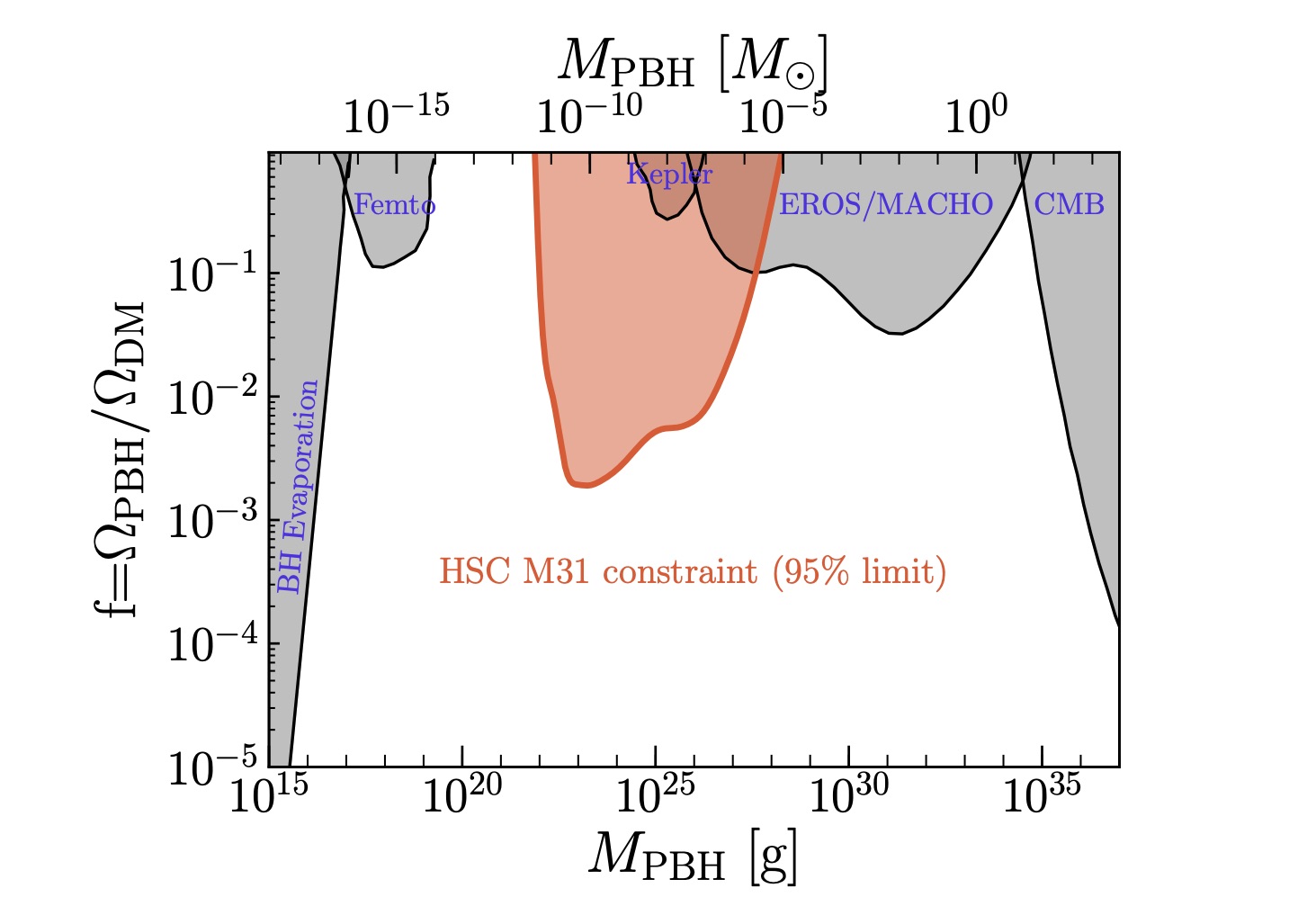}
	\caption{Proposed dark matter fraction with respect to Primordial Black Holes. The potential fraction of dark matter PBHs might consitute, $f_{\mathrm{PBH}}$, is shown relative to PBH mass, M$_{\mathrm{PBH}}$. Some of the strongest constraints can be placed using evaporation signals from PBHs. Plot is Figure~5 from Ref.~\citenum{Niikura:2017zjd}.}	
	\label{fig:pbh-dm}
\end{figure}

 
While there are no known processes in the current Universe that can create black holes with masses less than $\sim 1~ M_{\mathrm{Sun}}$, the chaotic conditions in the early Universe were conducive to the formation of black holes with masses ranging from the Planck mass to supermassive black holes~\cite{Carr:2009jm}. These windows into the first moments of our Universe's environment are called Primordial Black Holes (PBHs). PBH production in the early Universe would have broad observable consequences spanning the largest distance scales---including influencing the development of large-scale structure in the Universe and the primordial power spectrum~\cite{Silk2000, Carr:2019yxo, Emami:2017fiy}---to the smallest scales--- e.g., enhancing local dark-matter clustering~\cite{Clesse:2016vqa}. The detection of PBHs would drastically constrain our understanding of the physics of the early universe. Even just this monumental reward motivates the search for signs of PBHs across the multimessenger landscape, such as gravitational wave detection and gamma-ray and neutrino signatures of PBH evaporation. In the present Universe, PBHs in certain mass ranges may also constitute a non-negligible fraction of dark matter~\cite{Carr:2009jm, Carr:2016drx}. Since the existence of stellar-mass black holes was recently confirmed during the first observational run of Advanced LIGO~\cite{Abbott2017}, there has been a resurgence in support for a PBH component of the total dark matter energy density (e.g., Refs.~\citenum{Niikura:2017zjd,Garc_a_Bellido_2017, Tada:2019amh, Carr:2019yxo}; see Figure~\ref{fig:pbh-dm}).\\ 

\noindent\textbf{Detection of Evaporation Particle Signatures from PBHs: }
The prediction that a black hole will thermally radiate (evaporate) with a blackbody temperature inversely proportional to its mass was first calculated by Hawking by using a convolution of quantum field theory, General Relativity, and thermodynamics~\cite{Hawking1974}. The emitted radiation consists of all fundamental particles with masses less than $\sim$T$_\mathrm{BH}$~\cite{MacGibbon1990}. For black holes in the stellar mass range and above, Hawking radiation is nearly negligible. However, for lower-mass PBHs, this process dominates their evolution over time~\cite{Glicenstein:2013vha}. PBHs with initial masses of $\sim$10$^{14}$--10$^{15}$~g should be expiring today producing short bursts lasting a few seconds of high-energy radiation in the GeV--TeV energy range~\cite{MacGibbon2008, Ukwatta:2015tza}, making their final moments an ideal phenomenon to observe with current space-based gamma-ray telescopes (e.g., \textit{Fermi}-LAT~\cite{FermiPBH1,FermiPBH2}; Section~\ref{sec-currentGammaRay}), next-generation space-based gamma-ray telescopes (e.g., AMEGO~\cite{2020SPIE11444E..31K}; Section~\ref{sec-nextGenGammaRay}), current neutrino observatories (e.g., IceCube~\cite{Dave:2019epr}), current ground-based gamma-ray telescopes (e.g., HAWC~\cite{HAWCPBH}, H.E.S.S.~\cite{Tavernier2021HESSPBH}, and VERITAS~\cite{Archambault:2017asc}; Section~\ref{sec-curGB-GRT}), and future ground-based gamma-ray telescopes (e.g., SWGO~\cite{SWGOPBH}, CTA~\cite{CTAConsortium:2018tzg}; Section~\ref{sec-futGB-GRT}). While this mass regime is not currently a candidate for PBHs as dark matter (as shown by Figure~\ref{fig:pbh-dm}), confirmation of a PBH signal from any size would lend significant credence to that dark-matter model.\\ 

\noindent\textbf{Detection of Gravitational Wave Signatures from PBHs: }
Gravitational wave (GW) signals offer another, incredibly promising tool in the search for PBHs. Should a GW signal (be it from a merger or from a stochastic GW background) be detected where standard black hole formation channels are not present would be unambiguous support for the existence of PBHs. However, any GW merger event could involve a PBH as one of the progenitors (e.g., Refs.~\citenum{Takhistov:2017bpt,Tsai:2020hpi}), thus requiring a statistical study of the black hole merger population to distinguish between standard, astrophysical black holes and PBHs. Thankfully, this kind of analysis would be low-cost, as the raw data needed to perform such a study is already being gathered by current ground-based GW interferometers (e.g., LIGO~\cite{Harry:2010zz}, Virgo~\cite{VIRGO:2014yos}, KAGRA~\cite{Aso:2013eba}; Section~\ref{sec-currentGW}) and will likewise be as easily undertaken by third-generation ground-based GW interferometers (e.g., Advanced LIGO, Advanced Virgo; Section~\ref{sec-3rdGenGW}) and planned space-based GW detectors (e.g., LISA~\cite{Danzmann_1996}; Section~\ref{sec-spacedBasedGW}). The dedicated analysis requirements would only be to turn the data into population constraints.


\section{Dark Matter Detection}
\label{sec-DM}

\noindent
 \chapterauthor[]{Tracy R. Slatyer} \orcidlink{0000-0001-9699-9047}
 \\
 \begin{affils}
    \chapteraffil[]{Center for Theoretical Physics, Massachusetts Institute of Technology, Cambridge, MA, USA}
 \end{affils}
 
 The nature of dark matter is one of the great fundamental puzzles of particle physics and cosmology. If dark matter is some new particle (or an ensemble of new particles), its annihilations and decays could produce visible particles over a wide range of energy scales, which subsequently decay producing a range of visible secondary particles. There are long-standing searches for such signals in photons, cosmic rays, and neutrinos, and future experiments offer the prospect of significantly improved sensitivity. The Snowmass 2021 white paper on ``The Landscape of Cosmic-ray and High-energy Photon Probes of Particle Dark Matter" \cite{SnowmassCF1WP5} discusses the landscape of funded and proposed future probes of gamma-ray and cosmic-ray signals from dark matter, whereas the Snowmass 2021 white paper on ``Cosmogenic Dark Matter and Exotic Particle Searches" \cite{SnowmassNFCosmogenic} discusses neutrino-based indirect searches. Searches in these different channels are highly complementary, as we do not know which (if any) Standard Model particles would be produced by dark matter decay or annihilation. For example, production of quarks and gluons leads to hadronization with subsequent copious production of gamma rays, neutrinos, and (for sufficiently high masses) antiprotons and antinuclei. Production of electrons or muons leads to strong signals in searches for cosmic-ray positrons. Dark matter decaying or annihilating predominantly into neutrinos can be well-constrained by high-energy neutrino telescopes, and at high dark matter masses, also by photon and cosmic-ray searches sensitive to radiation of weak gauge bosons from the neutrinos. Combining constraints from all these channels allows us to avoid blind spots in sensitivity, and probe the lifetime or annihilation rate of dark matter in a way that is applicable to the broadest possible range of scenarios.

Sufficiently heavy dark matter could generically produce signals spanning these channels if it decays or annihilates; the Snowmass 2021 white paper on ``Snowmass2021 Cosmic Frontier: Ultra-heavy Particle Dark Matter" \cite{SnowmassCF1WP8} discusses a broad range of searches for such ultra-heavy dark matter, across a range of messengers and energies. Models of ultra-heavy dark matter may also feature modifications to the early-universe cosmology that simultaneously ensure the correct dark matter abundance and yield interesting gravitational wave signals. 

More generally, the presence of dark matter or dark sectors could have striking effects on gravitational wave signatures from black holes and other compact objects, as discussed in the Snowmass white paper on ``Dark Matter In Extreme Astrophysical Environments" \cite{SnowmassCF3Extreme}. If dark matter itself consists of primordial black holes, as explored in 
the Snowmass white paper on ``Primordial Black Hole Dark Matter" \cite{SnowmassCF3PrimordialBlackHoles}, gravitational waves from mergers may provide a smoking-gun signal for such a population, while electromagnetic signatures could reveal the Hawking radiation of asteroid-mass black holes.

Searches for dark matter often rely critically on an understanding of astrophysical backgrounds (Section \ref{sec-DiffuseBackgrounds}) or systems; poorly understood systematic errors associated with multimessenger astrophysics can be the major limiting factor for sensitivity to dark matter signals. The Snowmass white paper on ``Synergies Between Dark Matter Searches and Multiwavelength/Multimessenger Astrophysics" \cite{SnowmassCF1WP7} describes a range of areas where improvements in our understanding of the relevant astrophysics may yield significant dividends in sensitivity to dark matter. Examples include the characterization of astrophysical neutrino fluxes as a background for direct-detection experiments, and the use of cosmic-ray measurements to inform background modeling for dark matter searches in both gamma rays and cosmic rays.

Especially in the event of a possible detection of dark matter in an astrophysical data set, searches for multimessenger counterpart signals will be crucial in determining whether the apparent detection is truly associated with dark matter, and if so, the properties of the dark matter. The Snowmass 2021 white paper on ``Puzzling Excesses in Dark Matter Searches and How to Resolve Them" \cite{SnowmassCF1WP6} discusses several {\it current} examples of such possible signals, which have been observed in dark matter searches but are not yet fully understood (and may reflect poorly-understood backgrounds rather than true signals). Multimessenger observations and combined analyses can play an important role in conclusively resolving these puzzles, both by improving our understanding of relevant backgrounds and by identifying (or excluding) predicted counterpart signals. As one example, if the Galactic Center excess (Section \ref{sec-GammaBackground}) detected in GeV-scale gamma rays (Section \ref{sec-currentGammaRay}) originates from dark matter annihilation, counterpart signals would be expected in cosmic-ray antiprotons, antideuterons, and/or positrons; on the other hand, if it has a non-dark-matter origin in a new population of pulsars (Section \ref{sec-pulsars}), that origin could be confirmed by photon searches at radio, X-ray and gamma-ray frequencies and (in the future) observations of the Galactic stochastic gravitational wave background (Section \ref{sec-GWbackground}).

\clearpage
\section{Lorentz Invariance Violation}
\label{sec-LIV}
\noindent
 \chapterauthor[1,2]{Kristi Engel}\orcidlink{0000-0001-5737-1820}
 \chapterauthor[2,3]{J. Patrick Harding}\orcidlink{0000-0001-9844-2648}
 \chapterauthor[4]{Humberto Martínez-Huerta}\orcidlink{0000-0001-7714-0704}
 \\
 \begin{affils}
    \chapteraffil[1]{University of Maryland, College Park, College Park, MD 20742, USA}
    \chapteraffil[2]{Los Alamos National Laboratory, Los Alamos, NM, 87545, USA}
    \chapteraffil[3]{Michigan State University, East Lansing, MI, 48824, USA}
    \chapteraffil[4]{Universidad de Monterrey, San Pedro Garza Garc\'ia NL, 66238, Mexico}
 \end{affils}
 
Precise measurements of very-high-energy photons can be used to test the Lorentz symmetry~\cite{LHAASO:2021opi,HAWC:2019gui,Martinez-Huerta:2016azo,Vasileiou:2013vra}. From the point of view of quantum field theory, the Lorentz invariance is one of the main symmetries that govern the Standard Model of elementary particles--- the idea that the laws of physics are the same for all observers. However, proposed grand unified theories combing the fundamental sources so far suggest that our understanding of space-time is incomplete and that fundamental modifications to the Lorentz symmetry must be made to account for quantum effects~\cite{ADDAZI2022103948}. Therefore, some Lorentz Invariance Violation (LIV) at a high enough energy scale is both motivated and expected as a possible consequence of theories beyond the Standard Model, such as quantum gravity or string theory~\cite{ALFARO, QG1,Bluhm:2013mu, Calcagni:2016zqv,Colladay:1998fq,QG4,QG5, Kostelecky:1988zi,NAMBU, Pot}.

One consequence of LIV is that photons of sufficient energy are unstable and decay over short timescales~\cite{Martinez-Huerta:2016azo}. This means that high-energy photons from astrophysical objects cannot travel far from their source, with the photons decaying well before they can arrive at Earth. In the photon sector, LIV is usually parameterized as an isotropic correction to the photon dispersion relation: 
\begin{equation}\label{LIV:eq1}
E_{\gamma}^2 - p_{\gamma}^2 = \pm \frac{E_{\gamma}^{n+2}}{(E_{LIV}^{(n)})^n}\enspace,
\end{equation}
where $E_\gamma$, $p_\gamma$ are the photon energy and momentum, respectively, and $E_{\rm LIV}^{(n)}$ is the LIV energy scale at leading order n, which can be related with the coefficients of the underlying theory or with the Plank or the Quantum Gravity energy scales~\cite{ADDAZI2022103948}.

Above a certain photon energy threshold, the superluminal effects in Eq.~(\ref{LIV:eq1}) allow the photon decay, $\gamma \rightarrow e^-e^+$. This process can be so efficient that no photons above the threshold should reach the Earth from astrophysical distances. Hence, a direct limit to $E_{\rm LIV}^{(n)}$ can be established by the observation of high-energy photons with energy $E_\gamma$, given by, 
\begin{equation}
    E_{\rm LIV}^{(n)} > E_{\gamma} \left[ \frac{E_{\gamma}^2 - 4m_{e^-}^2}{4m_{e^-}^2}\right]^{1/n}\enspace.
\end{equation}
Constraints to the LIV energy scale have been established by looking at the highest-energy photons from the Crab nebula, eHWC J1825-134, and LHAASO J2032+4102~\cite{LHAASO:2021opi,HAWC:2019gui}. However, higher limits are expected from continued observations of even more high-energy sources, such as RXJ1713.7-3946, with upcoming observatories including the Cherenkov Telescope Array (CTA~\cite{CTAConsortium:2018tzg}) and the Southern Wide-field Gamma-ray Observatory (SWGO~\citenum{Albert:2019afb,Hinton:2021rvp,Schoorlemmer:2019gee}; see Section~\ref{sec-SWGO}).

The higher the energy of a detected gamma ray and the narrower its energy uncertainty, the more stringent the constraints on $E_{\rm LIV}^{(n)}$ would be. Thus, instruments optimized at the highest energies, such as SWGO, LHAASO~\cite{2019arXiv190502773B}, and CTA would be optimal instruments to search for LIV signatures. 

\begin{figure}[ht]
    \includegraphics[width=0.5\textwidth]{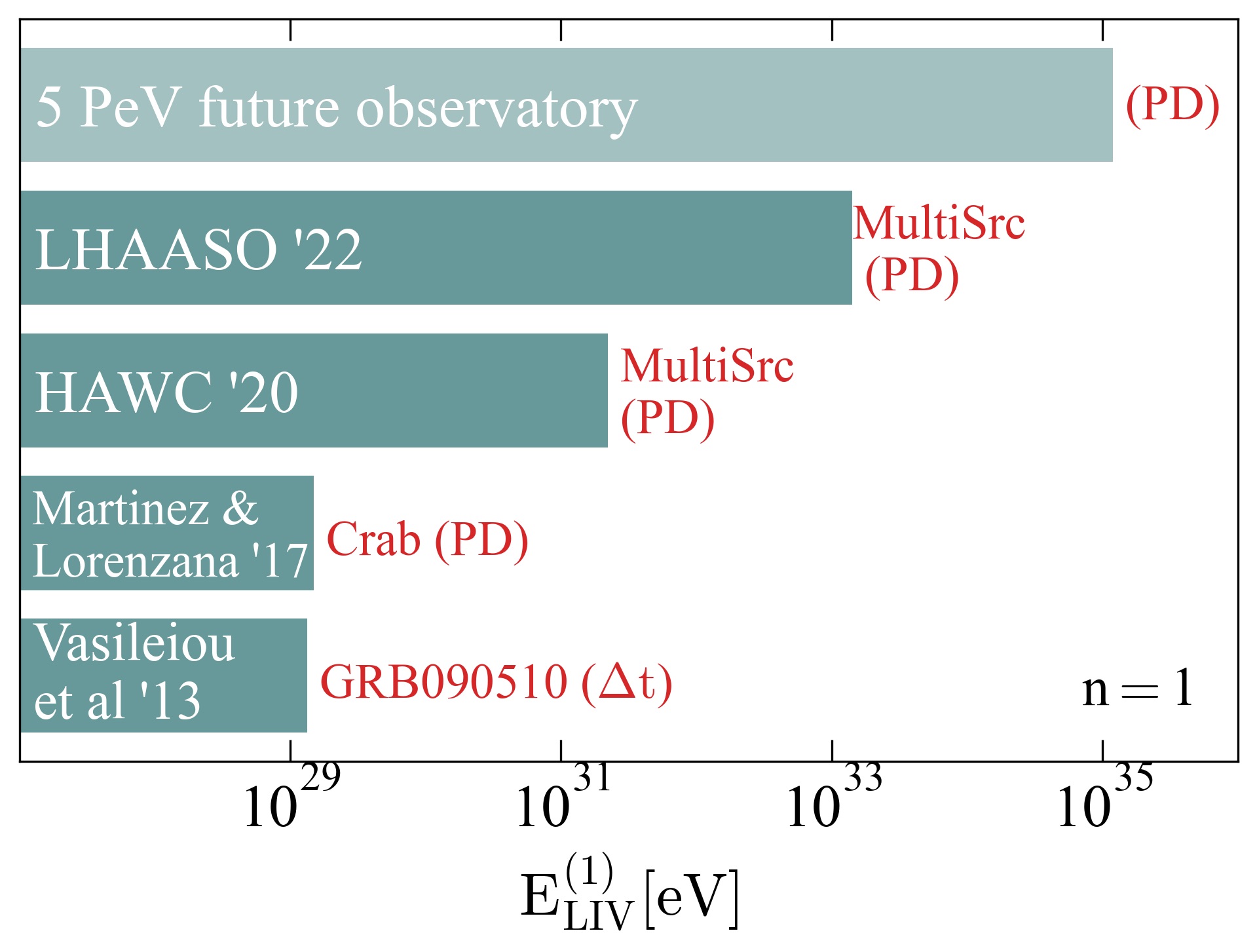}\includegraphics[width=0.5\textwidth]{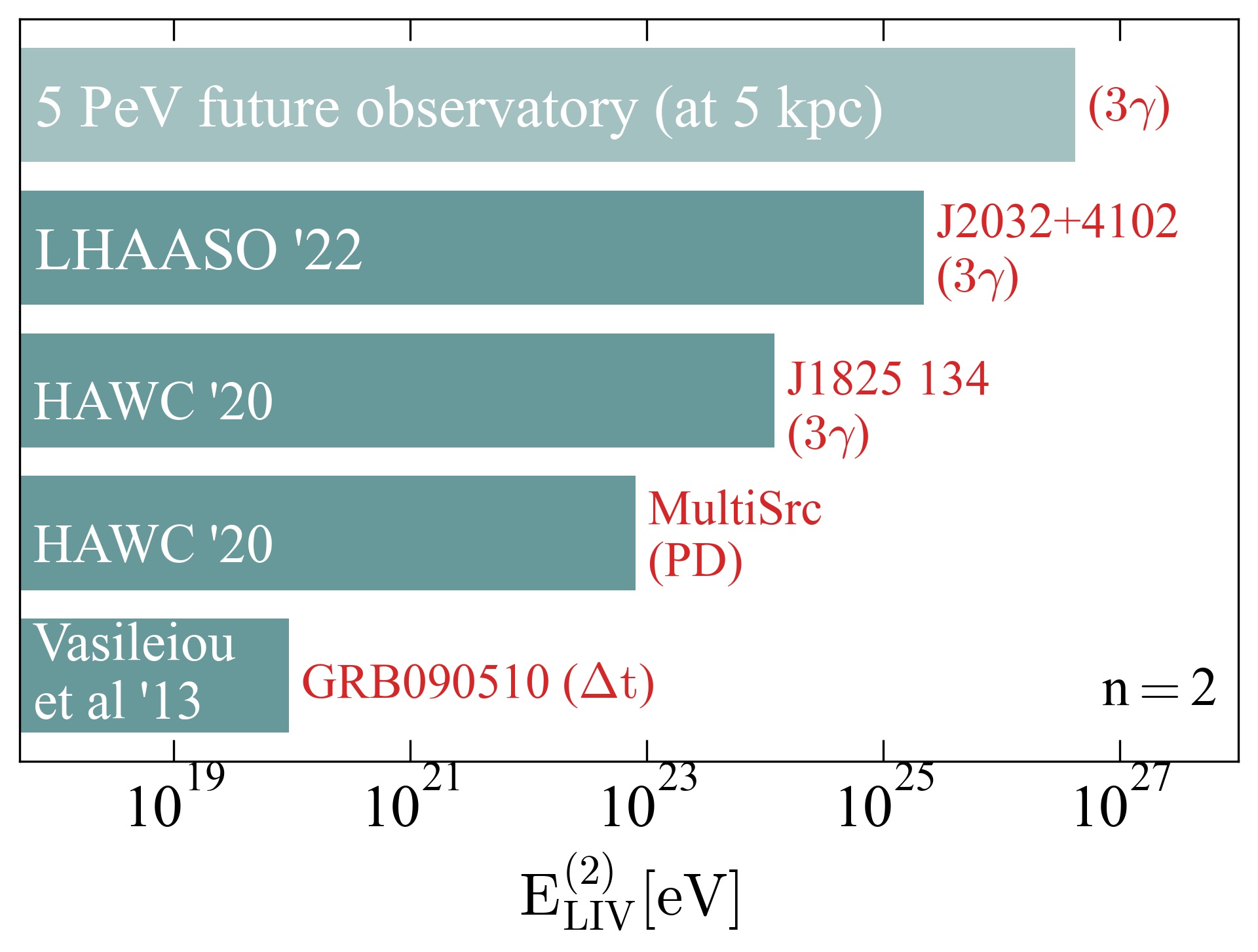}
    \caption{The strongest $E_{\rm LIV}$ limits from LIV searches with energy-dependent time delays ($\Delta$t), photon splitting into 3 photons ($3\gamma$),
    and photon decay (PD) into electron-positron pairs (from bottom to top, see Refs.~\citenum{LHAASO:2021opi,HAWC:2019gui,Martinez-Huerta:2016azo,Vasileiou:2013vra}). Results are shown both for leading CPT-violating (n~=~1) and CPT-conserving (n~=~2) LIV terms. The leading limits from LHAASO are based on constraining photons above 1~PeV in a Galactic source. If a future observatory were able to improve the photon constraints from such sources to energies \textgreater5~PeV, these limits would be further improved by orders of magnitude (shown in light teal at the top). }
    \label{fig:LIV}
\end{figure}





\chapter{Multimessenger Synergies in Particle Astrophysics}
\label{sec-astrophysics}

Natural particle accelerators invite us to hone our theories and discover new physics if only we commit to recording them. Gamma-rays bring us information from the reaches of extragalactic space with minimal interaction or loss of information. Together with neutrinos and gravitational waves they send signatures of specific types and rates of particle acceleration in active galactic nuclei, compact binaries, stellar dances, deaths, and diffuse backgrounds. Multimessenger astrophysics is at the core of the most fundamental physical questions of our time, in laboratories already set up (but irreproducible on Earth). The reason to prefer a laboratory setting is to control the environment. In astrophysics, we are not the creators, only the observers, and as such, it is our task to understand the experimental setup in order to extract the science. This chapter focuses on the astrophysical objects that form the experimental setup for fundamental physics with cosmic multimessenger sources. 




\section{Active Galactic Nuclei}
\label{sec-astroAGN}

Active galaxies contain an actively accreting supermassive black hole. This active galactic nucleus (AGN) is by definition at least 100 times brighter than all of the starlight in that galaxy combined. AGN are the most numerous extragalactic source in X-rays and gamma-rays and the energy transfer from their rotational momentum to their environment is significant in the energy budget of the universe. AGN, both jetted and unjetted seem likely sources of astrophysical neutrinos, suggesting that important discoveries in particle astrophysics may be made in winds or coronae in addition to the jet environment.

Astronomers and physicists have spent decades understanding the nature of the host galaxies and the extreme environment at the core of AGN, but questions remain about how they accelerate particles, especially in the case of blazars, which are the most powerful sustained source in the universe. These are the largest particle accelerators, operating at the highest energies, but to understand their messages about fundamental physics, we first need to understand their ``experimental setup" which means gaining an understanding of their composition, mechanics, and components. For example, the multimessenger blazar TXS 0506+056 has a jet probably composed of electrons and protons, which are accelerated at different rates due to their mass discrepancy. Understanding the particles and how they accelerate is important to understanding how the neutrinos are produced, which is a fundamentally multimessenger question with broad implications for fundamental particle physics. 

\subsection{Particle Acceleration in Jetted Active Galactic Nuclei}
\label{sec-jettedAGN}

\noindent
 \chapterauthor[1,2]{Haocheng Zhang}\orcidlink{0000-0001-9826-1759}
 \\
 \begin{affils}
    \chapteraffil[1]{NASA Postdoctoral Program Fellow}
    \chapteraffil[2]{NASA Goddard Space Flight Center\\
Greenbelt, MD 20771, USA}
 \end{affils}
 
 Relativistic jets from active galactic nuclei (AGN) are among the most powerful cosmic particle accelerators. They are collimated plasma outflows powered by strong accretion onto the supermassive black hole at the center of the AGN \cite{Blandford1977,Li2006,Lyubarsky2010,Tchekhovskoy2016}. These jets exhibit highly variable emission across the entire electromagnetic spectrum, from radio up to TeV $\gamma$-rays. Their emission is nonthermal-dominated, with variability time scales as short as a few minutes \cite{Ackermann2016,Aharonian2007,Albert2007}, indicating extreme particle acceleration in very localized regions. Blazars, which are relativistic jets pointing very close to our line of sight, are the most numerous extragalactic gamma-ray sources \cite{Abdollahi2020}. Recently, the detection of a very high energy neutrino event by IceCube in coincidence with the blazar TXS~0506+056 gamma-ray flare by {\it Fermi} strongly suggests that AGN jets can be sources of extragalactic cosmic rays and neutrinos \cite{IceCube:2018dnn}. This discovery puts AGN jets at the center of the multi-messenger astronomy, which will be one of the most important and fruitful research field in the next decade. AGN jets thus can be ideal cosmic laboratories for particle acceleration and interactions under extreme physical conditions.

It is generally believed that AGN jets are launched as magnetically dominated plasma outflows at the central engine. However, the particle acceleration mechanism that leads to the variable multi-wavelength emission remains not well understood. Shocks in relativistic jets can efficiently accelerate particles via the diffusive shock acceleration (DSA) mechanism if the emission region is kinetically dominated \cite{Achterberg2001,Kirk2000,Spitkovsky2008,Summerlin2012}. This scenario can generally explain the observed multi-wavelength spectra and variability \cite{Marscher1985,Joshi2007,Spada2001,Chen2014,Zhang2016,Larionov2013}, but it requires that the jet quickly dissipates its initial magnetic energy for bulk acceleration of the jet, in which lacks solid theoretical and observational evidence so far. If the jet remains highly magnetized in the multi-wavelength emission region, then magnetic instabilities, in particular, magnetic reconnection can be the primary driver for particle acceleration. This plasma physical process can dissipate the magnetic energy between two oppositely directed magnetic field lines that come too close to each other. Such conditions are likely present in kink-unstable jets or striped jets \cite{Begelman1998,Giannios2006,Giannios2019}. Recent particle-in-cell simulations with both pair and proton-electron plasma reveal efficient acceleration of electrons and protons into power-law distributions, consistent with observations \cite{Guo2014,Guo2016,Sironi2014,Petropoulou2019,Werner2016,Werner2018,Zhang2018,Zhang2021,Hosking2020,Christie2019,Melzani2014,Cerutti2012}. Additionally, turbulence can widely exist in magnetized jets. Recent numerical studies have shown acceleration of nonthermal particles via magnetic reconnection or stochastic scattering due to fluctuations in magnetic fields \cite{Zhdankin2013,Zhdankin2017,Baring2017,Comisso2018}. Turbulence can also explain typical observations \cite{Marscher2014,Guo2017,Macdonald2018,Baring2017}. While the three mechanisms are associated with distinct physical conditions and evolution of the jet, so far they cannot be distinguished by observations. The key issue here is that existing theoretical studies often can only interpret a few events or some aspects of observational data. On the other hand, multi-wavelength monitoring data are not always simultaneous and lack MeV gamma-ray coverage. Key advances in both observations and theories are therefore necessary in the next decade. The Cherenkov Telescope Array (CTA \cite{2019scta.book.....C}; Section \ref{sec-futureAirShower}) will be an essential component to observe emission in the TeV gamma-ray energies, which result from the most energetic particles in AGN jets. MeV gamma-ray telescopes like All-sky Medium Energy Gamma-ray Observatory (AMEGO \cite{McEnery2019}; Section \ref{sec-nextGenGammaRay}) will cover the long-standing gap in the MeV gamma-ray band. Theoretically, although recent combined PIC and radiation transfer simulations can study AGN jet emission under first principles, they are on much smaller physical scales than the realistic jet emission region. On the other hand, magnetohydrodynamic (MHD) simulations lose fundamental particle acceleration and feedback processes. Hybrid simulations with both MHD and PIC are necessary to understand the complex and mutually connected plasma dynamics and particle acceleration in jets. Combined with radiation transfer simulations and detailed analyses of statistical patterns in multi-wavelength radiation and polarization signatures, we can arrive to a consistent physical description. Such simulations will be computationally expensive, thus further supports in high-performance CPU and GPU clusters are required.

The broadband AGN spectrum generally shows a two-hump shape. The first hump from radio to optical, in some cases up to soft X-ray band, is due to synchrotron by ultrarelativistic electrons in a partially ordered magnetic field. This is evident by the observed radio to optical polarization signatures \cite{Marscher2008,Abdo2010,Lyutikov2005,Zhang2015}. In some sources with a bright accretion disk, a thermal component can show up in the optical to ultraviolet bands, often referred to as the big blue bump. The origin of the second hump from X-ray to gamma-ray band is still under debate. The leptonic model suggests that this high-energy hump results from inverse Compton scattering by the same electrons that produce the first hump \cite{Dermer1992,Sikora1994,Boettcher1998,Ghisellini1996,Chen2014}. The hadronic model, however, proposes that if protons can be accelerated to very high energies, they can dominate the high-energy hump via hadronic interactions or proton synchrotron \cite{Mannheim1992,Mucke2001,Cerruti2015,Diltz2015,Petropoulou2015}. In this scenario, the total jet power is typically much higher than the leptonic model, often involving stronger magnetic field and particle acceleration. Additionally, the hadronic model naturally predicts the acceleration of cosmic rays and production of neutrinos. Nonetheless, current theories typically model the broadband spectrum with one-zone stationary models which have serious degeneracy in parameters and cannot distinguish the leptonic and hadronic models \cite{Boettcher2013,Boettcher2019}. Time-dependent simulations including all relevant physical processes are necessary to distinguish the two models. Three aspects in observation are essential to further our knowledge in the next decade. First, high-energy neutrinos associated to jet flaring activities are the smoking-gun of hadronic interactions in jets. So far there was only one such event detected by  IceCube and {\it Fermi} \cite{IceCube:2018dnn}, in which models suggest that the MeV gamma-ray band is the key to understand the neutrino production in jets. Support for the upgrade of IceCube ( IceCube-Gen2 \cite{IceCube2014}) will significantly increase the detection rates of such events. Support for all-sky MeV gamma-ray monitoring telescopes with large effective area and time resolution, such as AMEGO, is crucial for advancing our knowledge on cosmic rays and neutrinos from jets. Second, X-ray and MeV gamma-ray polarization is the other smoking-gun of cosmic rays and neutrinos in jets \cite{Zhang2013,Paliya2018,Zhang2019}. The synchrotron emission by protons and/or hadronic cascading pairs can be as highly polarized as the low-energy hump, which can be detected by future high-energy polarimeters such as IXPE \cite{Weisskopf2018}, AMEGO, and AdEPT \cite{Hunter2014}. In particular, high MeV polarization is a unique signature that the proton synchrotron dominates the high-energy hump, which can probe the acceleration of ultra-high-energy cosmic rays (UHECRs). Finally, hadronic cascades can produce an additional component in the TeV gamma-rays, which can be highly variable as well. Support for CTA will be essential to diagnose this component. Theoretically, time-dependent radiation transfer simulations that include multi-wavelength polarization and neutrinos are critical to understand the particle acceleration and high-energy radiation mechanisms. These simulations should be based on self-consistent physical description of particle acceleration and magnetic field evolution.



\subsection{Plasma Phenomenology in Jetted Active Galaxies}
\label{sec-jettedAGN-plasma}
\noindent
 \chapterauthor[1,2]{Athina Meli}\orcidlink{0000-0002-3064-5307}
 \\
 \begin{affils}
    \chapteraffil[1]{North Carolina Agricultural and Technical State University, Greensboro, NC 27411, USA}
    \chapteraffil[2]{Universite de Liege, 4000 Liege, Belgium}
 \end{affils}

 Plasma is one of the four fundamental states of matter while omnipresent throughout the Cosmos. Astrophysical plasmas in the form of relativistic jets are observed in many astrophysical energetic sources, e.g. pulsars, Gamma-ray Bursts (GRB) and Active Galactic Nuclei (AGN)~\cite[e.g.][]{hawley15, cerutti17a, blandford19}.
 Our understanding of the formation of jets, their interaction with the interstellar and intergalactic medium, and the consequent observable properties such as polarization and spectra from these astrophysical events still remain limited~\cite{macdonald18}.  

Jet outflows are commonly thought to be dynamically hot (relativistic) magnetized plasma flows that are launched, accelerated, and collimated in regions where the Poynting flux dominates over the particle (matter) flux~\cite[e.g.,][]{blandford77,Aloy:1999ai,2017ComAC...4....1P}. This scenario involves a helical, large-scale magnetic field structure in some AGN jets providing a unique signature in the form of observed asymmetries across the jet width, particularly in the polarization~\cite[e.g.][]{Liang81,Aloy:1999ai,clausenBrown11}. Large-scale, ordered magnetic fields have been invoked to explain the launching, acceleration, and collimation of relativistic jets from the central nuclear region of an active galaxy~\cite[e.g.,][]{2009ASPC..402..342M}
or collapsing and merging stars (neutron star and black hole)~\cite[e.g.][]{Piran:2005qu}. 

Despite extensive observational and theoretical investigations, including simulation studies, our understanding of the jet formations, interaction and evolution in an ambient plasma, and consequently their observable properties, such as the time-dependent flux and polarity~\cite[e.g.,][]{MacDonald:2016iwi}), remain quite limited. Also, the magnetic field structure and particle composition of the jets are still not well constrained observationally.

From observations we know that the morphology of relativistic AGN jets is very large and the macroscopic views of jets are described  very well by reduced magnetohydrodynamics (RMHD) simulations~\cite[e.g.,][]{galaxies7010024}). However, the magnetohydrodynamics (MHD) approach is not able to include the dynamics of particles, thus their acceleration in jets cannot be investigated. The approach of Particle-in-Cell simulations in this point play an important role studying the cosmic-ray acceleration and the radiation from accelerated particles from AGN jets.

The associated accretion disk and X-ray emissions observed from a plethora of high energy sources, suggest that there might be combinations of different associated mechanisms. The transfer of the enormous amount of energy transferred from a generating black hole (i.e. core of an AGN) to a jetted plasma can be explained via  two early theories: (i) The Blandford-Znajek process~\cite{blandford77}, which describes how the energy from magnetic fields in relativistic jets is extracted from around an AGN accretion disk by the magnetic fields' dragging and twisting as black hole spins, which as a consequence launches relativistic material by the tightening of the magnetic field lines; and (ii) Punsly and Coroniti~\cite{punsly90} argued that the steady-state solutions of Blandford and Znajek, where the inertia  of plasma particles was completely ignored while their electric charges remained accounted as if in a perfectly conducting medium, were lacking causal connectivity and therefore could not hold in a time-dependent framework. Therefore, as a counter theory, the work of Ref.~\citenum{punsly90} proposed an alternative where the inertia of plasma particles were paramount, which was resembling the theory of the so called ``Penrose-mechanism''~\cite{penrose69}, and where the energy is extracted from a rotating black hole by frame dragging.

Most of the AGN jets are collimated and most of them extend between several thousands up to millions of parsecs, \cite[e.g.,][]{blandford19}. Observations show that jets are symbiotic to the activity of central compact objects in AGN~\cite[e.g.,][]{EventHorizonTelescope:2021iqj}, as well as GRBs~\cite[e.g.,][]{Ruiz:2017due}
and pulsars~\cite[e.g.,][]{hawley15}. The circular polarization (CP; measured as Stokes parameter V) in the radio continuum emission from AGN jets provides a powerful diagnostic for deducing magnetic structure and the jet's particle composition because, unlike linear polarization (LP), CP is expected to remain almost completely unmodified by external screens~\cite[e.g.,][]{OSullivan:2013dvn,MacDonald:2016iwi}. 

Among the highly energetic jetted sources, two of them---the GRBs and blazars (the latter being a class of AGN with a relativistic jet directed nearly towards an observer)---produce the brightest electromagnetic phenomena.
Relativistic jets exhibit a wide range of plasma phenomena, such as generation or decay of magnetic fields, turbulence, magnetic reconnection and propagation in the interstellar or intergalactic medium. In the dynamic environment of jetted sources, it is theorized that particle acceleration occurs via different mechanisms which may be able to achieve the highest level of energies resulting in the observed cosmic-ray spectrum. Especially favourable nowadays is the the magnetic reconnection which takes place in a short time, accelerating rapidly cosmic-rays.

It is important to note that AGN jets interact with the plasma environment of this source, while plasma instabilities and shocks occur along the jet's axis which are responsible for the acceleration of cosmic rays. In these jets MHD instabilities, the kinetic Kelvin-Helmhotz (kKHI) and the kink instability (KI) can additionally operate~\cite{Birkinshaw84,birkinshaw1996instabilities,stone1994numerical} 
contributing to the injection of pre-accelerated cosmic rays. Extensive computational simulations have shown that in magnetized or even unmagnetized jets,  plasma instabilities occur~\cite{universe7110450} such as the known Weibel instability (WI) which mediates relativistic shocks, and which occasionally results in the acceleration of cosmic rays via diffusive and stochastic acceleration mechanisms.

Recent Particle-in-Cell simulations explore the WI, kKHI and MI in slab models of jets and simulation studies focus on the evolution of more realistic jet schemes, like the cylindrical jets in helical magnetic-field geometries~\cite{sironi2013maximum,2015PhRvE..92b1101A,Ardaneh_2016,nishikawa2019,meli21,universe7110450}. 
Note that except Fermi (diffusive) acceleration, events of magnetic reconnection in AGN jets seems to be a viable cause of cosmic-ray acceleration especially in flaring events~ \cite[e.g.,][]{PhysRevLett.95.095001,Oka:2008iw,2011NatPh...7..539D,Kagan_2013,2013AGUFMSM13A2120W,karimabadi2014link,sironi2014relativistic,guo2015particle,guo2016efficient,Guo:2016yfq,Nishikawa_2016,nishikawa2016microscopic,nishikawa2017microscopic,nishikawa2020rapid,meli21}.

\subsection{Unjetted Active Galactic Nuclei}
\label{sec-unjettedAGN}

\noindent
 \chapterauthor[1,2,3]{Yoshiyuki  Inoue}\orcidlink{0000-0002-7272-1136}
 \\
 \begin{affils}
    \chapteraffil[1]{Osaka University, Toyonaka, Osaka 560-0043, Japan}
    \chapteraffil[2]{Interdisciplinary Theoretical \& Mathematical Science Program (iTHEMS), RIKEN, \\ 2-1 Hirosawa, Saitama 351-0198, Japan}
    \chapteraffil[3]{Kavli Institute for the Physics and Mathematics of the Universe (WPI), UTIAS, \\ The University of Tokyo, Kashiwa, Chiba 277-8583, Japan}
 \end{affils}
 
 Because of its tremendous power, relativistic jets of AGNs are one of the most plausible sites in the Universe to generate intense multi-messenger signals, as described in the previous section. Here, however, $\sim90$\% of AGNs are classified as radio-quiet AGNs \citep{Best2005MNRAS.362...25B} which do not possess strong jet activity. Even without strong jets, the deep gravitational potential of the central SMBHs in AGNs can still anchor many other plausible multi-messenger signals production sites such as corona, disk wind, and hot accretion flow. The recent rapid evolution of multi-messenger observational networks has already seen the tip of this iceberg. One example is the detection of a hint of neutrino signals from NGC~1068, a well-known nearby type-2 Seyfert galaxy  \citep{IceCube2020PhRvL.124e1103A}, which are proposed as the coronal origin \citep{Inoue2020ApJ...891L..33I, Murase2020PhRvL.125a1101M, Anchordoqui2021arXiv210212409A}. This subsection outlines our current understandings of expected multi-messenger signals from unjetted AGNs.

Back in the 1980s, to explain X-ray spectra of unjetted AGNs, non-thermal activity at the center of AGNs had been extensively discussed, such as pair cascade models \citep[e.g.,][]{Zdziarski1986, Kazanas1986}. These investigations tossed a coin to unjetted AGNs as cosmic-ray factories \citep{Sikora1987, Begelman1990, Stecker1991PhRvL..66.2697S, Stecker1992}. However, in the 1990s, the detection of the X-ray spectral cutoffs \citep[e.g.,][]{Madejski1995, Zdziarski2000} and non-detection of unjetted AGNs in the gamma-ray band \citep[e.g.,][]{Lin1993} ruled out the pair cascade scenario as a dominant source for X-ray emission. Currently, it is widely believed that Comptonization at a moderately optically-thick hot plasma above an accretion disk, namely corona, predominantly generates the observed X-rays \citep{Katz1976, 1977A&A....59..111B, Pozdniakov1977, Galeev1979, Takahara1979, Sunyaev1980}. Although non-thermal activity in unjetted AGNs is revealed to be minor, neutrino signals from the central AGN regions have been theoretically investigated under the constraints of thermal X-ray observations \citep{Stecker2005, Stecker2013, Kalashev2015, Murase2020PhRvL.125a1101M, Kheirandish2021ApJ...922...45K, Anchordoqui2021arXiv210212409A}. Here, recent millimeter ALMA observations of nearby Seyferts detected non-thermal coronal synchrotron emission \citep{Inoue2018, Inoue2020ApJ...891L..33I}. These observations not only suggest that AGN coronae possess non-thermal activity but also enable us to determine the size and the magnetic field of AGN coronae \citep{Inoue2014}. By combining recent non-thermal millimeter and thermal X-ray observations, multi-messenger signals from AGN coronae have been revisited \citep{Inoue2019ApJ...880...40I, Gutierrez2021A&A...649A..87G}. These models can reproduce the hint of neutrino signals from NGC~1068 \citep{Inoue2020ApJ...891L..33I}. However, several other models are also proposed concurrently, such as the interaction of broad-line-region clouds with accretion disk \citep{Muller2020A&A...636A..92M} and galactic cosmic-ray halo \citep{Recchia2021ApJ...914..135R}. As the hint of signals has already been reported \citep{IceCube2020PhRvL.124e1103A}, future dense and deep multi-messenger observations will be able to elucidate the nature of multi-messenger activity of AGN coronae.

About a half of nearby unjetted AGNs have a disk outflow with a velocity of $\sim0.1c$, so-called ultra-fast outflows (UFOs) \citep{Tombesi2010A&A...521A..57T}. Although the launching mechanism of UFOs is still under debate \citep{Proga2000ApJ...543..686P, Proga2004ApJ...616..688P, Fukumura2010ApJ...715..636F, Fukumura2015ApJ...805...17F, Nomura2016PASJ...68...16N}, strong shock occurs during the interaction of such fast and powerful winds with the interstellar medium of their host galaxies \citep{2012MNRAS.425..605F}. This shock would accelerate high-energy cosmic rays and produce significant multi-messenger signals \citep{Wang2016JCAP...12..012W, Wang2017PhRvD..95f3007W, Liu2018ApJ...858....9L}. Multi-messenger signals are also expected from AGNs in low accretion rate regime, the hot accretion flow, where accretion disk becomes hot and optically thin accretion flow on the contrary to the standard accretion disk where the disk is cool and optically think at relatively high accretion rate \citep{Ichimaru1977ApJ...214..840I, Narayan1994ApJ...428L..13N, Narayan1995ApJ...444..231N, Abramowicz1995ApJ...438L..37A, Yuan2014ARA&A..52..529Y}. Efficient particle acceleration can be operated in this hot accretion flow plasma and could result in significant multi-messenger signals \citep[see, e.g.,][]{Kimura2015ApJ...806..159K}. However, gamma-ray or neutrino signals are not firmly established yet in either wind or low-luminosity AGNs. The next generation multi-messenger observatories would see these classes of objects.

\section{Tidal Disruption Events} 
\label{sec-TDE}

\noindent
 \chapterauthor[]{Robert Stein}\orcidlink{0000-0003-2434-0387}
 \\
 \begin{affils}
    \chapteraffil[]{California Institute of Technology, Pasadena, CA 91125, USA}
 \end{affils}

Tidal Disruption Events (TDEs) occur when stars pass sufficiently close to an supermassive black hole (SMBH), where the tidal force exerted by the SMBH causes the star to disintegrate.
The stellar debris is ultimately accreted onto the SMBH, and generates an accompanying electromagnetic flare. TDEs were first proposed to exist in 1988 \cite{rees_tde_88}, but in the subsequent two decades only eight candidates were found \cite{gezari_21}.
Fortunately, systematic searches for TDEs have been steadily increased the underlying TDE detection efficiency in recent years~\cite[e.g.,][]{van_velzen_21}, with a new TDE now discovered every 3-4 weeks. We are now firmly in the era of TDE population science, for example with the tentative emergence of spectral subclasses \cite{van_velzen_21}, and this provides us with the opportunity to study TDE multi-messenger properties.

TDEs have long been suggested as possible sources of cosmic rays and high-energy neutrinos \cite{farrar_09}, with possible emission zones including relativistic jets, winds or outflows, accretion discs or disc coronae~\cite[see ][for a recent review]{hayasaki_21}. These models can be tested with electromagnetic follow-up observations of high-energy ($>$100 TeV) neutrinos detected by IceCube (Section \ref{sec-iceOceanDetectors}). An optical follow-up program  with the Zwicky Transient Facility (ZTF; see also Section \ref{sec-opticalFollowup}) identified the TDE AT20129dsg as the likely source of neutrino IC191001A \cite{2021NatAs...5..510S}, and candidate TDE AT2019fdr as the likely source of neutrino IC200530A \cite{at2019fdr}. This represents the first direct evidence of multi-messenger TDE emission, and accompanying modelling confirmed that conditions in these sources were consistent with requirements for the detection of a high-energy neutrinos \cite{at2019dsg_winter_21, at2019dsg_murase_20, at2019dsg_liu_21, at2019fdr}. An archival search has since identified AT2019aalc as a third candidate neutrino-TDE, with a combined statistical significance of 3.7$\sigma$ \cite{van_velzen_neutrino_21}. A complimentary probe of TDE emission, searching for archival cross-correlation with neutrinos at $\sim$10 TeV energies, constrained the overall contribution of the TDE population to no more than 39\% of the astrophysical neutrino flux under the assumption of an unbroken E$^{-2.5}$ power law \cite{stein_icrc_19}. Taken together, these results suggests that TDEs contribute a subdominant component of the astrophysical neutrino flux. 

Fully understanding particle acceleration in TDEs, including a precision measurement of the TDE neutrino spectrum, would require many more associations to be found. The advent of more sensitive neutrino telescopes, in particular IceCube-Gen2 (see Section \ref{sec-iceOceanDetectors}), will enable this to be addressed with much much greater statistics. Another avenue of investigation is the search for the predicted gamma-ray emission of TDEs, which will be probed by the Cherenkov Telescope Array (CTA; see Section \ref{sec-futureAirShower}). 

Completing the multi-messenger quartet, TDEs have also been predicted to emit gravitational waves. Any GW emission would however be weak, so a detection would only be expected for a particularly nearby TDE. The probability of detecting a TDE over the lifetime of planned detector LISA has been estimated to be just $\sim$1-10\%, so any detection of TDEs will likely have to wait for even more sensitive generations of GW detectors \cite{pfister_22}. \\

\section{Massive Compact Object Binaries}
\label{sec-astroLargerBinaries}

This section explores the observable signatures and their implications for black holes above $10^5$ solar masses as they interact with similarly sized black holes and also stellar mass objects which may produce electromagnetic signatures in addition to gravitational waves. 

\subsection{Massive Black Hole Binaries}
\label{sec-largerBH-BH}

\noindent
 \chapterauthor[]{Scott C. Noble}\orcidlink{0000-0003-3547-8306}
 \\
 \begin{affils}
    \chapteraffil[]{Gravitational Astrophysics Laboratory, NASA Goddard Space Flight Center, Greenbelt, Maryland 20771, USA}
 \end{affils}
Every year several pairs of massive black holes (MBHs), with masses
$\sim 10^5 \msun - 10^9 \msun$, should merge somewhere in the
universe, leaving behind a still more massive single black hole
\citep{Klein2016,Katz2020}.  These MBHs, originating from galactic
nuclei in separate host galaxies, are brought together through
galactic mergers \citep{2001ApJ...558..535M,2013ARA&A..51..511K},
dynamic friction from stars
\citep{2002ApJ...567..817H,2004ApJ...606..788M,2006astro.ph..1520H,2012ApJ...745...83A,2017PhRvD..95h4032R}
and gas
\citep{escala-04,Dotti:2005kq,Cuadra09,2013CQGra..30x4008M,2015ApJ...810..139H},
and eventually become gravitationally bound to each other to form a
MBH binary (MBHB).  These systems are extremely challenging to
observe, but are of significant interest because they are likely the
most distant gravitational wave sources we can hope to detect, and
complementary photon and gravitational wave data could provide
uniquely powerful diagnostics of these events
\citep{Baker:2019nct,Kelley2019,2019BAAS...51c.112K,2019NatRP...1..585M,Mangiagli2020,2019NewAR..8601525D,2021arXiv210903262B}.
Because of their masses, the gravitational radiation from MBHBs must
be detected using observatories in space \citep{1976ApJ...204L...1T},
which is why they are prime targets for the space-based
ESA-lead/NASA-assisted LISA (Laser Interferometric Space Antenna)
mission \citep{Baker:2019nia} (Section \ref{sec-spacedBasedGW}) and pulsar timing arrays \citep{2021arXiv210110081V} (Section \ref{sec-pulsarArrays}).  In addition, the
consequences of such mergers for galactic evolution are profound,
including strong correlations between the galaxies and the (merged)
central black holes.

At any point in their evolution, the MBHB may be accreting ambient gas
at sufficient rate to launch jets rich with
particle emission, send out powerful winds that may be driven
magnetically or via radiation pressure, and be sufficiently bright and
broadband to be seen at high redshift, just like single AGN
\citep{1980Natur.287..307B} (Section \ref{sec-astroAGN}).  In fact, the confluence of binaries with
galactic mergers may mean that they are more likely to reside in
gaseous environments and have sufficient fuel to ignite activity
\citep{2016ApJ...823..107F,2017ApJ...848..126S,2018PASJ...70S..37G}, though
maybe not \citep{2011ApJ...726...57C,2016ApJ...830..156M}. Therefore, the key
difference between single AGN and binaries is the imprint of the
binary's dynamical gravitational environment on the particle and
electromagnetic emission \citep{2011CQGra..28i4021S}.  MBHBs relevant
to multi-messenger astrophysics, i.e. emitting detectable
gravitational radiation, are not expected to be spatially resolvable
in the foreseeable future due to their necessarily close separations
and likely cosmological distances.  Hence, electromagnetic/particle
identification of MBHB systems requires matching theoretical
expectations to observed phenomena in their light curves, spectra, and
polarization.  Knowing this, astronomers have surveyed the sky looking
for ``smoking gun'' signatures of MBHBs
\citep{2017ApJ...848..126S,2019ApJ...875..117P,Liu2019,Chen2020,2019NewAR..8601525D,Liao2021}.  Prior to merger, the orbiting black
holes may carry their own gas, leading to a multitude of spectral
effects, such as Doppler shifts between narrow-line (circumbinary) and
broad-line (black hole centered) emission
\citep{2012ApJS..201...23E,2014ApJ...789..140L},
such as broad Fe K$\alpha$-line features
\citep{2001A&A...377...17Y,Sesana2012,2013MNRAS.432.1468M,2014AdSpR..54.1448J}.
Binaries of mass-ratio near unity are thought to carve out a cavity in
the accretion flow at 2 to 3 times the binary separation
\citep{MM08,Shi12,Noble12,DOrazio13,Gold14,Farris14b,Miranda2017}, which
distinguishes the outer part of the flow as the circumbinary disk
\citep{MM08}.  Within the cavity, accretion is maintained at the same
rate \citep{Shi15,Miranda2017}, though now via non-axisymmetric
accretion streams stemming from the circumbinary disk to mini-disks
oribiting each black hole
\citep{Cuadra09,Farris14a,Bowen2018,Bowen2019,Paschalidis2021,2021arXiv210901307C},
and at declining yet significant rates as the binary inspirals close
to merger \citep{Noble12}.

Although double-peeked broad-lines were once thought to be possibly
due to the presence of a binary, the consensus is that both peaks
originate from the same central source
\citep{1997ApJ...490..216E,2016ApJ...817...42L}.
A multi-temperature black body spectrum is thought to arise from the
disk-like components of the flow (circumbinary and mini-disk) much
like an AGN; the only difference here is the presence of a ``notch,''
or drop, in the spectral energy distribution power due to less
dissipation---and therefore emission---occuring in the ballistic
accretion streams \citep{RoedigKM14}. Simulations of the thermal
emission confirm the presence of the notch
feature\citep{Farris14b,Farris15,dAscoli2018,Gutierrez2022}, though
not to the same significance as originally predicted. At the same
accretion rates and total black hole mass, the thermal spectrum of a
MBHB resembles that of a single MBH, but with noticeably weaker UV
emission \citep{Gutierrez2022}.

Purported observations of periodic emission from AGN have been
reported
\citep{2015Natur.518...74G,2015ApJ...803L..16L,Charisi2016,2022arXiv220111633J}
(though see \citep{Liu2018,ZhuThrane2020}), and from BL Lac systems
OJ~287 \citep{1988ApJ...325..628S,1996ApJ...460..207L}, PG~1553+113
\citep{2015ApJ...813L..41A,2018ApJ...854...11T}, PKS~2131-021 \citep{ONeill2022} which
are particularly relevant to multi-messenger particle astrophysics
studies of MBHBs as their emission is jet related. OJ 287 has been
observed for more than a century at optical wavelengths and has
maintained a fairly consistent $\sim 12$-year flaring cycle.  Notably,
PG 1553+113 shows signs of quasi-periodic emission at gamma-ray
wavelengths, as well as in the radio and optical bands.  Most of the
models for the BL Lac binary candidates involve the jet launching from
a more massive primary black hole perturbed by a less massive
secondary black hole.  The non-jet binary candidates showing
periodic phenomena can be explained a number of ways, including
modulated accretion from an orbiting overdense feature in the
circumbinary disk called the ``lump''
\citep{MM08,Shi12,Noble12,DOrazio13,Munoz2016,Ragusa2016,Ragusa2017,Miranda2017,Derdzinski2019,Moody2019,Mosta2019,Gutierrez2022},
or Doppler modulation from the orbital motion of the binary 
that may be augmented by strong lensing from the black holes passing
near the line of sight
\citep{2018MNRAS.474.2975D,2020MNRAS.495.4061H,2021MNRAS.508.2524K,2021arXiv211205828D}.
Extensive surveys have turned up few reliable sources
\citep{Liu2019,Chen2020,Liao2021}, though more are expected with the
Vera Rubin observatory.  Searches in the time domain are frustrated by
the fact that AGN typically exhibit red-noise temporal power spectra
and so must be observed for $\sim 5$ cycles to convincingly identify a
periodic signal from the noise \citep{2016MNRAS.461.3145V}.

MBHB mergers are also expected to exhibit novel observational
signatures.  Leading up to merger, environmental plasma may accrete
ordered magnetic field onto the black holes and help establish a
Blanford-Znajek-like outflow, leading to binary jets, Poynting and
synchrotron flux that grows with the increasing orbital velocity of
the binary, and flux that peaks at the time of merger which gives a
clear merger signature \citep{2012PhRvL.109v1102F,2014PhRvD..89f4060G,Farris15,2017PhRvD..96l3003K,2021PhRvD.103j3022C,2022arXiv220208282C}.
However, it is unclear if the Poynting flux is efficiently converted
to EM/particle emission to be observable over other radiative
processes occurring simultaneously in the system.  The merger of
these two jets may induce internal shocks that ultimately generate
high-energy EM and neutrino emission
\citep{2020PhRvD.102h3013Y,2021ApJ...911L..15Y}.  After merger, the
circumbinary disk is expected to heat up via internal shocks arising
from, primarily, the sudden loss of central mass from the radiated
gravitational wave energy and, secondarily, from the kick from the
remnant black hole attaining linear momentum from the merger
\citep{2005ApJ...622L..93M,2008ApJ...684..835S,2010ApJ...714..404T,Robinson:2010ui,Ponce:2011kv}.

\subsection{Intermediate and Extreme Mass Ratio Inspirals}
\label{sec-Inspirals}

\noindent
 \chapterauthor[ ]{Zachary Nasipak}
 \\
 \begin{affils}
    \chapteraffil[ ]{NASA Goddard Space Flight Center}
 \end{affils}
 
 
 Massive black holes (MBHs) can also form binaries within nuclear star clusters by capturing smaller bodies from the surrounding stellar cusp \citep{Amaro07}. If the small body is a compact object---such as a white dwarf (WD), neutron star (NS), or stellar-mass black hole---then it can survive tidal forces near the MBH and slowly inspiral due to gravitational wave (GW) emission. These \textit{extreme-mass-ratio inspirals} (EMRIs) can undergo $\gtrsim 10^4$ orbital cycles before merging with the MBH, producing mHz GW signals that endure for months to even years \cite{Amaro07, Berry19}. This makes EMRIs promising GW sources for future space-based observatories (Section \ref{sec-spacedBasedGW}), such as the Laser Interferometer Space Antenna (LISA) \cite{Amaro17, Baker:2019nia}. Closely related mHz GW sources are \textit{intermediate-mass-ratio inspirals} (IMRIs), which can form between intermediate mass black holes (IMBHs) and stellar compact objects. IMRIs have the exciting potential to be observed by both ground (Sections \ref{sec-currentGW} \& \ref{sec-3rdGenGW}) and space-based detectors (Section \ref{sec-spacedBasedGW}) \cite{Amaro07}.

EMRIs and IMRIs also have the potential to be unique multimessenger sources \cite{Baker:2019nct, Eracleous19}. WDs and Helium-rich stellar cores captured by MBHs or IMBHs can produce electromagnetic (EM) counterparts if the smaller bodies become tidally disrupted and stripped of their mass \cite{Sesana08, Bogdanovic14, MacLeod14}. EMRIs composed of one or more main sequence stars will also experience tidally disruption, leading to EM flares \cite{Rees88, Eracleous19} or, possibly, quasi-periodic eruptions \cite{Metzger21}. Alternatively, a highly eccentric WD may become so tidally compressed during a close periastron passage that it detonates, generating an electromagnetic flare and a potential neutrino flux \cite{MacLeod16, Eracleous19}. However, the GW emission from these latter two scenarios (EMRIs with main sequence stars or highly eccentric WDs) will be so weak that their GW signals will only be observable with next-generation detectors (Section \ref{sec-3rdGenGW}) or if the systems reside in the Local Group \cite{Metzger21, Pfister22}. If a MBH has an accretion disk, then an inspiraling compact object, or even an inspiraling IMBH, can disrupt the surrounding material and alter emission lines from the luminous disk \cite{McKernan13, Baker:2019nct}. EMRIs can also be dual radio and GW sources if the small compact object is a millisecond pulsar. Altogether, observing EMRIs or IMRIs via these multiple windows of the universe will unveil new insights into the nature of MBHs and their surrounding dense stellar environments.

\section{Stellar Mass Compact Object Binaries}
\label{sec-astroStellarBinaries}

This section explores the implications for additional multimessenger observations of stellar mass binary systems and connects with observational methodologies. 



\subsection{Neutron Star-Neutron Star}
\label{sec-NS-NS}

\noindent
 \chapterauthor[1,2,3,4]{Cecilia Chirenti}\orcidlink{0000-0003-2759-1368}
 \\
 \begin{affils}
    \chapteraffil[1]{Department of Astronomy, University of Maryland, College Park, Maryland 20742, USA}
    \chapteraffil[2]{Astroparticle Physics Laboratory, NASA Goddard Space Flight Center, Greenbelt, Maryland 20771, USA}
    \chapteraffil[3]{Center for Research and Exploration in Space Science and Technology, NASA Goddard Space Flight Center, Greenbelt, Maryland 20771, USA}
    \chapteraffil[4]{Center for Mathematics, Computation and Cognition, UFABC, Santo Andr{\`e}, SP 09210-580, Brazil}
 \end{affils}

Binary neutron star systems have been known in our galaxy from radio observations since the discovery of PSR B1913+16 \cite{1975ApJ...195L..51H}. For this reason, they have been considered as guaranteed sources of gravitational waves even before LIGO reached the necessary sensitivity for the first detections \cite{2003Natur.426..531B,2010CQGra..27q3001A}. The observation of binary neutron star mergers in gravitational waves is of direct importance to particle physics, due to the influence of the neutron star equation of state (EOS) in the gravitational waveform \cite{2009PhRvD..79l4033R,2014PhRvL.113i1104T}. The gravitational wave signal of a neutron star-neutron star (NSNS) merger will carry distinct information on the NS EOS during the different stages of the coalescence:

\begin{itemize}
\item Inspiral: both neutron stars can be tidally deformed as they inspiral closer together, causing a dephasing of the gravitational waveform when compared to a binary black hole merger \cite{2010PhRvD..81l3016H}; additionally dynamical tides can be excited as characteristic modes of oscillation of the fluid in the binary components \cite{2017ApJ...837...67C,2019PhRvD.100b1501S}.
\item Merger and post-merger: information on the maximum mass supported by the EOS can be obtained from details of the merger and associated short GRB \cite{2015ApJ...812...24F,2015ApJ...808..186L,2018PhRvD..97b1501R}, and the merger remnant can be characterized by the oscillations in the ringdown (post-merger) waveform \cite{2012PhRvD..86f3001B,2019ApJ...884L..16C}.
\end{itemize}

Constraints on the tidal deformability can be translated to constraints on the radius of the NS \cite{2018PhRvL.120q2703A}, adding to radius estimates from X-ray observations \cite{2016ARA&A..54..401O} such as those from NICER \cite{2019ApJ...887L..24M,2019ApJ...887L..21R} and current theoretical investigations on the NS EOS, as well as recent laboratory results \cite{2016PhR...621..127L}. Next-generation ground-based gravitational wave detectors such as the Einstein Telescope \cite{2010CQGra..27s4002P} and the Cosmic Explorer \cite{2017CQGra..34d4001A} will be needed to explore the wealth of information from the NS modes of oscillation in the 1 -- few kHz range. Alternatively, a dedicated high frequency GW observatory has been proposed, called Neutron Star Extreme Matter Observatory (NEMO) \cite{2020PASA...37...47A}.

So far two NSNS mergers have been reported by LIGO and Virgo: GW170817 \cite{2017PhRvL.119p1101A} and GW190425 \cite{2020ApJ...892L...3A}. The first event inaugurated the era of multimessenger astronomy with gravitational waves, with a nearly coincident short GRB detected by Fermi and INTEGRAL \cite{2017ApJ...848L..13A} and an intensive campaign of follow-up observations from X-rays to radio \cite{2017ApJ...848L..12A}. Unfortunately, the same did not happen with the second event: poorer sky localization made it hard to identify an EM counterpart, which would be in any case fainter due to the larger distance to this source; moreover, detection of the short GRB is serendipitous, since off-axis sources are so much weaker. 

This shows the extraordinary potential for the observations of NSNS mergers. Now, the community is  even better prepared to respond to a similar event. The rates for NSNS mergers are still rather uncertain, but currently estimates are $13 - 1900 {\rm Gpc}^{-3} {\rm yr}^{-1}$ \cite{2021arXiv211103634T}, which could result in few -- tens of NSNS mergers observed within a $160 - 190 {\rm Mpc}$ range during O4.

\subsection{Neutron Star-Black Hole}
\label{sec-NS-BH}

\noindent
 \chapterauthor[]{Francois Foucart}\orcidlink{0000-0003-4617-4738}
 \\
 \begin{affils}
    \chapteraffil[]{Department of Physics and Astronomy, University of New Hampshire, 9 Library Way, Durham New Hampshire 03824, USA}
 \end{affils}

The mixed neutron star-black hole (NSBH) binary mergers are the latest systems observed through gravitational waves by the LIGO/Virgo/KAGRA collaboration. Two NSBH mergers were observed in January 2020 (GW200105, GW200115)~\cite{LIGOScientific:2021qlt}, while more uncertain candidates NSBH mergers were announced in the gravitational wave catalogue GWTC-3~\cite{LIGOScientific:2021djp}. The rate of NSBH merger remains fairly uncertain, $(7.8-140)\,{\rm Gpc^{-3} yr^{-1}}$ \cite{LIGOScientific:2021psn}, but sufficient to expect tens of additional observations by current detectors. Next generation ground detectors (Einstein Telescope, Cosmic Explorer) will observed NSBH systems up to cosmological distances, and the closest NSBH mergers with signal-to-noise ratio allowing for high-accuracy measurements of the mass and spin of compact objects, and the properties of the dense matter forming neutron star's cores~\cite{Maggiore:2019uih,Evans:2021gyd} -- at least if sufficiently accurate waveform models are constructed by the time these detectors become operational.

Like NSNS mergers, NSBH mergers have the potential to be powerful multimessenger sources. All NSBH mergers emit gravitational waves. Additionally, in some NSBH mergers the neutron star is tidally disrupted by its black hole companion before being captured by that black hole. Then, neutron rich matter is ejected into the surrounding interstellar medium, enriching the Universe in heavy elements~\cite{1976ApJ...210..549L} and powering optical/infrared transients days to weeks after the merger~\cite{Li:1998bw,Roberts2011,Kasen:2013xka,Tanaka:2013ana} and radio emission month to years after the merger~\cite{Hotokezaka:2016}. Disrupting NSBH mergers may also be the engine behind some short gamma-ray bursts (SGRBs)~\cite{eichler:89,1992ApJ...395L..83N,moch:93,Faber:2006tx,Meszaros2006,2007NJPh....9...17L,Nakar2007,Paschalidis2014,Christie:2019lim}, the associated emission of high-energy particles, and possibly seconds-long x-ray plateaus observed in the afterglow of some SGRBs that may be associated to fallback material in NSBH mergers~\cite{2010MNRAS.402.2771M,Desai:2018rbc}. Non-disrupting NSBH binaries, on the other hand, have gravitational wave signals mostly indistinguishable from black hole binaries~\cite{Lackey2011,Foucart:2013psa}, and are not expected to power bright post-merger electromagnetic signals. They are useful probe of the mass and spin distribution of compact objects, but their electromagnetic emission is likely limited to hard-to-detect and/or weaker pre-merger signals~\cite{TsangEtAl:2012,PaschalidisEtAl:2013,Schnittman:2017nhg}. As a result, the question of whether a neutron star is disrupted or not during a NSBH merger is maybe the most important characteristic of these systems. A low black hole mass, high black hole spin, large neutron star radius, or large orbital eccentricity all favor disruption~\cite{1976ApJ...210..549L,Rosswog:2005su,Faber:2006tx,Etienne:2008re,Foucart:2013psa,kyutoku:2015,Kawaguchi:2016,Foucart:2018rjc}. One of the most interesting aspect of NSBH binaries is that the simple existence of a multimessenger signal already provides us with valuable information about the properties of the merging objects by imposing a simple, well-understood cut on the allowed parameters of the binary~\cite{Pannarale:2010vs,Foucart2012,Foucart:2018rjc}. 

Another important difference between NSNS and NSBH mergers it that a single disrupting NSBH merger likely ejects close to its orbital plane $\sim (0.01-0.1)M_\odot$ of cold, fast, neutron rich matter~\cite{Kawaguchi:2016,Kruger:2020gig}. A comparable amount of hotter, slower, less neutron-rich ejecta is produced during the subsequent disk evolution~\cite{Fernandez2013,Siegel:2017nub,Fernandez:2018kax,Miller:2019dpt}. This differs noticeably from NSNS mergers, for which the first type of ejecta typically has mass $\leq 0.01M_\odot$. As a result, one disrupting NSBH mergers will contribute significantly more to the formation of the heaviest r-process elements than a NSNS merger. Its post-merger electromagnetic emission is also likely to be redder, and to evolve more slowly~\cite{2014ApJ...780...31T,Barbieri:2019bdq,Kawaguchi:2020osi}. As current population models favor volumetric rates for NSBH mergers significantly lower than for NSNS mergers, NSBH mergers however probably contribute less to heavy-element nucleosynthesis than NSNS systems~\cite{Chen:2021fro}.

\subsection{Black Hole-Black Hole}
\label{sec-BH-BH}

\noindent
 \chapterauthor[1,2,3,4]{Cecilia Chirenti}\orcidlink{0000-0003-2759-1368}
 \\
 \begin{affils}
    \chapteraffil[1]{Department of Astronomy, University of Maryland, College Park, Maryland 20742, USA}
    \chapteraffil[2]{Astroparticle Physics Laboratory, NASA Goddard Space Flight Center, Greenbelt, Maryland 20771, USA}
    \chapteraffil[3]{Center for Research and Exploration in Space Science and Technology, NASA Goddard Space Flight Center, Greenbelt, Maryland 20771, USA}
    \chapteraffil[4]{Center for Mathematics, Computation and Cognition, UFABC, Santo Andr{\`e}, SP 09210-580, Brazil}
 \end{affils}

One of the first surprises from the gravitational wave (GW) detections by LIGO was the existence of a whole population of black holes (BHs) with masses of tens of solar masses, higher than the inferred masses of the BHs observed in X-rays \cite{LIGOScientific.116.061102}. The most numerous of the LIGO sources, with nearly one hundred events reported so far, BHBH mergers have created the field of GW astronomy \cite{LIGOScientific:2018mvr,LIGOScientific:2020niy,2021arXiv211103606T}. The continued observation of such events can provide information on stellar evolution and the possible existence of BH mass gaps, and eventually distinguish between different binary formation channels \cite{2020ApJ...896L..44A,Abbott:2020tfl,2021arXiv211103634T}. Additionally, important tests of fundamental physics can be performed by constraining alternative theories of gravity and different models of exotic compact objects (black hole alternatives) \cite{2013LRR....16....9Y,TheLIGOScientific:2016src,2016PhRvD..94h4016C,Berti:2018vdi,Abbott:2020jks}. It is also expected that constraints on possible dark matter candidates, such as axions, can possibly come from future GW detections, also with future GW space detector LISA \cite{2011PhRvD..83d4026A,amaroseoane2017laser}. A GRB detection associated with the first GW detection, GW150914, was claimed, but the lack of other coincident GRB detections indicates that it could have been unrelated \cite{2016ApJ...826L...6C}. Electromagnetic (EM) counterparts of stellar mass BHBH mergers are not physically ruled out, but might require very extraordinary circumstances and even then might not be detectable at realistic distances. The situation is of course expected to be very different for supermassive BHBH mergers, where circumbinary accretion disks may provide interesting EM counterparts.

\section{Other Transients}
\label{sec-OtherTransients}

This section emphasizes the contributions of additional transient phenomena to multimessenger science.

\subsection{Core-Collapse Supernovae and Long Gamma-Ray Bursts}
\label{sec-CCSNeLongGRBs}


\noindent
 \chapterauthor[1]{Christopher L. Fryer}
 \chapterauthor[2]{Eric Burns}
 \\
 \begin{affils}
    \chapteraffil[1]{Los Alamos National Laboratory, Los Alamos, NM, 87545, USA}
    \chapteraffil[2]{Louisiana State University, Baton Rouge, LA, 70803, USA}
 \end{affils}
 
SN 1987A was the first multimessenger transient being detected first in MeV neutrinos and then in optical light. This event greatly set our modern understanding of the engine of core-collapse supernova and forthcoming multimessenger facilities promise greater advancements.
Neutrinos and Gravitational Waves provide the most direct diagnostic of stellar collapse and the supernova engine.  Both provide insight into the progenitor structure and the equation of state~\citep{2017ApJ...851...62K,2020ApJ...898..139W}, the nature of the convection~\citep{2011LRR....14....1F,2021MNRAS.500..696N,2009A&A...496..475M} and the role of rotation~\citep{10.1093/mnrasl/sly008,1997A&A...317..140M,2004ApJ...609..288F}, and the neutrino physics~\citep{2017hsn..book.1605R,2006PhRvD..74l3004D,2010ARNPS..60..569D,2016NCimR..39....1M}.  These two diagnostics are sensitive to different aspects of the engine and its physics:  e.g. gravitational wave signals are particularly sensitive to the rotation whereas neutrino signals probe neutrino physics such as neutrino flavor oscillations.  
But neutrinos and gravitational waves are not the only way to probe the nature of the supernova engine.  To date, the strongest observational constraints on the asymmetries in the supernova engine has been supernova-remnant observations of the distribution of elements produced in the central engine ~\citep{2014Natur.506..339G,2017ApJ...834...19G}.  Observed through the hard X-ray/$\gamma$-rays emitted in nuclear decay, these remnant distributions provide a clean probe of the engine asymmetries. 
Asymmetries in the central engine can be proposed by the velocity distribution of compact remnants (assuming the velocities are produced by asymmetric ejecta from large-scale convection~\citep{1995PhR...256..117H,2004ApJ...601L.175F,2017ApJ...837...84J}).  The relative velocities of neutron star and black hole systems can also help determine whether these kicks are produced through asymmetric ejecta or asymmetric neutrino emission~\citep{2006ApJS..163..335F}.  The compact remnant distribution, measured in a broad range of binaries:  X-ray and pulsar (radio) binaries or gravitational wave merger events, can constrain the growth time of convection~\citep{2012ApJ...749...91F}.  The spin distribution of these remnants further constrains the role of rotation in the explosion~\citep{2020A&A...636A.104B}.
A number of less direct (or more complicated) observations Broader nucleosynthetic yield measurements (in supernova light-curves, supernova remnants and galactic chemical evolution)~\citep{} and prompt supernova emission~\citep{2021MNRAS.508.5766I,2021arXiv211201432B} probe both the stellar structure and explosion properties. The list of diagnostics that contributes to our understanding of the core-collapse engine is immense.  And, by combining all of these diagnostics, we are able to disentangle the many physical effects behind the core-collapse supernova engine.

A rare class of core-collapse supernovae are collapsars: fast rotating core-collapse events allowing for supereddington accretion powering bipolar ultrarelativistic jets that ultimately release long gamma-rays bursts (GRBs). GRBs were thought to be promising sources of UHECRs \citep{vietri1995acceleration,waxman1995cosmological} as their energy-dense jets should always have some level of baryon content \citep[e.g.][]{lei2013hyperaccreting}. Owing to the difficulties in associating UHECR to their origin, searches for associated neutrinos were expected to prove UHECR arise from GRBs, as both particles arise from generic photohadronic production.

Deep searches have never robustly associated these signals with GRBs \citep{abbasi2011limits,aartsen2015search}, suggesting very low baryon loading in GRB jets. Alternatively, it can be explained as the dissipatation radius being far larger than previously thought \citep[e.g.,][]{ICMART_2}. These non-detections led to suggestions that choked \acp{LGRB}, where the jet fails to breakout through the massive star, may be significant sources of neutrinos \citep[e.g.,][]{Meszaros:2001ms,senno2016choked}. Improved high-energy neutrino telescopes will either associated neutrinos to GRBs or continue to advance understanding of particle acceleration in the most extreme regime.

\subsection{Fast Radio Bursts}
\label{sec-FRBs}

\noindent
 \chapterauthor[1]{Elijah Willox}\orcidlink{0000-0002-6623-0277}
 \\
 \begin{affils}
    \chapteraffil[1]{University of Maryland, College Park, College Park, MD 20742, USA}
 \end{affils}
 Fast Radio bursts (FRBs) were first discovered in 2007 \cite{Lorimer:2007qn} and now there have been just under 800 bursts detected by multiple experiments \cite{TNS}. Fast radio bursts (FRBs) are a class of short-duration, high fluence transients in radio wavelengths, with some sources observed to repeat, while others are apparent single-burst events. In response to this newly discovered class of events, many observatories have taken lessons from the history of the GRB field \cite{Zhang2020}. Multiwavelength observations are not only informative, but critical to the identification of FRB sources. Optical follow-ups have already identified the source galaxies of a few of these sources \cite{Nicastro:2021cxs}, and X-ray and gamma ray data is being analyzed to provide more insight into the sources of FRBs. The recent discovery of an FRB from the galactic magnetar SGR1935+2154, with simultaneous hard X-ray emission provides new insight to FRB mechanics, and is evidence of the benefit provided by multi-messenger studies \cite{Kirsten:2020yin}. In the coming years the number of recorded FRBs will increase dramatically and observations at all wavelengths will provide more insights on this still mysterious class of transients.

\subsection{Supernova Remnants}
\label{sec-SNRs}

\noindent
 \chapterauthor[1]{Miroslav D. Filipovi\'c}\orcidlink{0000-0002-4990-9288}
 \chapterauthor[2]{Robert Brose}\orcidlink{0000-0002-8312-6930}
 \chapterauthor[1]{Shi Dai}\orcidlink{0000-0002-9618-2499}
 \\
 \begin{affils}
    \chapteraffil[1]{Western Sydney University, Locked Bag 1797, Penrith, NSW 2751, Australia}
    \chapteraffil[2]{Dublin Institute for Advanced Studies, Astronomy \& Astrophysics Section, 31 Fitzwilliam Place, D02 XF86 Dublin 2, Ireland}
 \end{affils}


The expanding shock of stellar ejecta from Supernovae (SNe) explosions sweep up and enrich the surrounding interstellar medium (ISM). This expanding shock front and swept up material is known as an supernova remnant (SNR) and is a strong source of synchrotron emission at radio frequencies. The detection of non-thermal X-ray emission from about a dozen Galactic SNRs confirmed that electrons get efficiently accelerated in these objects \citep{2004A&A...414..545C}. 

SNRs are mainly studied via the photons that emit from radio to gamma-ray energies as any charged particle that they might release get scattered in the ISM and are not back-tractable to their origin. Hence, the study of SNRs is closely connected to the search of the sources of Galactic Cosmic Rays (CRs) -- so protons, heavier nuclei and electrons with energies up to at least $10^{15}\,$eV -- and their detection with space and ground-based detectors at Earth. Theoretically, the acceleration of protons and nuclei via diffusive shock acceleration (first-order Fermi acceleration) at the expanding shock of the SNR, whereby particles are trapped by the magnetic fields and cross over the shock multiple times, gaining energy in the process, was predicted for a long time \cite[e.g.][]{book1,book2}. The acceleration of electrons up to $\approx20\,$TeV in SNRs was known since the 1970s' but recently, the gamma-ray spectra of the Galactic remnants IC443 and W44 revealed a low-energy cutoff, characteristic for gamma-rays originating from decaying neutral Pions that get created by a population of highly-relativistic protons (or nuclei) \citep{Fermi.2013a}. The same process that produces the neutral pions (and subsequently gamma-rays) will also produce charged pions that decay into neutrinos, which constitute the third messenger by which SNRs can be studied. Enormous synergies arise when information from these messengers gets combined.  

There are about a dozen historical SNe observed in our Galaxy over the past 2\,000 years and almost 350 Galactic SNRs known to exist \cite{2019JApA...40...36G}. Previous studies of the Large Magellanic Cloud (LMC) SNRs revealed 71 confirmed objects and 19 candidates \cite{2017ApJS..230....2B,2016A&A...585A.162M,2021MNRAS.504..326M,2021MNRAS.500.2336Y,2021arXiv211100446K} while in somewhat smaller neighbouring Small Magellanic Cloud we found 21 bona fide SNRs with couple of additional candidates \cite{2019A&A...631A.127M}. Also, extensive search for SNRs are performed in the M\,31 and M\,33 \cite{2014SerAJ.189...15G,2018ApJ...855..140L} as well as other nearby galaxies \cite{2021MNRAS.507.6020K}. Finally, we now discovered a possible first intergalactic SNR located in between the Milky Way and the LMC \cite{2022MNRAS.tmp..324F}. 

SNRs heat and ionise the ambient ISM and distribute the chemical elements that were processed in the progenitor’s interior and in the supernova into the ISM. In addition, electrons and nuclei are accelerated in the shock waves to highly relativistic energies and are responsible for a considerable fraction of the energy density in the Universe. The ratio of chemical abundances of the accelerated CRs represents the chemistry of the environment into which the SNRs expand. The $^{22}$Ne/$^{20}$Ne ratio observed in CRs is for instance a factor of 5 higher than in the solar wind \citep{2006NewAR..50..516B}, pointing to a sizeable contribution of CRs being accelerated from win-material of massive stars. At the same time, is the acceleration of heavier nuclei from the material surrounding massive stars affecting the gamma-ray signal that has to be expected at the hadronic low-energy cutoff \citep{2020APh...12302490B, 2020MNRAS.497.3581D}. The precise measurements of nuclei-ratios \cite{2020PhRvL.124u1102A} reveal features in the abundance-ratios that either point to necessary modification of our models for the Galactic CR propagation or at particularities of the acceleration-process itself. For instance, a possible selection-effect on the accelerated CRs at a SNR shock-front \citep{2019ApJ...872..108H} was only detectable based on the AMS-02 spectra but not from direct observations of SNRs via photons. Further, composition-measurements of CRs around $10^{15}\,$eV might point to a different spectral behaviour of protons and heavier nuclei \citep{2019arXiv191003721S}, a finding that needs to be further explored by direct CR-measurements for instance by Tibet AS$\gamma$ \citep{2011AdSpR..47..629T}, LHAASO \citep{2020JPhCS1342a2009C} and SWGO \citep{2019arXiv190208429A}, deeper observations in the electromagnetic spectrum of SNRs and theoretical models that are able to accommodate all these findings. 

Similarly to SNRs, highly relativistic particles have been detected in superbubbles \cite[e.g., 30~Dor~C;][]{2017ApJ...843...61S,2015A&A...573A..73K}, which are interstellar structures created by the combination of stellar winds of massive stars and their supernovae. However, the underlying physics such as particle injection, magnetic field configuration and amplification, and the escape of particles from the shock regions requires further investigation. Magnetic fields in SNRs are most likely a complex mixture of interstellar magnetic fields, relic fields of the progenitor, fields modified and enhanced by turbulence in the shock regions, and fields excited by relativistic particles. Therefore, various new generations of high spatial resolution, high sensitivity, and high spectral resolution multimessenger observations are necessary to address these challenges.

A sub-field of SNR-studies that will extraordinarily benefit from multimessenger efforts is the investigation of very young SNRs as the sources of the highest-energy CRs. While evolved Galactic remnants should produce neutrino-signals, the expected fluxes are too low to be directly measured with current or even next-generation instruments. Further, the expected neutrino-energy is limited by the parent-proton energies of below $\approx20\,$TeV \citep{2020ApJ...894...51A}. However, acceleration theory points to SNRs with ages of less than 20 years expanding in very dense circumstellar environments as potential source of CRs up to $10^{15}\,$eV \citep{2013MNRAS.431..415B}. These objects might be powerful gamma-ray emitters even though a sizeable part of the gamma-ray emission gets absorbed close to the source by $\gamma\gamma$-absorption \citep{2020MNRAS.494.2760C}. However, the neutrinos produced in these objects -- not by the SN explosion itself (link to CC-SN section), but by the particles accelerated in the shock-fronts -- can freely escape the sources and might be detected by the next-generation neutrino facilities like IceCube Gen-II. Further, the proposed Einstein-telescope GW observatory will be able to detect the GW-signal from core-collapse supernovae (CC-SN) at distances up to 4~Mpc. This will help to constrain the explosion-process of CC-SN and be utterly important as a trigger for ground and space-based targeted observations across the electromagnetic spectrum. A precise and early localisation of such events is essential to detect the faint gamma-ray signal that has to be detected days to weeks after the explosion in time. Further constraints on the explosion mechanism impact our understanding of the shock dynamic, that are crucial for the particle acceleration itself.

Similarly, will the LISA mission be sensitive to close-in Galactic white-dwarf (WD) binaries? These systems are one type of progenitors for Type~Ia SN and the remnants that are formed from these explosions. Understanding the properties of Galactic WD binaries will help to put the observations of Galactic SNRs in an appropriate context in understanding the conditions of the explosion, the resulting shock-dynamics and the particle-acceleration that arises there-of. 

For SNRs with a young pulsar born inside, another extremely energetic phenomenon is the so-called pulsar wind nebula (PWN). PWNe are generally believed to be powered by relativistic winds generated by the central pulsar inside the shell of a SNR. They show rich wide-band emissions from radio to infrared and from optical to X-ray and gamma-ray sources\cite{2006ARA&A..44...17G}. PWNe can also be efficient TeV gamma-ray emitters, for example the Crab Nebula is a well-known source of TeV gamma-rays \cite{1989ApJ...342..379W}. Future large ground-based telescopes, such as Cherenkov Telescope Array (CTA) \cite{2019scta.book.....C}, will have the sensitivity to detect gamma-ray emission from PWNe at even higher energies and from a large sample of PWNe. Recently, the detection of UHE photons by LHAASO revealed the possible presence of an additional hadronic component at the highest gamma-ray energies in the emission from pulsars \citep{2021Sci...373..425L, 2022ApJ...925...85H}. This additional component will also produce a neutrino signal that might be detectable with next-generation neutrino experiments and more data from existing facilities. These will allow us to understand the radiation mechanism and probe the magnetic field of PWNe, which can be used in modelling and interpreting other nebular structures.

A nearby (5-100~pc) explosion of SN and it's consequent remnant expansion might have a profound effect on our life and existence. Supernovae (and SNRs) distribute the products of stellar burning, which are the raw materials of life. However, don't stand too close: they adversely affect nearby life-friendly planets, bathing them in high-energy radiation, cosmic rays, neutrinos, gravitational waves and ejecta. 
This is primarily done via their impact on Earth's ozone layer, which is plausibly responsible for the irradiation and destruction of surface sea life, causing mass extinction events. SNe and SNRs emit enough extreme and high energy radiation to strip planetary atmospheres at few tens of parsec-scale distances. New observational data and improved theoretical models of the high energy SNRs in this `extreme Universe' could illuminate the history of life on Earth (and further), and aid the search for potentially life-hosting planets in our Galaxy. The extreme Universe including its SNRs puts bounds on the spaces in which life -- and particularly complex life -- can exist. This leads to a complementary approach to the study of life in the Universe. Traditional astrobiology concentrates on finding places with a high prior probability of finding life ('follow the water'). By studying the Extreme Universe (SNRs), we can instead start to exclude parts of the Universe from consideration.

There are currently a number of observational studies of SNRs using today’s state-of-art gamma-ray (HESS), X-ray (Chandra, XMM, eROSITA) and radio telescopes (ATCA, LOFAR, eVLA) and will continue our efforts with upcoming telescopes like CTA, IceCube \& KM3NeT \cite{2021EPJC...81..445A}, gravitational wave observatory \cite{2021SerAJ.203....1L} and the SKA precursors, including synergistic programmes such as MeerKAT-ASKAP-MWA. SKA pathfinders' observations in radio at low frequencies with moderate-to-high sensitivity will detect new SNRs in our Galaxy and the Magellanic Clouds, which are either old and too faint, young and too small, or located in a too confusing environment and have thus not been detected yet. In addition, the SKA pathfinders' observations will also allow high-resolution polarimetry and are key to the study of the energetics of accelerated particles as well as the magnetic field strength and configurations. Future gamma-ray studies will provide answers to the long-standing question in high energy astrophysics: Where do cosmic rays come from? The gamma-ray emission seen from some middle-aged SNRs is now known to be from distant populations of cosmic-rays (probably accelerated locally) interacting with gas, but there is still much work to be done in accounting for the Galactic cosmic-ray flux. Young PeV gamma-ray SNRs (a.k.a. PeVatrons) require different techniques to address the question of cosmic-ray acceleration. We particularly expect that the CTA, LHAASO and SWGO \citep{2019arXiv190208429A} will allow us to do this.

\subsection{Pulsars and Magnetars}
\label{sec-pulsars}

\noindent
 \chapterauthor[1,2,3]{Zorawar Wadiasingh}\orcidlink{0000-0002-9249-0515}
 \\
 \begin{affils}
    \chapteraffil[1]{University of Maryland, College Park, Maryland 20742, USA}
    \chapteraffil[2]{NASA Goddard Space Flight Center,\\ Greenbelt, MD 20771, USA}
    \chapteraffil[3]{Center for Research and Exploration in Space Science and Technology, NASA/GSFC, Greenbelt, Maryland 20771, USA}
 \end{affils}
 
 Magnetars are a topical subclass of neutron stars with surface fields exceeding $10^{10}$~Tesla, a regime where exotic QED processes may operate\citep{2008A&ARv..15..225M,turolla15:mag,2017ARA&A..55..261K,2006RPPh...69.2631H}. Magnetars in our galaxy are largely observed through their X-ray/gamma-ray emission via bursts and persistent emission phenomenology. This radiation is powered by the dissipation of their strong fields. For multimessenger studies of magnetars, magnetar bursts (particularly a subclass known as ``giant flares") offer the best prospects both in the neutrino and gravitational wave sectors. Giant flares are relatively rare events (roughly once per two decades in our galaxy) that involve energetics exceeding $10^{45}$~erg \citep{1979Natur.282..587M,1980Natur.287..122R,Hurley-1999-Nature,Feroci-1999-ApJ,Hurley-2005-Nature,Palmer-2005-Nature}. In contrast, more than ${\cal O}(10^3)$ common lower energy recurrent ``short bursts" are observed by gamma-ray burst detectors per decade from nearby magnetars. These, too, have some prospects for potential multimessenger signals. Both giant flares and short bursts exhibit as impulsive events transpiring on timescales much less than a second. Recent observations favor a very low altitude origin for both short bursts and giant flares, one that involves the neutron star crust such that these events can excite global seismic oscillations \citep[e.g.,][]{2006ApJ...653..593S,2006ApJ...637L.117W,2014ApJ...787..128H,2014ApJ...795..114H,2019ApJ...871...95M,2022ApJ...924..136Y,2022ApJ...924L..27Y}. If neutron star f-modes are excited in these bursts \citep{2001MNRAS.327..639I,2011MNRAS.418..659L,PhysRevD.83.104014}, third generational gravitational detectors observations  of the nearby universe will attain highly constraining levels for magnetar models \citep{2021ApJ...918...80M}. Detection of gravitational waves would enable astroseismology \citep{1998MNRAS.299.1059A}, constraints on the neutron star equation of state, and speed of gravity measurements. 

In April 2020, a magnetar giant flare was observed from the nearby Sculptor galaxy \citep{Roberts-2021-Nature,2021Natur.589..211S}. Subsequently, it was demonstrated \cite{Burns-2021-ApJL} that magnetars giant flares cosmological volumetric rate is high and that they constitute the most prolific class of extragalactic gamma-ray bursts. The giant flare was also accompanied by a GeV afterglow, consistent with an ultrarelativistic outflow impacting local ambient medium and accelerating particles via diffusive shock acceleration. Analogous to jets and shocks in canonical cosmological gamma-ray bursts, such shocks ought to also produce neutrinos by p-$\gamma$ interactions. Moreover, the magnetospheres of magnetars may produce neutrinos if sufficiently high voltages (to accelerate protons) are realized \citep{2003ApJ...595..346Z,2008JCAP...08..025H,2017A&A...603A..76G}. For giant flares and short bursts, TeV to PeV voltages may be realized if the magnetosphere is charge-starved to a triggering impulsive disturbance (e.g. starquake) \citep[][]{Wadiasingh-2019-ApJ,2020ApJ...891...82W} -- incidentally, similar conditions may be a necessary condition for producing a fast radio burst. The total diffuse flux of neutrinos from such events is expected to be low, yet temporal/spatial coincidences may enhance detection significance.



\subsection{Pulsar Halos}
\label{sec-Halos}

\noindent
 \chapterauthor[]{Mattia Di Mauro}\orcidlink{0000-0003-2759-5625}
 \\
 \begin{affils}
    \chapteraffil[]{Istituto Nazionale di Fisica Nucleare, Sezione di Torino, Via P. Giuria 1, 10125 Torino, Italy}
 \end{affils}
The HAWC Collaboration reported the detection of few-degrees-extended $\gamma$-ray emission at TeV energies around the Geminga and Monogem pulsars~\citep{Abeysekara:2017science}. Very recently, the LHAASO experiment reported the detection of an extended $\gamma-$ray emission around the pulsar PSR~J0622+3749~\citep{LHAASO:2021crt} at energies $E>25$~TeV as well.
The existence of these $\gamma$-ray structures, called halos, has been predicted a while ago by Ref.~\citep{Felix-book-2004vhec.book.....A}. $\gamma$-ray halos are the result of electrons and positrons ($e^{\pm}$) accelerated at the pulsar's wind termination shock and propagating diffusively in the turbulent interstellar medium (ISM) and inverse Compton scattering (ICS) on the interstellar radiation field.

The small angular size of the $\gamma$-ray halos around the PSR~J0622+3749, Monogem, and Geminga pulsars led to the conclusion that the cosmic-ray (CR) diffusion was inhibited within few tens of pc from the pulsar, and consequently the energy dependent CR diffusion coefficient, $D(E)$, should be smaller, by at least two orders of magnitudes, than the {\it nominal} value used in conventional models of propagation of Galactic CRs  \cite{Abeysekara:2017science}. 
Since then, the suppression of the diffusion coefficient around pulsars has become a popular hypothesis \cite{Hooper:2017gtd,Tang:2018wyr,Fang:2018qco,DiMauro:2019yvh,DiMauro:2019hwn, Giacinti-2019nbu}, but so far no convincing theoretical explanation of this effect has been proposed (see, e.g.,~\cite{Lopez-Coto-Giacinti:2017pbk, Evoli-Morlino-2018PhRvD..98f3017E, Liu-Yan-2019zyj}).

Very recently, Ref.~\cite{Recchia:2021kty} has shown that the conclusion about the inhibited diffusion is driven by the wrong assumption that particles propagate diffusively right away after the injection without taking into account the ballistic propagation. However, the particles first move ballistically until they travel a distance approximately equal to the typical diffusion length, $\lambda_c(E) = 3 D(E)/c$, after which they start to scatter efficiently on the inhomogeneities of the magnetic field and undergo diffusion.
Ref.~\cite{Recchia:2021kty} examined the extended emission around the Geminga, Monogem, and PSR~J0622+3749 pulsars considering the transition from the quasi-ballistic, valid for the most recently injected particles, to the diffusive transport regime and found a good match with the data for typical interstellar values of the diffusion coefficient without the need to invoke a strong suppression of the diffusion coefficient.

\section{Diffuse Backgrounds}
\label{sec-DiffuseBackgrounds}

Diffuse astrophysical backgrounds arise in all of the messengers not just due to the limitations of current detectors, but as an indication of large scale and diffuse structure in the universe. These diffuse backgrounds are studied extensively for individual messengers, but future insights to the origin of the cosmos may arise from considering their similarities and collaboration across diffuse working groups for each messenger. 

\subsection{Introducing the Diffuse Gamma-Ray Background}
\label{sec-GammaBackground-Mora}
\noindent
 \chapterauthor[1]{Mora Durocher}\orcidlink{0000-0003-2169-0306}
 \\
 \begin{affils}
    \chapteraffil[1]{Los Alamos National Laboratory, Los Alamos, NM, 87545, USA}
 \end{affils}
 
 Mora (Pat Harding's postdoc) has promised this to Kristi. We can combine it with Michela's contriibution once it is received. 
 
 The high-energy Diffuse Gamma-Ray Background (DGRB) is an isotropic gamma-ray emission representing the superposition of uncorrelated gamma-ray sources. It is believed to originate from extragalactic objects which are too faint or too diffuse to be resolved, such as active galactic nuclei, starburst galaxies \citep{DGRB_starburst} and gamma-ray bursts \citep{EG-components}. Understanding the origin of the DGRB would help understand the nature of high-energy astrophysical objects and would help with searches for physics beyond the Standard Model. Dark matter annihilation and decay (Section \ref{sec-DM}) are expected to spread across the sky with a nearly isotropic distribution, in addition to isolated dense regions. Due to Earth's location near the middle of the Milky Way's dark matter halo, this would produce a galactic contribution to the DGRB \citep{DGRB_DM,EGB_DM}. Some studies have set limits on isotropic emissions from dark matter interactions \citep{DM_IC,DM_LAT}, while other studies  have observed  or constrained the DGRB \citep{casa-mia_graph,grapes_graph,fermi_graph,DGRB_harding} and its anisotropies \citep{DGRB_anistro}. In general, astrophysical pion decays produce neutrinos as well as gamma rays. A relation between the gamma ray flux and the neutrino flux has been established \citep{gamma_nu,AMONTeam:2020otr} thus inviting potentially significant multi-messenger studies, such as constraining the origin of TeV-PeV isotropic neutrino flux detected by IceCube \citep{IceCube:2016umi}.

\subsection{Phenomenology of the Diffuse Gamma-Ray Background}
\label{sec-GammaBackground}

 \noindent
 \chapterauthor[1,2,3]{Michela Negro}\orcidlink{0000-0002-6548-5622}
 \\
 \begin{affils}
    \chapteraffil[1]{University of Maryland, Baltimore County, Baltimore, MD 21250, USA}
    \chapteraffil[2]{NASA Goddard Space Flight Center, Greenbelt, MD 20771, USA}
    \chapteraffil[3]{Center for Research and Exploration in Space Science and Technology, NASA/GSFC, Greenbelt, MD 20771, USA}
 \end{affils}

The diffuse gamma-ray background is defined as a smooth residual component of the measured gamma-ray emission emerging after the subtraction of known sources of gamma-rays, such as point-like and extended sources (both Galactic and extragalactic) and the Galactic diffuse emission produced by energetic cosmic rays interacting with the interstellar medium and radiation fields in our Galaxy. 
The gamma-ray background defined above, appears in literature with different names. The list includes: IGRB (isotropic gamma-ray background), DGRB (diffuse gamma-ray background), CGB (cosmic gamma-ray background), and UGRB (unresolved gamma-ray background). Sometimes it is mistaken with the extragalactic gamma-ray background (EGB), which typically include both the gamma-ray background and the gamma-ray emission from extragalactic sources. Hereafter we will use the acronym UGRB, as the term ``unresolved'' is the most comprehensive to collectively describe anything that may contribute to the gamma-ray background.

The intensity of the UGRB is \textit{nearly} isotropic, and this kind of topology can be easily explained by the cumulative emission of randomly distributed gamma-ray sources whose flux is below the sensitivity of the observing instrument.
The first measurement of the UGRB intensity spectrum, (as opposed to the EGB, which was measured already by EGRET and SAS-2) was performed by the \textit{Fermi} Large Area Telescope (LAT, \cite{0004-637X-697-2-1071}) collaboration in 2010 \citep{Abdo:2010nz}, and then updated in 2015 \citep{2015IGRB}, which still represents the most updated measurement of the UGRB intensity between 100 MeV and 800 GeV. Despite the extensive interpretation campaign, the exact composition of the UGRB emission and the relative contributions from different populations of sources remains one of the main unanswered questions of gamma-ray astrophysics.
Contribution from well-known extragalactic astrophysical source populations, such as blazars \citep{AndoCluAPS:2007, 2011ApJ...736...40S} and misaligned AGNs (mAGNs) \citep{2011ApJ...733...66I, 2014ApJ...780..161D} is guaranteed, them being quite rare objects generally speaking, but the brightest and the most numerous seen in gamma-rays. Also, a non-negligible contribution (even dominant, according to some models) is expected from SFGs \citep{MARoth2021,2011ApJ...736...40S, 2010ApJ...722L.199F, Linden2017, 2011ApJ...728..158M}, which are not very bright in the gamma-band but extremely abundant in the Universe. Minor contributions from an unresolved population of Galactic millisecond pulsars (MSPs) can be expected \citep{SiegalGaskins:2010mp, Calore:2014oga}, as well as from galaxy clusters \citep{2002MNRAS.337..199M, 2003ApJ...585..128K, AndoCluAPS:2007}, Type Ia supernovae \citep{2005PhRvD..71l1301A, 2006PhRvD..73j3518R}, and GRBs \citep{Casanova:2008zz, 2008ApJ...689.1150A}. Furthermore, more exotic scenarios may contribute as well \citep{Ando:2009fp, Calore:2013yia, Ajello:2015mfa, 2016PhRvD..94l3005F, Zechlin:2017uzo, korsmeier2022}: despite a huge current experimental effort aimed to search for evidence of annihilating or decaying particles of dark matter (DM) through the detection of gamma-rays (primarily or secondarily produced), no signal has been robustly associated with DM up to now, so if present it is most probably unresolved and contributes to the UGRB. 
Anyway, the interest in finding a definitive answer is attributable to the need to constrain the faint end of the luminosity function of the UGRB contributors, which could also tell something about the cosmological evolution of the classes of objects involved. Being these objects too faint to be resolved individually, the study of the UGRB may represent the only source of information about them, at least until a new, more sensitive instrument will improve upon the \textit{Fermi}-LAT observations.

In addition to the intensity spectrum of the UGRB one can extract valuable information from the study of the angular scale and the amplitude of the intensity fluctuation field of the UGRB. Several spatial autocorrelation analyses have been performed throughout the \textit{Fermi}-LAT survey \citep{Ackermann:2012uf, 2016PhRvD..94l3005F, 2018PhRvL.121x1101A}, every time unveiling fainter components (by resolving and removing more sources). Recently, the latest measurement of the anisotropy energy spectrum has been interpreted in terms of blazars \citep{Manconi:2019ynl, korsmeier2022}, showing how this population can account for 100\% of the measured anisotropy and be consistent with the resolve population of blazars. In particular, Ref.~\cite{korsmeier2022}, show how flat spectrum radio quasars contributes more at lower energies (having on average steeper spectra) of the anisotropy energy spectrum, while BL Lacs dominates above $\sim$5 GeV 
. The constrain on the blazar contribution to the UGRB emission derived from the anisotropy spectrum are more stringent than those derived considering only the intensity spectrum: even if representing the 100\% of the anisotropy, the unresolved blazars can account only for a fraction of the UGRB intensity spectrum 
. Nevertheless, such a contribution is guaranteed, and any other additional contributor (e.g. SFG), should not overshoot the total measured intensity. 

Other works considered the photon count statistics and the one-point probability distribution functions of the expected UGRB components to constrain their contribution \citep{2011ApJ...738..181M, 2016ApJS..225...18Z, 2016ApJ...826L..31Z, 2016ApJ...832..117L, DiMauro:2017ing, Marcotulli2020, Manconi:2019ynl}. These works set constraints on the unresolved blazar population which are compatible to those resulting from the anisotropy study.

One final remark regards the possibility to characterized the gamma-ray background composition by exploiting a multiwavelength and, possibly (in the future) multimessenger approach. Since the majority of the unresolved emission of the UGRB is extragalactic, we expect a certain level of cross-correlation signal with the large-scale structure (LSS) tracers of the Universe. Cross-correlation studies involving the UGRB have been done considering the special distribution of galaxies \citep{xia11, Cuoco:2017bpv, Ammazzalorso:2018evf}, galaxy cluster catalogs \citep{Branchini:2016glc, 2017arXiv170809385L, 2017arXiv170900416L, 2018PASJ...70S..25M},  cosmic shear from weak lensing \citep{Camera:2012cj, Camera:2014rja, Shirasaki:2014noa, 2019arXiv190713484A}, and lensing potential of the cosmic-microwave background \citep{Fornengo:2014cya}. By exploiting the redshift and/or band-dependent luminosity distributions available in some galaxy catalogs, a tomographic study of the UGRB is possible \cite[see e.g.,][]{Cuoco:2017bpv,Ammazzalorso:2018evf}. Such studies allow the characterization of the UGRB sources in terms of time-evolution, star formation activity, and masses of the objects. Future wide-field deep surveys, such as Nancy Roman Observatory, will be fundamental to push forward our understanding of the UGRB and its evolution.

From a multimessenger point of view, of particular interest is the relation between the very-high-energy astrophysical neutrinos \citep{IC2015} and the gamma-ray emission from extragalactic objects. The observation of a neutrino event by IceCube\footnote{\url{https://icecube.wisc.edu}} in temporal and spacial coincidence with a gamma-ray flare from the BL Lac TXS~0506+056 \citep{IceCube:2018dnn}, suggested that blazars are good candidates to contribute to the neutrino astrophysical background. However no many more of these events have been observed from gamma-ray detected blazars. Also SFG have been suggested as contributors to the astrophysical neutrino flux \cite[see e.g.,][]{Bechtol:2015uqb}. Looking for connections between very-high-energy neutrinos and the UGRB might shed some light on the origin of the astrophysical neutrinos observed by IceCube. Spatial cross-correlation analyses (both with UGRB and LSS tracers, as the one attempted in Ref~\cite{Ke2020}), might be a useful tool for the future, once (and if) the neutrino event localization accuracy will significantly improve.

\subsection{Diffuse Astrophysical Neutrino Background}
\label{sec-DiffuseNeutrino}

 \noindent
 \chapterauthor[]{Samalka Anandagoda}\orcidlink{0000-0002-5490-2689}
 \\
 \begin{affils}
    \chapteraffil[]{Clemson University, Department of Physics \& Astronomy, Clemson, SC 29634-0978}
 \end{affils}

The detection of 25 neutrinos from the Type II supernova in the Large Magellanic Cloud (LMC), the event named SN1987A \citep{hirata_1987,bionta_1987,alekseev_1987}, is believed to be the beginning point of multi-messenger astronomy. This event marked the first time neutrinos were detected from a massive star undergoing core-collapse. The Electromagnetic observations of this event along with the distribution of the recorded neutrinos in time and energy confirmed the formation of a hot proto neutron star (PNS) in a core collapse supernova (ccSN) \citep{janka_2007}, highlighting the importance of a multi-messenger approach when addressing astrophysical questions. The underlying mechanisms of these collapsing stars are still poorly understood and neutrinos are the ideal candidate to reveal such information about the core. This is mainly because neutrinos are weakly interacting particles and they are able to escape from the dense core revealing its properties. In order to reveal the dynamics of these exotic environments a signal with high statistics is required which will be possible when a galactic supernova occurs due to sensitivity limitations of current neutrino detectors \citep{abe_2021_upperlim}. However, the galactic supernova rate remains quite low at $\leq$ 3 per century \citep{adams_2013} and much larger neutrino detectors will be required to detect supernovae neutrinos from nearby galaxies (1-10 Mpc). Another avenue to study these explosions is available through the detection of the Diffuse Supernova Neutrino Background (DSNB) which constitutes of MeV neutrinos from all past core-collapse supernovae. 

\begin{figure}[hbt!]
    \centering
    \includegraphics[width=1.0\textwidth]{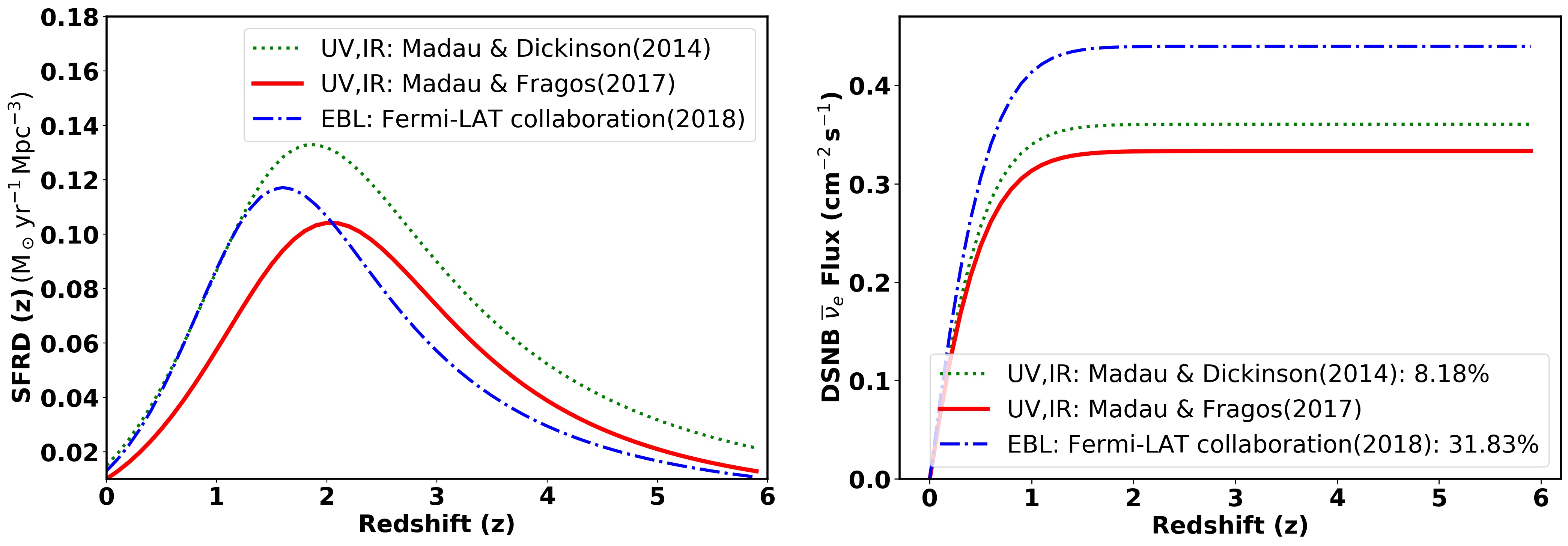}
  \caption{The DSNB $\overline{\nu}_e$ flux integrated over the 19.3 - 35 MeV energy range is shown as a function of redshift in the right plot. The corresponding SFRD models used to obtain these DSNB fluxes are shown in the left plot. The colors are kept consistent based on the SFRD model used.}
  \label{fig:DSNB_SFRD_dependence}
\end{figure}

As the DSNB is comprised of neutrinos from past core collapse supernovae over the history of the universe, it carries rich information such as the cosmic core-collapse supernova rate and supernova neutrino emission \citep{beacom_2010}. The dependency of the DSNB flux on the core-collapse supernova rate and the use of a future DSNB detection to place constraints on the underlying star formation rate density (SFRD) model have been discussed by various groups in the literature \citep{ando_2004,strigari_2005,hopkins_2006,mathews_2014,nakazato_2015,anandagoda_2020,riya_2021}. Modeling of the SFRD is done using various tracers like UV, IR continuum and H$\alpha$ emission \citep{madau_2014} along with luminosity functions and Initial Mass Function (IMF) which is subject to uncertainties. Mild discrepancies remain among various SFRD models based on the type of tracer and the method used to determine them. This highlights the need for alternative methods that are independent of each other to constrain the SFRD models. For example, utilizing different SFRD models to calculate the $\overline{\nu}_e$ DSNB flux, while keeping other parameters constant, it is found that the DSNB flux varies by $\approx$30\% over the energy range of 19.3 MeV - 35 MeV shown in figure~\ref{fig:DSNB_SFRD_dependence} \citep[see also][]{anandagoda_2020}. Hence a well determined DSNB flux measurement can in turn be used to constrain the SFRD model using a method similar to the one illustrated in \citep{fermi_lat_2018}. 

\begin{figure}[hbt!]
    \centering
    \includegraphics[width=0.8\textwidth]{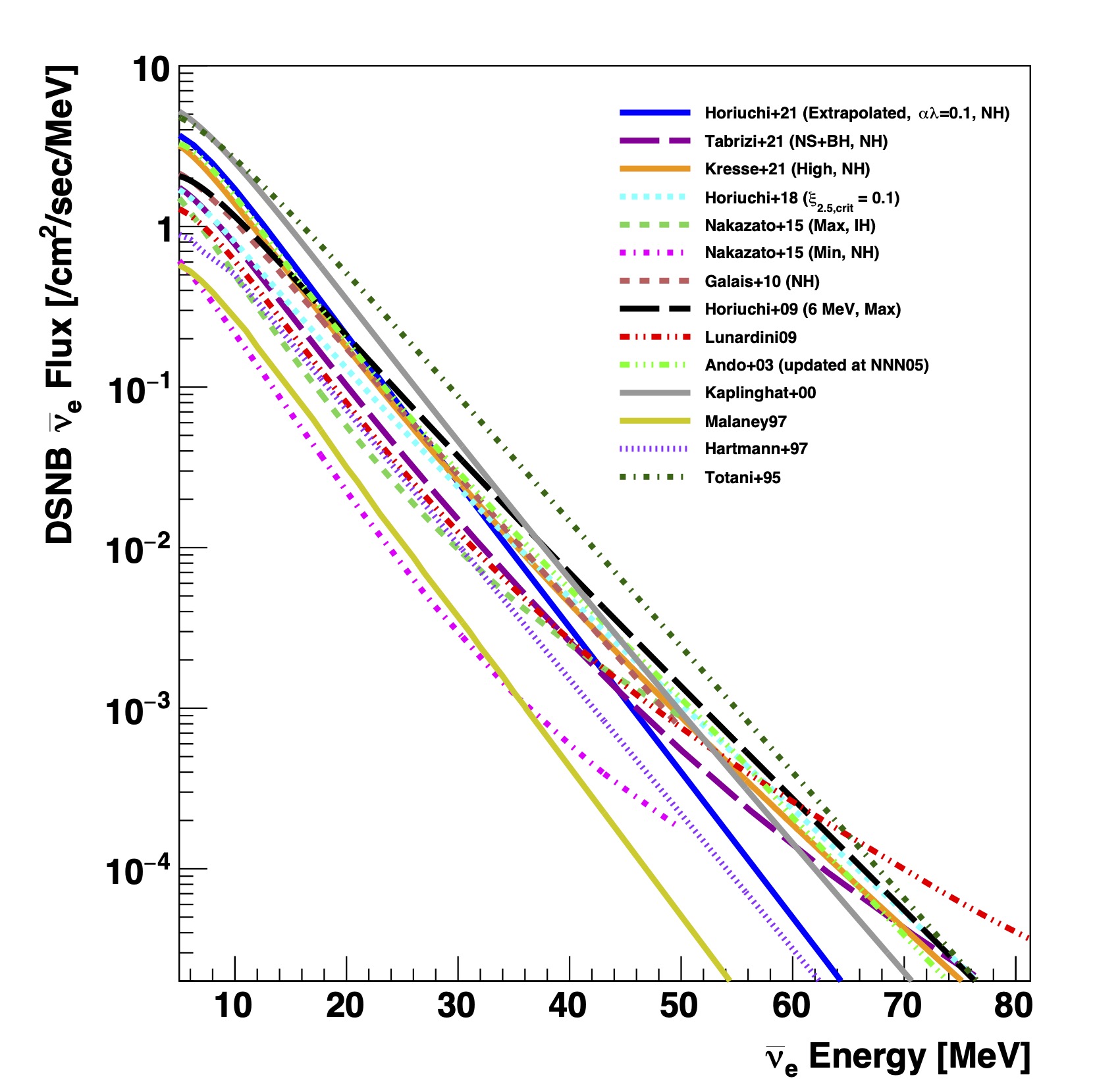}
    \caption{DSNB $\overline{\nu}_e$ theoretical flux predictions from various models in the literature. Figure obtained from \cite{abe_2021_upperlim}. Displayed here are models by \cite{horiuchi_2021,tabrizi_2021,kresse_2021,horiuchi_2018,nakazato_2015,galais_2010,horiuchi_2009,lunardini_2009,ando_2003,kaplinghat_2000,malaney_1997,hartmann_1997,totani_1995}. Note the change in the spectral shape of the DSNB due to various physical effects such as failed supernovae \citep{lunardini_2009}, neutrino flavor conversions \citep{ando_2003}, etc.}
    \label{fig:DSNB_models}
\end{figure}

The spectral shape of the DSNB is affected by a wide range of physical effects \citep{abe_2021_hyper} and can be used as a tool to probe the intrinsic core-collapse supernova source spectrum. For an example, black-hole forming supernovae (failed supernovae) make the DSNB spectrum harder \citep{lunardini_2009} (see Figure~\ref{fig:DSNB_models}). Even though the DSNB has not been detected yet, the upper flux limits of the DSNB set by Super-Kamiokande experiment \citep{abe_2021_upperlim} are already close to the theoretical predictions. Furthermore, the discovery prospects of the DSNB in the next decade are promising with the gadolinium enhanced Super-Kamiokande (SK-Gd) detector \citep{beacom_vagins_2004,SK-GD_2021} which will have reduced backgrounds and low energy thresholds making the detection of low energy events (>10 MeV) possible \citep{beacom_2010,nakazato_2015}. Data taking in the SK-Gd configuration started in late 2020 \citep{li_2022prospects} and a statistical evidence (3$\sigma$) of the DSNB signal is expected within 10 years of running time. Other experiments include, Jiangmen Underground Neutrino Observatory \citep[JUNO;][]{JUNO_2015} and Hyper-Kamiokande \citep{abe_HK_2011}. A future DSNB measurement will no doubt be an exciting discovery and in convergence with electromagnetic observations would provide valuable insight into the core-collapse supernova physics, various physical processes as well as the star formation history of the universe.

\subsection{Stochastic Gravitational Wave Background}
\label{sec-GWbackground}

 \noindent
 \chapterauthor[]{Patrick M. Meyers}\orcidlink{0000-0002-2689-0190}
 \\
 \begin{affils}
    \chapteraffil[]{Theoretical Astrophysics Group, California Institute of Technology, Pasadena, CA 91125, USA}
 \end{affils}

While individual transient detections of gravitational-wave events pile up, efforts continue towards a detection of a gravitational-wave background (GWB). Exciting possibilities lie in a detection of a GWB from cosmological sources. Here we highlight four promising possibilities for multimessenger science with a GWB from unresolved point sources over the next decade. We do not cover observations of individual white dwarf binary systems, as well as improved white dwarf binary modelling, which will help characterize the confusion noise that will limit sensitivity to individual events in a large section of the LISA frequency band. While LISA won't fly until 2034, the groundwork for the multimessenger science that will come from its detection of a GWB from unresolved white dwarf binaries will be set by observing campaigns and modelling that is done over the next decade.

\textbf{Constraining star formation history and binary black hole formation and evolution}
In the next decade, a detection of a GWB from unresolved binary black hole mergers is plausible~\cite{KAGRA:2021kbb}.
The amplitude of the background can immediately be combined with individual CBC detections to constrain
the merger rate as a function of redshift~\cite{Callister:2020arv,KAGRA:2021kbb}.
Even a null result could offer significant information,
including the redshift when the merger rate peaks~\cite{Callister:2020arv,KAGRA:2021kbb}.
Additionally, GWB results can also then be used to constrain the distribution of time delay between formation and merger,
and the formation metallicity of binary black holes that merge in the LIGO/Virgo frequency bands~\cite{Safarzadeh:2020qru,Fishbach:2021mhp}.
Similar studies have also been performed for primordial black holes~\cite{Mukherjee:2021ags,Mukherjee:2021itf},
which are proposed candidates for dark matter.

\textbf{Constraining galactic neutron star population properties}
Pulsar surveys with ultra-wideband receivers and next-generation radio telescopes
like SKA and DSA-2000 promise to detect new, fainter galactic pulsars.
These surveys will further constrain the spatial distribution of pulsars in the galaxy
and the fraction of pulsars that spin with millisecond periods.
A gravitational-wave background from unresolved, continuously emitting galactic neutron stars
could be detectable with third-generation gravitational-wave detectors.
Specifically, sensitivity estimates presented in~\cite{Talukder:2014eba}, e.g. for LIGO A+ upgrades,
start to become comparable to the potential minimum neutron star ellipticity discussed in ~\cite{Woan:2018tey},
while third generation detector projections are even more promising.
Sensitivity to this background is significantly improved by having reliable templates
for the spatial distribution of pulsars and the expected shape of the frequency spectrum\footnote{The shape of the frequency spectrum depends on the number of sources emitting at a given frequency.
The emission frequency is typically expected to be a harmonic of the rotation frequency.}
~\cite{Thrane:2009fp,Talukder:2010yd,Talukder:2014eba}.
A detection of the GWB from unresolved neutron stars could be used to constrain,
e.g. the average ellipticity of pulsars contributing
to the background, in the simplest model of a non-axisymmetric neutron star with a ``mountain''. 
Other mechanisms for gravitational-wave emission have also been explored, and could still be detectable~\cite{Lasky:2013jfa}.
A review of potential gravitational-wave emission mechanisms from individual neutron stars can
be found in~\cite{Glampedakis:2017nqy}.

\textbf{Cross-correlating GWB maps with Galaxy catalogues}
A large interest has been taken in the expected anisotropy of the astrophysical gravitational-wave background~\cite{CONTALDI20179,PhysRevD.96.103019,PhysRevD.97.123527,PhysRevD.98.063501,PhysRevLett.122.111101,PhysRevLett.120.231101,PhysRevD.100.063004,PhysRevD.101.103513,PhysRevD.101.081301} as a potential secondary probe of large-scale structure in the Universe, and recent searches have placed limits on the level of anisotropy~\cite{Renzini:2019vmt,KAGRA:2021mth}. However, the initial measurement of an anisotropic background will be dominated by shot noise~\cite{Jenkins:2019uzp,Jenkins:2019nks,Renzini:2021iim}. Most methods to properly map the anisotropic background involve either waiting for more data~\cite{Jenkins:2019nks}, or cross-correlating the GWB maps with galaxy catalogues~\cite{PhysRevD.96.103019,PhysRevD.100.063004,Yang:2020usq} or CMB lensing maps~\cite{PhysRevD.102.043513}. A detection and reliable map of the GWB would be ground-breaking in its own right, but is likely not possible until third generation detectors like Einstein Telescope or Cosmic Explorer. However, it is also a useful tool for the ``budgeting'' problem of stochastic backgrounds~\cite{Martinovic:2020hru,Biscoveanu:2020gds,Contaldi:2020rht} -- a measurement of the level of contribution of the GWB that is correlated with large-scale structure could be used separate the astrophysical GWB from cosmological backgrounds that would not necessarily be expected to trace large-scale structure.

\textbf{Multimessenger constraints on supermassive black hole binary populations}
In recent years, pulsar timing arrays (PTAs; Section \ref{sec-pulsarArrays}) like NANOGrav, Parkes Pulsar Timing Array, European PTA, and the International PTA have shown evidence for a ``common spectrum process'' in pulsar timing data~\cite{NANOGrav:2020bcs,Goncharov:2021oub,Chen:2021rqp,Antoniadis:2022pcn}. While the expected correlations between pulsars that would be expected from a true GWB detection have not been measured, the estimated parameters of the common spectrum are consistent with a GWB from supermassive black hole binaries (SMBHBs; Section \ref{sec-largerBH-BH}). First, note that pulsar timing arrays are inherently multimessenger enterprises--improvement in our sensitivity to gravitational-waves come from improving radio telescopes we use to time pulsars (and the data can be used for myriad endeavours, including e.g.  testing general relativity, studies of the interstellar medium, etc.).
Limited information can be learned about the population of SMBHBs from a GWB detection alone, however, due to the covariance between numerous population parameters. A GWB detection can shed light on the mass distribution of SMBHBs~\cite{Rasskazov:2016jjk}, but the spin distribution and the interaction between the binary its surrounding environment (e.g. dynamical friction~\cite{Merritt:2004gc,AntoniniMerritt2012}, stellar loss cone scattering~\cite{MikkolaValtonen1992,Quinlan:1996vp}, and viscous circumbinary disk interactions~\cite{Begelman:1980vb,Kocsis:2010xa}) play a large role in how the binary orbit evolves with time, and therefore the amplitude of the GWB as a function of frequency.
Observations of individual ``GW precursor'' binaries, e.g. active galactic nuclei within a few kpc of one another, are accessible to large electromagnetic surveys~\cite{Wen:2008hw}. 
Measuring the properties of these systems can provide statistics on the systems that make up the binary population that would contribute to the GWB detectable by PTAs, and help break degeneracies in parameters we cannot measure with a GWB alone.
 
In addition to the science from an unresovled GWB from SMBHBs (Section \ref{sec-largerBH-BH}), the most promising multimessenger science to be had with pulsar timing array observations almost certainly lies in observing both a continuous gravitational-wave signal from an individual supermassive black hole binary systems, combined with radio observations of the same system. See, e.g.~\cite{Burke-Spolaor:2018bvk} for a complete discussion.

\textbf{Other potential sources} We have highlighted four of the most promising targets for multimessenger science with a GWB from unresolved point sources. There are others, as well. For example, a GWB from magnetars (Section \ref{sec-pulsars})~\cite{Wu:2013xfa,Chowdhury:2021vqn}, or from unresolved supernovae ~\cite{Buonanno:2004tp,Zhu:2010af,Crocker:2015taa,Crocker:2017agi,Finkel:2021zgf}. A GWB from boson clouds around a black hole with a superradiant instability can constrain the mass of ultralight bosons, a generic prediction of many beyond-standard-model theories~\cite{Tsukada:2018mbp,Tsukada:2020lgt}. These constraints rely on external measurements of the population characteristics of black holes, which will continue to improve through GW and electromagnetic observations.

\chapter{The Current and Future Multimessenger Network}
\label{chap-MMNetwork}








\section{Real-Time Alert Network Coordination}
\label{sec-alertnetworks}

\subsection{Astrophysical Multimessenger Observatory Network}
\label{sec-AMON}

 \noindent
 \chapterauthor[]{Hugo Alberto Ayala Solares}
 \\
 \begin{affils}
    \chapteraffil[]{Department of Physics, Pennsylvania State University, State College, PA 16801, USA}
 \end{affils}

\textit{AMON}: The Astrophysical Multimessenger Observatory Network (AMON) is a cyber-infrastructure developed to perform real-time coincidence analysis for multimessenger astrophysics.
AMON accepts sub-threshold events, which are data that are below the point-source analysis thresholds for individual observatories. Well-reconstructed events can be below these thresholds and hence unusable by the individual observatory. However, with the use of careful statistical analyses, AMON enables combining the datasets from different observatories and can recover these astrophysical events for point source analysis \citep{Smith:2012eu,AyalaSolares:2019iiy}.

AMON has started sending real-time alerts from its Neutrino-Electromagnetic (NuEM) channel.  The channel consists of data combinations from HAWC and IceCube \citep{AMONTeam:2020otr}; and \textit{Fermi}-LAT and ANTARES \citep{AMON:2019zxe}. The coincidence alerts with low false-alarm rates ($<4$ per year) are sent to the Galactic Coordinates Network. AMON also works as a pass-through system, delivering the IceCube Gold, Bronze and Cascade events; as well as the HAWC Burst-like alerts. 

The next steps for AMON are to increase the number of analyses in its NuEM channel and to connect to the GW network in order to perform coincidence analyses between high-energy gamma-ray and neutrino data. AMON is also partnering with SCiMMA to distribute alerts and receive events from public streams that can help in the search for multi-messenger sources.

\section{Facilities}
\label{sec-facilities}


\subsection{Pulsar Timing Arrays}
\label{sec-pulsarArrays}

\noindent
 \chapterauthor[1,2,]{Joris P. W. Verbiest}\daggerfootnote{Supported by the Deutsche Forschungsgemeinschaft(DFG) through the Heisenberg programme (Project No. 433075039).}\orcidlink{0000-0002-4088-896X}
 \chapterauthor[3]{Stefan Os\l{}owski}\orcidlink{0000-0003-0289-0732}
 \\
 \begin{affils}
    \chapteraffil[1]{Fakult{\"a}t f{\"u}r Physik, Universit{\"a}t Bielefeld, Postfach 100131, 33501 Bielefeld, Germany}
    \chapteraffil[2]{Max-Planck-Institut für Radioastronomie, Auf dem H{\"u}gel 69, 53121 Bonn, Germany}
    \chapteraffil[3]{Manly Astrophysics, 15/41-42 East Esplanade, Manly, NSW 2095, Australia}
 \end{affils}

Pulsar Timing Arrays (PTAs) are experiments that exploit the unrivalled timing precision of millisecond pulsars to detect nanohertz gravitational waves (GWs). Presently, four such experiments have been set up around the globe: NANOGrav in North America~\cite{2013ApJ...762...94D}, the PPTA in Australia~\cite{2013PASA...30...17M}, the EPTA in Europe~\cite{2016MNRAS.458.3341D}, and the InPTA in India~\cite{2018JApA...39...51J}, all of which collaborate in the International PTA~\cite{2016MNRAS.458.1267V}. In addition, two emergent PTAs have recently been formed in South Africa and China~\cite{2020SCPMA..6329531L}. The most likely GWs expected to be detected by these experiments are those emanating from supermassive black-hole binaries (SMBHBs), although waves from cosmic strings, the early Universe, primordial black holes, or some dark-matter models have also been predicted.

The sensitivity of PTAs has been steadily increasing over the past decade, reaching a sensitivity that is already informative for models of galaxy evolution and supermassivee black hole (SMBH) population models~\cite{2013Sci...342..334S}. Indeed, the most recent analyses by NANOGrav~\cite{2020ApJ...905L..34A}, EPTA~\cite{2022MNRAS.509.5538C}, PPTA~\cite{2021ApJ...917L..19G}, and IPTA~\cite{2022MNRAS.510.4873A} have already identified signals similar in character to those expected from a background of SMBHBs, although the confirmation that this signal is caused by GWs requires a higher signal-to-noise ratio (S/N) and is still pending.

In this intermediate-S/N regime, the sensitivity of a PTA to GWs is dominated by the number of pulsars in the array, although the timing precision and data set length also play a role~\cite{2013CQGra..30v4015S}. The recent commencement of PTA experiments at the Square Kilometre Array (SKA) pathfinder telescopes, \textit{MeerKAT}~\cite{2016mks..confE..11B} and \textit{FAST}~\cite{2020SCPMA..6329531L}, hold great promise. Their supreme sensitivity enables numerous new pulsar discoveries and improved timing precision of all pulsars. By the inclusion of such novel telescopes and by continuing to expand the timing baseline of the present projects, sensitivity continues to be gained. Beyond this, the introduction of the SKA, which has recently commenced construction, would provide a further boost in GW sensitivity for PTAs.

Given the presence of the "GW-like" signal in current PTA data, a first statistically significant detection of the stochastic GW background from SMBHBs is presently anticipated in the near future~\cite{2021ApJ...911L..34P}. In the following decade, further sensitivity improvements and extended baselines imply a detailed characterisation of the spectral properties of this GW background will become possible, and, due to enhanced resolution in the GW spectrum, a detection of individual GW sources (or, initially, anisotropies in the GW background) will become ever more likely. Such a detection would naturally enable multi-messenger studies focused on the identification and characterisation of the host galaxy. To allow localization, a high-S/N detection of the GW source is of paramount importance~\cite{2013CQGra..30v4004E}. Such multi-wavelength identification would aid in breaking the chirp-mass/luminosity-distance degeneracy and will furthermore place unique constraints on the SMBHB formation efficiency, which is highly uncertain~\cite{2019A&ARv..27....5B}. Also, in rare cases, a host identification could be used to provide an independent measure of the mass of the graviton~\cite{2003PhRvD..67b4015C}. A full review of further potential multi-wavelength studies of GW sources in the nanohertz band can be found in Ref.~\citenum{2019A&ARv..27....5B}.

Since PTA research will require highly sensitive pulsar surveys to be undertaken in the coming decade, a different type of multi-messenger astronomy will be enabled. A sizeable number of ultracompact binary neutron stars is expected to be detected~\cite{2021ApJ...912...22P}. The binary properties of these systems will be easily determined through pulsar timing, while their GW emission should be readily detectable in the mHz GW band by space interferometers, enabling unique high-precision tests of gravity and neutron-star properties~\cite{2020MNRAS.493.5408T}.

\subsection{Current Ground-Based Gravitational Wave Interferometers and Upgrades}
\label{sec-currentGW}

\chapterauthor[1,2]{Salvatore Vitale}\orcidlink{0000-0003-2700-0767}
 \\
 \begin{affils}
    \chapteraffil[1]{Kavli Institute for Astrophysics and Space Research and Department of Physics, Massachusetts Institute of Technology, Cambridge, MA 02139, USA}
    \chapteraffil[2]{LIGO Laboratory, Massachusetts Institute of Technology, Cambridge, MA 02139, USA}
 \end{affils}

 Modern ground-based gravitational-wave (GW) detectors are kilometer-scale Michelson interferometers with Fabry-Perot cavities, which are sensitive to relative arm displacements of the order of $10^{-22}$. The current network consists of the two advanced LIGO observatories in the US~\cite{Harry:2010zz} and advanced Virgo in Italy~\cite{VIRGO:2014yos}. A fourth detector---KAGRA~\cite{Aso:2013eba}---has been recently constructed in Japan, while another LIGO detector will be built in India and become operational after 2025~\cite{IndigoProposal}.

Unlike most electromagnetic (EM) telescopes, GW observatories are all-sky instruments, and thus a single GW site cannot provide information about the sky location of a source~\cite{KAGRA:2013rdx}. Two geographically separated sites yield limited information, usually constraining the location of a source in an annular region of hundreds or thousands of square degrees~\cite{Singer:2014qca}. A network of at least three sites is required to localize sources to better than 100~square degrees, while larger networks can bring the sky location for the best sources to less than 10~square degrees. The need to precisely localize multimessenger sources of GWs is thus one of the main reasons why it is desirable to have several ground-based GW detectors online.  

The advanced LIGO detectors have been taking data since 2015, with advanced Virgo joining in 2017. These observatories are sensitive in the audio band ($\sim$10--2000~Hz), and thus target lighter sources that pulsar timing arrays or LISA. To date, roughly 100 GWs from the merger of two compact objects, black holes and neutron stars, have been detected in LIGO-Virgo data~\cite{LIGOScientific:2021djp,LIGOScientific:2021usb,LIGOScientific:2020ibl,LIGOScientific:2018mvr,Nitz:2020oeq,Nitz:2021uxj,Venumadhav:2019lyq,Zackay:2019btq}. These include the discovery of GW170817~\cite{LIGOScientific:2017vwq}, the merger of two neutron stars that was also detected in the EM band, at all frequencies, and for which the host galaxy was discovered~\cite{LIGOScientific:2017ync,Coulter2017,2018ApJ...856L..18M,2017ApJ...848L..20M,2017ApJ...848L..17C,2017Sci...358.1565E}. Other potentially EM-bright sources such as other binary neutron star (BNS) mergers, as well as mergers of neutron stars with black holes, have been discovered in LIGO/Virgo data, but no EM counterparts have reported~\cite{LIGOScientific:2021qlt}. Most of the sources detected to date have been binary black hole mergers~\cite{LIGOScientific:2021psn}. The merger of two stellar mass black holes is not usually expected to produce observable EM counterparts, though it has been suggested that mergers happening in gas-rich environments such as Active Galactic Nuclei (AGN) disks might be EM-bright~\cite{McKernan:2019hqs}.

In the next few years, both the sensitivity of the detectors and the size of the network will increase~\cite{KAGRA:2013rdx}. In their latest observing run (O3), which ended in early 2020, the LIGO detectors could observe the merger of two neutron stars up $\sim$120~Mpc away, whereas Virgo had an horizon distance of 50~Mpc~\cite{KAGRA:2013rdx}. The fourth observing run (O4) is scheduled to start in late 2022 and last for one year. LIGO and Virgo will have a horizon distance of $\sim$160--190~Mpc (90--120~Mpc). KAGRA is expected to join O4 with an horizon distance of 25--130~Mpc~\cite{KAGRA:2013rdx}. Since the number of detections scales like the cube of the horizon distance in the local universe, one can project that roughly one BNS merger will be detected every month, and half of those will have sky localization uncertainties smaller than $\sim$30~square degrees. Improved low-frequency sensitivity, as well as progress in low-latency searches for compact binaries open the possibility of \textit{pre}-merger alerts, which might be circulated to the broader community before the two neutron stars merge~\cite{Magee:2021xdx,Sachdev:2020lfd}. A few neutron star black hole mergers should also be detected in O4, with slightly worse sky localization, owing to the smaller bandwidth of heavier sources. The fifth observing run is currently schedule to start in 2025, with target BNS horizons of $330$~Mpc for the US- and India-based LIGO, $150-260$~Mpc for Virgo, and better than $130$~Mpc for KAGRA. Such a network would detect tens of EM-bright mergers per year, many of which localized to better than 10 square degrees~\cite{KAGRA:2013rdx}.

Even though all of the GWs detected to date have been generated by the merger of two compact objects, there exist other potential sources detectable with other messengers~\cite{LIGOScientific:2019ryq}. Galactic core-collapse supernovae are expected to emit detectable GWs, and would naturally detectable also in the EM and neutrino bands. Although rare, such an event would be extremely consequential. 

\subsection{Third-Generation Ground-Based Gravitational Wave Interferometers}
\label{sec-3rdGenGW}

\chapterauthor[1]{Duncan A. Brown}\orcidlink{0000-0002-9180-5765}
\chapterauthor[2,3]{Salvatore Vitale}\orcidlink{0000-0003-2700-0767}
\\
 \begin{affils}
    \chapteraffil[1]{Department of Physics, Syracuse University, Syracuse, NY 13244, USA}
    \chapteraffil[2]{Kavli Institute for Astrophysics and Space Research and Department of Physics, Massachusetts Institute of Technology, Cambridge, Massachusetts 02139, USA}
    \chapteraffil[3]{LIGO Laboratory, Massachusetts Institute of Technology, Cambridge, Massachusetts 02139, USA}
 \end{affils}

The gravitational-wave discoveries by Advanced LIGO\footnote{https://www.ligo.caltech.edu/} and Advanced Virgo\footnote{https://www.virgo-gw.eu/} have opened a new window on the universe. There is significant international interest in, and mobilization toward, developing the next generation of ground-based gravitational-wave observatories capable of observing gravitational waves throughout the history of star formation and using gravitational waves to study fundamental physics. A community study of the potential for a network of such third-generation observatories (and its synergy with other types of gravitational-wave observatories and electromagnetic and astro-particle observatories) have been undertaken by the Gravitational-Wave International Committee (GWIC) and summarized in a
series of white papers~\cite{GWIC3GDocs}.

\begin{figure}[th]
    \centering
    \hspace*{-0.53cm}\includegraphics[width=0.68\textwidth]{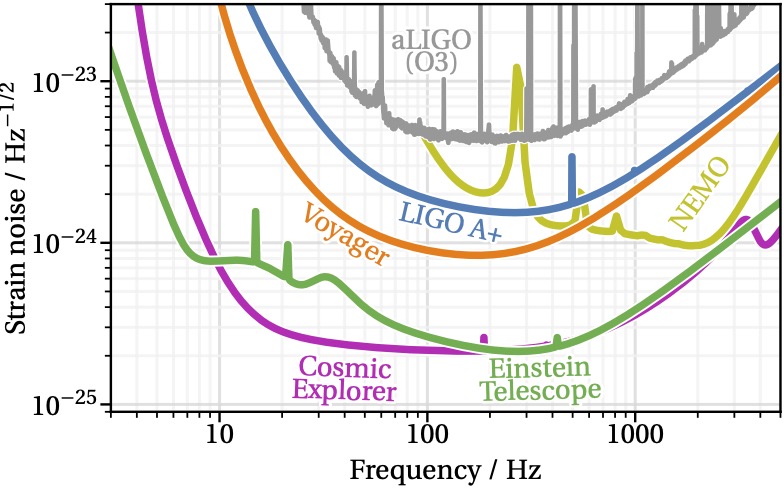}
    \caption{Amplitude spectral densities of detector noise for Cosmic
      Explorer, the current (O3) and upgraded (A+) sensitivities of Advanced
        LIGO, LIGO Voyager, the proposed Australian NEMO detector~\cite{Distefano:2006bi}, and the three paired detectors of the triangular Einstein Telescope. At each frequency, the noise is referred to the strain produced by a source with optimal orientation and polarization.}
\label{fig:gw_noise_curves}
\end{figure}

In the U.S., the proposed Cosmic Explorer observtory is designed to have ten times the sesnitivity of Advanced LIGO and will push the reach of gravitational-wave astronomy towards the edge of the observable universe ($z \sim 100$)~\cite{Reitze:2019iox,Evans:2021gyd}. The European Einstein
Telescope proposal will offer a similar increase in observational reach~\cite{Punturo:2010zz}. Cosmic Explorer's increased sensitivity comes primarily from scaling up a detector that uses LIGO technology from 4~km to 40~km L-shaped arms. The Einstein Telescope will use advanced detector technologies in a 10~km triangular interferometer with $60^\circ$ angles built underground to minimize low-frequency noise. A proposal known as LIGO Voyager would upgrade the existing LIGO facilities to the limit of their observational reach using advanced detector technologies~\cite{LIGO:2020xsf}, although this design does not reach the senstivity of Cosmic Explorer and Einstein Telescope, which require new facilities. A comparison of the strain stensitivity of these proposed detectors, and the existing LIGO detectors, is shown in Figure~\ref{fig:gw_noise_curves}.

The third-generation ground-based gravitational-wave observatories will be a critical part of the multimessenger landscape in the coming decades. A network consisting of Cosmic Explorer in the U.S. and Einstein Telescope in Europe would detect $\gtrsim 10^5$ binary neutron stars per year, with a median redshift of $ \sim$1.5---close to the peak of star formation---and a horizon of $z\gtrsim 9$~\cite{Borhanian:2022czq}. Approximately 200 of these binary neutron stars would be localized every year to better than one square degree, enabling followup with telescopes with small fields of view~\cite{Borhanian:2022czq}. A factor of ten increase in the number of binary neutron star mergers detected to within a square degree could be obtained by building a second Cosmic Explorer observatory. The greater the separation between the detectors, the more precisely sources can be localized, making Australia a prime site for a possible second Cosmic Explorer detector, or for the proposed NEMO high-frequency third-generation detector~\cite{2019CQGra..36v5002H}. The improved low-frequency sensitivity of third-generation detectors allows them to detect and localize sources prior to merger. The triangular configuration of the Einstein Telescope allows it to localize sources without second observatory~\cite{Maggiore:2019uih}. The Einstein Telescope is able to detect 6 (2) sources per year at 5 (30) minutes before merger with a localization better than ten square degrees, however a full three-detector network with the operation of two Cosmic Explorer detectors and the Einstein Telescope would allow of order 10 sources per year to be localized localized to better than one square degree five minutes before the merger~\cite{Nitz:2021pbr}. These multimessenger observations would allow exploration of a wealth of fundamental physics. 

Neutron stars are excellent astrophysical laboratories for ultra-dense matter. The physics of the star's interior is encoded in the gravitational waves emitted when neutron stars coalesce~\cite{2008PhRvD..77b1502F,2010PhRvD..81l3016H,2012PhRvD..85l3007D,2020GReGr..52..109C}, allowing us to probe the fundamental properties and constituents of matter in a phase that is inaccessible to terrestrial experiments~\cite{2017JPhG...44j4002S}. After a binary neutron star merges, oscillations of the hot, extremely dense remnant produce postmerger gravitational radiation. This as-yet-undetected signal probes the unexplored high-density, finite-temperature region of the quantum chromodynamic (QCD) phase diagram where new forms of matter are most likely to appear~\cite{2009PhRvD..80l3009Y,2016PhRvL.117d2501K,2016EPJA...52...50O,2019PhRvL.122f1101M}. Cosmic Explorer and Einstein Telescope are well-suited to observing postmerger gravitational waves~\cite{2018PhRvD..98d4044M,2018PhRvD..97b4049Y,2019PhRvD..99j2004M} and, together with multimessenger observations of the merger remnant, their observations will shape theoretical models describing fundamental many-body nuclear interactions and answer questions about the composition of matter at its most extreme, such as whether quark matter is realized at high densities~\cite{2019PhRvL.122f1102B,2019PhRvL.122f1101M,2020PhRvD.102l3023B}.

Gravitational wave standard sirens are expected to play an important role in the context of cosmology. Gravitational waves allow measurement of the luminosity distance of the source and, together with redshift measurements, can probe the distance-redshift relation~\cite{Schutz:1986gp}. Mesurement of the Hubble parameter using standard sirens does not require a cosmic distance ladder and is model-independent: the absolute luminosity distance is directly calibrated by the theory of general relativity. Approximately fifty additional multi-messenger binary neutron star observations would be needed to resolve the tension between the Planck and R19 measurements of $H_0$ with a precision of 1--2$\%$~\cite{Nissanke:2013fka,Mortlock:2018azx}. The precision of third-generation detectors, combined with deep optical-to-near-infrared observatories, would allow third-generation observatories to resolve this tension.

Multi-messenger observations of binary neutron star mergers are a promising new environment to probe weakly interacting light particles. Immediately after the merger, these remnants reach temperatures in the 30--100~MeV range and densities above $10^{14}~\text{g/cm}^3$, similar to the proto-neutron stars formed in core-collapse supernovae that have been used to place constraints on a wide range of scenarios. The large temperature and density of a post-merger remnant makes them very efficient at producing feebly interacting dark sector particles, which can escape this environment and lead to observational signals~\cite{Dietrich:2019shr,Harris:2020qim,Diamond:2021ekg}. Dark photons with masses in the 1--100~MeV range would be copiously produced and, for a large range of unconstrained couplings, would lead to a very bright transient gamma-ray signal originating from the dark photon decay~\cite{Diamond:2021ekg}. The precision and early warning offered by next-generation detectors allows the use of the associated gravitational-wave signal as a trigger and a timing measurement to help distinguish signal from background fluctuations and allows for gamma-ray observatories with narrower fields of view to observe events. Observations of gravitational waves from neutron star mergers can allow exploration of an object with a non-negligible contribution from vacuum energy to their total mass. The presence of vacuum energy in the inner cores of neutron stars occurs in new QCD phases at large densities, with the vacuum energy appearing in the equation of state for a new phase. This, in turn, leads to a change in the internal structure of neutron stars and influences their tidal deformabilities, which are measurable in the gravitational-wave signals of merging neutron stars~\cite{Csaki:2018fls}.

The vast cosmological distances---redshifts in excess of $z\sim 20$---over which gravitational waves travel, will severely constrain violation of local Lorentz invariance and the graviton mass~\cite{Will:2014kxa}. Such violations or a non-zero graviton mass would cause dispersion in the observed waves and hence help to discover new physics predicted by certain quantum gravity theories. At the same time, propagation effects could also reveal the presence of large extra-spatial dimensions that lead to different values for the luminosity distance to a source, as inferred by gravitational-wave and electromagnetic observations~\cite{Belgacem:2018lbp, Pardo:2018ipy}, or cause birefringence of the waves predicted in certain formulations of string theory~\cite{Alexander:2009tp, Alexander:2017jmt}. The presence of additional polarizations predicted in certain modified theories of gravity, instead of the two degrees of freedom in general relativity, could also be explored by future detector networks~\cite{Will:2014kxa, Isi:2017fbj}.

Massive stars undergoing core-collapse supernova also generate gravitational waves from the dynamics of hot, high-density matter in their central regions. Cosmic Explorer and Einstein Telescope will be sensitive to supernovae within the Milky Way and its satellites, which are expected to occur once every few decades~\cite{2019PhRvD.100d3026S}. Core collapses should be common enough to have a reasonable chance of occurring during the few-decades-long lifetime of Cosmic Explorer. A core-collapse supernova seen by Cosmic Explorer will have a significantly larger signal-to-noise ratio than one seen by current gravitational-wave detectors, and could be detected by a contemporaneous neutrino detector like DUNE~\cite{2021EPJC...81..423A}, giving a spectacular multimessenger event. Detection of a core-collapse event in gravitational waves would provide a unique channel for observing the explosion's central engine~\cite{2020arXiv201004356A} and the equation of state of the newly formed protoneutron star~\cite{2018ApJ...861...10M}. Detection of a supernova would be spectacular, allowing measurement of the progenitor core's rotational energy and frequency measurements for oscillations driven by fallback onto the protoneutron star~\cite{2021PhRvD.103b3005A}.

\subsection{Space-Based Graviational Wave Detectors}
\label{sec-spacedBasedGW}

\noindent
\chapterauthor[1]{James Ira Thorpe}\orcidlink{0000-0001-9276-4312}
\chapterauthor[2]{Guido Mueller}
\chapterauthor[3]{David Shoemaker}\orcidlink{0000-0002-4147-2560}
\chapterauthor[4]{Kelly Holley-Bockelmann}\orcidlink{0000-0003-2227-1322}
 \\
 \begin{affils}
    \chapteraffil[1]{Gravitational Wave Laboratory, NASA Goddard Space Flight Center, Greenbelt, MD 20771, USA}
    \chapteraffil[2]{Department of Physics, University of Florida, Gainesville, FL 32611, USA}
    \chapteraffil[3]{LIGO Laboratory, Massachusetts Institute of Technology, Cambridge, MA 02139, USA}
    \chapteraffil[4]{Department of Physics and Astronomy, Vanderbilt University, Nashville, TN 37235, USA}
 \end{affils}

Earth-based gravitational wave observatories are typically designed to detect gravitational waves (GWs) at frequencies above $\sim$1~Hz due to the increase in both environmental and anthroprogenic noise levels at lower frequencies, as well as the impractically large antenna sizes needed to optimally couple to low-frequency GWs. Moving the observatory to space allows both of these issues to be addressed.  Specifically, designers of space-based observatories have focused on the millihertz frequency band, which is rich in both the number of sources and variety of source types. The most mature space mission concept is the Laser Interferometer Space Antenna (LISA)~\cite{Danzmann_1996}, which is currently in development as a Large-class mission of the European Space Agency with substantial contributions from NASA and many European member stages~\cite{L3Proposal}. LISA, which expects to be operational in the 2030s, will observe in the band between $0.1\,\textrm{mHz} < f < 1\,\textrm{Hz}$. LISA and other LISA-like concepts, such as China's Taiji and TianQin programs~\cite{Taiji,2016CQGra..33c5010L}, define the first generation of space-based GW observatories, which can expect to observe in the 2030s and 2040s. Further in the future, second generation facilities may expand capabilities in frequency, sensitivity, and angular resolution.

It is worth recognizing that one of the scientific motivations for developing an entirely new kind of instrument is the opportunity to examine old problems from a new perspective. In this sense, all of GW science is `multimessenger'---GW observations will yield direct information about populations of objects, including information such as masses and distances that are difficult to measure by other means---and this information will provide additional insight when combined with electromagnetic (EM) and particle observations of those same populations. For example, LISA will perform a census of massive black hole mergers into the Cosmic Dawn, while X-ray missions will measure growth through accretion onto single black holes; the combined information will paint a fuller picture of the relative importance of these massive black hole growth channels through time.

With that context, there are a number of anticipated milliHertz GW measurements for which there are opportunities for \textit{contemporaneous} multimessenger astronomy--- observations of the same physical system on human timescales with both GWs and other messengers. A white paper~\cite{Astro2020_LISA_MMA} was developed for the 2020 Decadal Survey of Astronomy and Astrophysics~\cite{NAP26141} summarizing the multi-messenger opportunities for LISA, both contextual and contemporaneous. Here we briefly summarize the latter category:

\begin{description}

\item[Cosmology with Standard Sirens]
One of the most vexing problems in modern astronomy is the apparent discrepancy between measurements of the Hubble flow made using standard candles versus those inferred from the Cosmic Microwave Background (CMB). GWs offer an opportunity to bring a third technique to the problem--- standard sirens~\cite{2005ApJ...629...15H}. Standard sirens take advantage of GWs ability to measure luminosity distances to chirping sources directly, without invoking the multi-rung distance ladder needed to infer distances to standard candles such as Type Ia supernova (SN). When combined with a redshift obtained from an EM measurement, this technique can yield an independent measurement of the Hubble constant. This technique was first demonstrated with the binary neutron star (NS) event GW170817~\cite{2021ApJ...909..218A}. While future binary NS events will improve the accuracy of this measurement, space-based GW observatories will allow the technique to be extended to much higher redshifts using GW observations of stellar-mass black holes at low redshifts ($z\lesssim 0.1$), extreme mass-ratio inspirals at moderate redshifts ($0.1 \lesssim z\lesssim 2$), and massive black hole mergers at high redshifts ($z \gtrsim 1$)~\cite{2017JPhCS.840a2029T}. It is worth noting that these sources are observable by GWs for hours to years, and this longer interval provides a better opportunity for coordinated EM/Particle observation campaigns. The primary challenge for this technique is identifying the EM counterpart to the GW event, so that both redshift and luminosity distance can be measured. Approaches range from coincident searches for EM signals produced by the event, to searches for the host galaxy using targeted surveys within the GW error volume, to correlation and statistical inference between a population of GW events and galaxy catalogs.

\item[Physics of Massive Black Hole Accretion]
The massive black holes (MBHs) that LISA and other space-based GW detectors will observe merging are found in a surprisingly large variety of galaxy hosts, including some with low-level Active Galactic Nuclei (AGN) activity. GW observations will provide direct information about the MBHs, including mass, spin, and orbital properties of this central engine, while EM observations will yield information about the physical properties of the material in the AGN disk--- temperatures, densities, and magnetic fields. The combined set of observations will allow dramatic improvement in modeling the detailed physics of accretion flow in AGN. As with the standard sirens, the challenge for this class of investigation lies in locating and identifying the host galaxy where the GW-observed merger is taking place. Since EM measurements coincident with the merger are required, extra emphasis is placed on the ability to identify the target \textit{prior} to the merger. As GW localization rapidly improves as the merger is approached, this places extra emphasis on rapid production of GW alerts (communications and data analysis), EM facilities with fast survey capability, and a robust and large time-domain database of the sky.   

\item[Astrophysics of Compact White Dwarf Binaries]
By far the most numerous population of millihertz GW sources are compact binary systems, predominantly white dwarf binaries, in the Milky Way. LISA is expected to individually resolve perhaps $20 \times 10^4$ of these systems, with millions more contributing to an unresolved background that will be detectable above the instrument noise floor. Unlike other sources, these galactic binaries have long evolutionary timescales and, as a result, are observed as persistent GW sources. This greatly reduces the difficulty in conducting multimessenger observations. In fact, roughly a dozen systems that will be detectable by LISA have already been identified through EM surveys~\cite{2018MNRAS.480..302K}, representing a population of \textit{guaranteed multi-messenger sources}. Additional surveys before the launch of LISA, such as those conducted by the Vera Rubin Telescope~\cite{2020ApJ...900..139F}, will add to this population. Once LISA launches, many more multimessenger systems will be added to the catalog using the reverse process--- identification of EM sources from GW triggers. The science opportunities for this population are vast. The sheer increase in the number of identified compact binary systems will help constrain models of the end states of stellar evolution. For some individual systems, GW measurements will be able to measure changes in the orbital frequency caused by GW emission, mass transfer, or a combination of the two. The combined constraints from GW and EM observations are improved by an order of magnitude over what either messenger can do alone~\cite{2019arXiv190305583L}. 
\end{description}

\begin{figure}[h]
\centering
\includegraphics[width=0.7\textwidth]{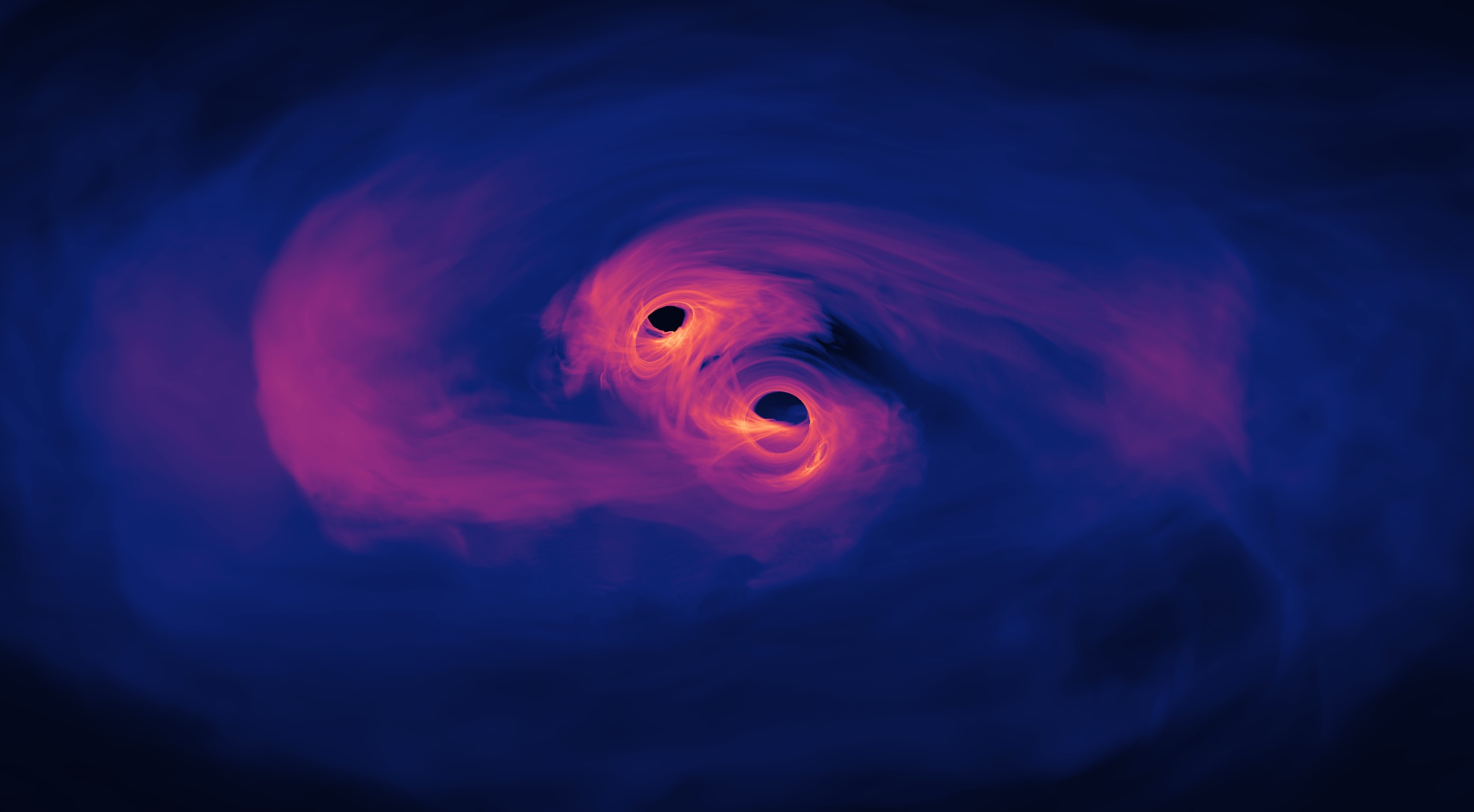}
\caption{Simulated electromagnetic emission from accreting binary massive black hole during the late inspiral phase. [NASA/GSFC]}
\end{figure}

As with development of ground-based GW detectors, LISA will be just the beginning of GW detectors in space~\cite{GWICRoadmap}. The landscape of potential second-generation detectors has been the subject of several recent white papers associated with the US Astronomy Decadal Survey~\cite{2019BAAS...51g.243M} and the European Space Agency's Voyage 2050 process\footnote{https://www.cosmos.esa.int/web/voyage-2050}. The opportunities for second generation detectors fall into three main categories: increasing the sensitivity of the detector in roughly the same measurement band, moving to higher~\cite{2021ExA....51.1427A} or lower~\cite{2021ExA....51.1333S} frequency bands relative to LISA, or increasing the number of baselines and improving the astrometric localization capabilities~\cite{2021ExA....51.1427A}. Advances in each of these areas has the potential to advance the scientific investigations outlined above.  

For example, a detector with sensitivity at lower frequencies than LISA would preferentially measure more MBH mergers and provide longer early-warning times for mergers of a given mass, both of which could significantly improve prospects for detecting a contemporaneous EM signal. Such a detector would also allow the study of the still-unsolved puzzle of the rapid emergence of high-z quasi-stellar objects (QSOs), with major implications on the evolution of supermassive black holes and galaxy formation. The observation of the myriad of galactic binaries with orbital periods below the LISA band would also expand our knowledge about their distribution and has the potential to uncover new sources such as binary brown dwarfs and exoplanets. Technologically, such an observatory is within reach using incremental evolution of LISA technologies, although the longer baselines may require somewhat larger telescopes or more powerful lasers depending on the precise sensitivity targets.

A mission covering the frequency gap between the Laser Interferometer Gravitational-wave Observatory (LIGO)~\cite{Harry_2010} and LISA ($.03\,\textrm{Hz}\lesssim f \lesssim 3\,\textrm{Hz}$) would allow observation of the mass gap between LISA's MBH mergers and LIGO's stellar-mass black hole mergers. Proposed missions in this frequency range also promise a sensitivity that could allow unprecedented tests of General Relativity (GR) or other beyond-the-Standard-Model physics. This is also the frequency range where white dwarfs in binary systems make contact, giving rise to an unprecedented early warning system for nearby supernovae. However, the shorter arms required to optimize GW coupling in this band require corresponding improvements in interferometric sensitivity by at least two to three orders of magnitude, a technological challenge which will need to be addressed as soon and which may require departure from the LISA architecture. 

A final category of improvement over LISA for future GW missions would be increasing angular resolution through the addition of multiple baselines or multiple constellations. Rather than relying on long-period modulations of the GW signal due to the orbit of the constellation, such a system would have vastly improved sky localization, especially in the early stages of a detection event. This could have an important impact on multi-messenger studies by allowing facilities with higher sensitivities, but smaller fields of view, to participate in the search process. Realizing such a mutli-detector network in space would likely require changes to the project management and engineering approach away from highly-specialized, highly-reliable spacecraft to easily manufacturable and potentially replaceable spacecraft, forming a robust and flexible network.

\subsection{Current Space-Based Gamma-Ray Telescopes}
\label{sec-currentGammaRay}

\chapterauthor[1]{Colleen A. Wilson-Hodge}
\chapterauthor[1]{Joshua Wood}
\chapterauthor[2]{Eric Burns}\orcidlink{0000-0002-2942-3379}
\\
\begin{affils}
   \chapteraffil[1]{NASA Marshall Space Flight Center, Huntsville, AL 35805}
   \chapteraffil[2]{Department of Physics \& Astronomy, Louisiana State University, Baton Rouge, LA 70803, USA}
\end{affils}


The two pillars of modern multimessenger astrophysics, GW170817 \cite{Abbott17} and TXS 0506+056 \cite{IceCube:2018dnn} were both made possible by gamma-ray discovery of the electromagnetic half of these events. The nuclear gamma-rays from the first multimessenger source in astronomy, SN 1987A, is still one of the strongest observational results enabling our current understanding of the supernova engine. Gamma-ray counterparts are the foundation for all modern multimessenger science. The essential gamma-ray capabilities that enable multimessenger science are a very wide field-of-view with good sensitivity, rapid response and alerts, good time resolution, broad energy coverage in the keV to GeV range, and good gamma-ray localizations. No single gamma-ray mission provides all of these capabilities.

A very wide field of view, as much of the sky as possible, is needed because the gamma-ray emission is short-lived and nearly coincident with the GW or neutrino emission. Detection of a gamma-ray counterpart is necessary to determine if various types of mergers produce counterparts and to determine the astrophysical origin of neutrinos. Rapid response and alerts enable follow-up of events to detect afterglow or kilonovae. Good time resolution is needed to correlate events and to measure the time difference between multimessenger events and gamma-ray emission. For gravitational wave events, the time delay can be used to measure the speed of gravity and to constrain properties of the jet and central engine. Broad energy coverage in the keV to GeV range enables measurements of energy spectra,  GRB energetics, and detections of flaring gamma-ray blazars. Good gamma-ray localizations increase the confidence of the associations with the multimessenger events and enable follow-up.

The Fermi Gamma-ray Space Telescope, launched in 2008, includes two instruments, the Gamma-ray Burst Monitor (GBM, \cite{Meegan09}) and the Large Area Telescope (LAT, \cite{Atwood09}). The GBM instrument is sensitive from 8 keV to 1MeV and has a time resolution of 2$\mu$s. GBM views the entire sky that is not occulted by the Earth and provides automated gamma-ray burst (GRB) triggers rapidly disseminated to the community through the Gamma-ray Burst Coordinates Network (GCN, \cite{Barthelmy1995}). These automated alerts include degree-scale burst localizations derived from the relative rates GBM detectors. Spectral analysis of GBM detected GRBs constrains the peak energy for most GRBs, providing a measure of the GRB energetics. On August 17, 2017, GBM independently detected and reported GRB 170817A \cite{Goldstein17}, the GRB associated with GW170817 \cite{Abbott17}, before the gravitational wave event was announced by LIGO. GBM has worked closely with the LVK to provide joint automated alerts in the next LVK observing run (O4). In addition to its automated triggers, GBM continues to provide sensitive sub-threshold searches for GRBs associated with gravitational waves \cite{Hamburg2020,Goldstein2019}. These sub-threshold searches can also be used to search for gamma-ray counterparts to neutrinos.  The LAT instrument on-board Fermi is a pair production telescope with silicon strip trackers, a Cesium Iodide calorimeter, and a plastic scintillator anticoincidence system. The LAT has a large field of view of about 2.4 steradians and the Fermi spacecraft is operated such that the LAT covers the entire sky approximately every three hours. The LAT data provide the arrival time, arrival direction, and energy for individual detected photons. Less than 10\% of the GRBs detected by GBM are also detected by the LAT \cite{Ajello2021}, however for those GRBs, the LAT provides much more precise localizations. A large number of high energy sources are monitored by the LAT,  including a flaring blazar, TXS 0506+056 associated with high-energy neutrinos \cite{IceCube:2018dnn}. 

The International Gamma-Ray Astrophysics Laboratory (INTEGRAL, \cite{Winkler2003}) was launched in 2002.  The thick Anti-Coincidence Shield surrounding the Spectrometer on INTEGRAL (SPI-ACS, \cite{vonKienlin2003}, is an effective nearly omni-directional GRB detector, reaching 0.7 m$^2$ above about 75 keV with a time resolution of 50 ms and a single energy band. The SPI-ACS detected GRB 170817A \cite{Savchenko2017}, associated with GW170817. The SPI-ACS has no localization capability. SPI-ACS announces GRB detections through the GCN.

The Neil Gehrels Swift Observatory \cite{Gehrels2004} was launched in 2004. Swift comprises three instruments, the Burst Alert telescope (BAT), the X-ray telescope (XRT), and the UV/Optical Telescope (UVOT). The BAT is sensitive to hard X-rays from 15-150 keV, with a 2 steradian field of view and arcminute localizations \cite{Barthelmy2004}. Swift sends out rapid alerts through the GCN. The Swift spacecraft can rapidly repoint in response to GRBs detected with the BAT, using the XRT (0.3-10 keV, \cite{Burrows2005}) to image and localize to arcsecond scales. The UVOT (170-600 nm, \cite{Roming2005} detected a UV counterpart to GW170817 \cite{Evans2017}, indicating that the event produced a hot blue kilonova. A new capability has been developed \cite{Tohuvavohu2020} to provide event level Swift BAT data on demand in response to GW events, GRB detections from GBM and other instruments, and other potentially exciting gamma-ray events. This provides arcminute localizations for GRBs detected in the subthreshold searches, and for non-detections, improves localizations by eliminating sky regions visible to Swift BAT. 

The InterPlanetary Network (IPN) is an international collaboration that combines information from multiple GRB monitors \cite{hurley2013interplanetary}. Over the last few decades the IPN has near perfect sky coverage and almost unity livetime, and utilizing the finite speed of light to triangulate bursts on the sky. It has been key in the discovery of soft gamma-ray repeaters (now known to be a key magnetar class), magnetar giant flares, the first counterpart to fast radio bursts, matching prompt GRBs to optically-discovered collapsars, and, thus far, providing upper limits for externally discovered events of interest including orphan afterglows, neutrinos, and more.

The current fleet of gamma-ray missions is quite old, ranging from 14-20 years, with some of the IPN missions being much older. Maintaining and improving these capabilities in the future is crucial to multimessenger astrophysics.  Strategic coordination between space and ground based assets, so that limited space missions overlap with ground facilities operating at their full sensitivity is key to the future success of multimessenger astrophysics.

\subsection{Next-Generation Space-Based Gamma-Ray Telescopes}
\label{sec-nextGenGammaRay}
\noindent
\chapterauthor[]{Carolyn Kierans}
\\
\begin{affils}
\chapteraffil[]{NASA Goddard Space Flight Center, Greenbelt, MD 20771, USA}
\end{affils}

There is a significant community effort to develop the next-generation gamma-ray mission with the goal of advancing multimessenger astrophysics. As described in Section~\ref{sec-currentGammaRay}, a wide field of view, rapid response, good timing resolution, broad energy coverage, good localization, and, of course, high sensitivity, are needed to detect and study the high-energy electromagnetic counterparts to multimessenger events. The fleet of missions in development and proposal stages satisfy different combinations of these required capabilities.

The past few years have seen a new fleet of small satellites (CubeSats and SmallSats) that are being developed for the detection of gamma-ray burst (GRB) prompt emission. These are simple scintillator instruments optimized for all-sky transient detections with no imaging capabilities, though rough localizations are achievable by analyzing the rate-differential in multiple on-board detectors~\cite{2015ApJS..216...32C}. Modeled off of \textit{Fermi}-GBM~\cite{Meegan2009}, BustCube~\cite{perkins2020}, Glowbug~\cite{Grove2020}, and StarBurst~\cite{StarBurst} are just a few of the many missions in this vein that are current being developed. With a simple detector design, these instruments are also being worked on by university student groups. With the increasing number of small missions dedicated to the all-sky monitoring of GRBs, the community is also organizing to maximize the science output of the arrays of instruments through collaboration\footnote{grbnanosats.net}.

To move beyond transient detections, a telescope capable of imaging allows for source localization, background reduction and rejection, and the study of steady-state sources or those with longer variability, such as flaring blazars. Imaging in gamma rays and the low end of the MeV range requires a Compton telescope and above $\sim$10~MeV, pair conversion dominates. The Compton Spectrometer and Imager (COSI) is a Small Explorer mission that was recently selected for launch in 2025~\cite{Tomsick:2021wed}. COSI is a wide-field telescope designed to survey the gamma-ray sky at 0.2--5~MeV. With excellent spectral resolution, and background rejection through Compton imaging, COSI will detect and localize short GRBs from merging neutrons stars and observe flaring blazers that are potential neutrino source counterparts.

The gamma-ray astrophysics community is actively pushing for a mission that can bring the \textit{Fermi} Large Area Telescope (LAT;~\cite{Atwood09}) capabilities to the MeV range where multimessenger sources are the brightest. While no large-scale gamma-ray missions have been selected, many mission concepts have been developed and proposed. Most of these missions combine Compton imaging and pair capabilities in the same instrument volume: AMEGO~\cite{2020SPIE11444E..31K} and AMEGO-X~\cite{2021arXiv210802860F}, GRAMS~\cite{ARAMAKI2020107}, and APT~\cite{Buckley2019Advanced}. These missions provide $\sim1^{\circ}$ localization for transients and a sensitive, wide field of view for source monitoring and observations. GECCO~\cite{Moiseev:2021sx} uses Compton imaging combined with a coded mask to achieve excellent angular resolution for more detailed studies of specific sources and the dense Galactic center region.

Another sought-after capability is gamma-ray polarization to constrain the mission models of GRBs and neutrino counterparts. Transient detectors, such as POLAR~\cite{2021arXiv210902977K} and LEAP~\cite{McConnell.2021}, are optimized for GRB observations. At higher energies, concepts like AdEPT~\cite{10.1117/12.2312732} will operate in the pair regime where polarization measurements probe the fundamental processes of particle acceleration in active astrophysical objects often associated with multimessenger sources. 

Here we have only highlighted a small subset of the current instrument development in the community as the drive to enable multi-messenger astrophysics strengthens. We encourage the interested reader to reference the Snowmass CF07 Gamma-Ray Experiments white paper \textit{The Future of Gamma-Ray Experiments in the MeV--EeV Range} for further details.

\subsection{Current Ground-Based Gamma-Ray Telescopes}
\label{sec-curGB-GRT}
\noindent
 \chapterauthor[1,2]{Brenda Dingus}\orcidlink{0000-0001-8451-7450}
 \chapterauthor[1]{Jordan Goodman}\orcidlink{0000-0002-9790-1299}
 \chapterauthor[1]{Brian Humensky}\orcidlink{0000-0002-1432-7771}
 \\
 \begin{affils}
    \chapteraffil[1]{University of Maryland, College Park, College Park, MD 20742, USA}
    \chapteraffil[2]{Los Alamos National Laboratory, Los Alamos, NM 87545, USA}
 \end{affils}
At the highest energies, the flux of gamma rays is too low to be detected with satellite sized detectors.  Two different technologies are used to detect gamma rays from the ground: observing the showers of particles in our atmosphere with large mirrors to detect Cherenkov light, or with high-altitude arrays of particle detectors.  The Imaging Atmospheric Cherenkov Telescopes (IACTs) observe the entire shower and therefore can measure gamma-ray energy as well as distinguish gamma rays from the more plentiful background of cosmic rays.  However, IACTs can operate only on clear nights and their field of view (FoV) is limited to a diameter of a few degrees. Extensive Air Shower (EAS) particle detectors operate continuously and have a wide FoV of about 2~sr---observing about $2\pi$~sr daily---but have a higher energy threshold and less efficient cosmic-ray rejection.  

The complementary nature of these two technologies requires both these technologies to probe multi-messenger phenomena.  EAS detectors---with 1/6 of the sky always observed---are uniquely capable of catching short, bright transients and providing long term, continual monitoring of the TeV sky. For example, the prompt emission from short Gamma Ray Bursts (GRBs; likely correlated with neutron star mergers that produce gravitation waves) has a duration {$<$1--2} seconds as well as the evaporation of Primordial Black Holes (PBHs) in which the highest energy emission is produced in the last fraction of a second.  EAS detectors can provide sub-degree localization and multimessenger triggers, as well as observations of multimessenger sources prior, during, and after another multimessenger trigger.  IACTs must slew their mirrors to the direction of a multimessenger trigger, taking tens of seconds; however, their better flux sensitivity and energy resolution is ideally suited to detecting the decaying afterglow emission of GRBs and measuring the temporal variation of energy spectra throughout an Active Galactic Nuclei (AGN) flare.  

Current IACTs  and EAS detectors have made important observations of multimessenger sources. The nature of neutrino as well as gravitational-wave sources is not well constrained, but both are likely particle accelerators which will produce gamma rays at energies which can be detected from the ground.  Gamma-ray observations of multimessenger sources can provide the link to identification with lower-energy sources as well as constrain both fundamental physics and particle acceleration mechanisms.  For example, the most distant transient sources provide strong constraints on Lorentz Invariance.  Both IACTs and EAS have observed flaring in AGN--- one AGN flare observed by an IACT was associated with an IceCube neutrino.  Also, IACTs and EAS detectors have observed gravitational-wave sources, placing upper limits on the highest energy gamma-ray emission.  IACTs have significantly detected several GRB afterglows, and EAS detectors have placed strong limits on the prompt emission.  

Three major arrays of IACTs are currently operating: H.E.S.S.~\cite{2006A&A...457..899A} in the Southern Hemisphere, and MAGIC~\cite{2016APh....72...76A} and VERITAS~\cite{2015ICRC...34..771P} in the Northern Hemisphere, with the latter two separated by about 90 degrees in longitude. All three have programs in place to respond rapidly to alerts regarding GRBs, astrophysical neutrinos, and gravitational-wave events, and can begin observations within anywhere from a few tens seconds to several minutes after a trigger. Their angular resolution (below 0.1~degree at 1~TeV) and background rejection provide deep instantaneous sensitivity, though some serendipity is required for an IACT to be in position to respond quickly to an interesting alert. Four long GRBs have been detected at 5~sigma or above by IACTs in recent years, including GRB190114C~\cite{MAGIC:2019lau,MAGIC:2019irs} and GRB201216C~\cite{2020GCN.29075....1B} by MAGIC and GRB180720B~\cite{Abdalla:2019dlr} and GRB190829A~\cite{2019GCN.25566....1D} by H.E.S.S.

Currently there are two major EAS detectors operating in the TeV--PeV range. They are HAWC in Mexico and LHAASO in China. Both continuously monitoring the Northern sky, but observe different regions of the sky due to their different longitudes.  Their data are recorded and can be searched in near real-time for transient events, sending alerts to other instruments, or their archival data can be searched to look for events that are later observed/reported by other detectors. These include AGN flares, GRBs, astrophysical neutrinos, gravitational-wave events, or even fast radio bursts. Since they observe the particles that reach the ground, they don’t require dark nights and therefore can observe the parts of the sky that are in daylight (which is approximately half the sky for half the year). 

HAWC has been in operation since 2015 and, up to now, has reported on flaring AGN and set limits on GRBs, neutrino events, PBHs, and gravitational waves. Its threshold is $\sim$1 TeV and has measured galactic-source spectra beyond several hundred TeV.  HAWC’s newest reconstruction algorithms have enabled a significantly lower threshold and wider FoV that should provide greater sensitivity to transients. LHAASO’s $\mathrm{km}^2$ has come online within the last two years and has already observed sources extending up to, and possibly beyond, a PeV. They have recently started operation with their water Cherenkov ponds (about four times the area of HAWC with better light collection), which will give them improved sensitivity below 1~TeV. LHAASO can be expected to operate for the foreseeable future.

\subsection{Future Ground-Based Gamma-Ray Telescopes}
\label{sec-futGB-GRT}
\noindent
 \chapterauthor[1,2]{Brenda Dingus}\orcidlink{0000-0001-8451-7450}
 \chapterauthor[1]{Jordan Goodman}\orcidlink{0000-0002-9790-1299}
 \chapterauthor[1]{Brian Humensky}\orcidlink{0000-0002-1432-7771}
 \\
 \begin{affils}
    \chapteraffil[1]{University of Maryland, College Park, College Park, MD 20742, USA}
    \chapteraffil[2]{Los Alamos National Laboratory, Los Alamos, NM, 87545, USA}
 \end{affils}
Future IACTs and EAS particle detectors will have more than an order of magnitude improved sensitivity over current observatories. Observatories are planned for both the Northern and Southern Hemisphere, which is required to catch the relatively rare multimessenger transients. There are two major international efforts planned--- the Cerenkov Telescope Array (CTA) and the Southern Wide-field Gamma-ray Observatory (SWGO). Both of these observatories were endorsed by the National Academy of Sciences Astro 2020 Particle Astrophysics Group panel. 

CTA~\cite{2019scta.book.....C} will consist of IACT arrays located at two sites--- a Northern site on La Palma, and a Southern site at the European Southern Observatory in Paranal, Chile. Telescopes for CTA are being designed in three size classes: large-size telescopes (LSTs) to cover low energy ranges (30~GeV--1~TeV), medium-size telescopes (MSTs) to cover medium energy ranges (0.1--10~TeV), and small-size telescopes (SSTs) to cover high energy ranges (1-- {$>100$}~TeV). The LSTs, in particular, are designed with a focus on rapid response to transients and will be capable of slewing to any point on the sky within 20 seconds. The LSTs will have a field of view (FoV) of 4.5~degrees, and while the MSTs will slew more slowly (60--90 seconds to reach any point on the sky), their 7--8 degree FoV and larger number (25 at the Southern site and 15 at the Northern site) provide the opportunity to survey large areas deeply and rapidly--- ideal for follow-up of GW events.

The EAS detector, SWGO~\cite{Albert:2019afb,Hinton:2021rvp,Schoorlemmer:2019gee} (see Section~\ref{sec-SWGO}), will be built at an altitude even higher than HAWC and LHAASO with a substantially larger water Cherenkov detector to optimize the detection of lower-energy gamma rays.  A lower energy threshold of detection allows more distant sources to be detected as lower-energy gamma rays are less attenuated by pair production with extragalactic infrared photons. With SWGO in the Southern Hemisphere and LHAASO in the Northern Hemisphere, there will be, over the course of the day, nearly full sky coverage at a threshold where transient events can be observed from $\sim$100~GeV to above 1~PeV.

\subsubsection{The Southern Wide-field Gamma-ray Observatory}
\label{sec-SWGO}
\chapterauthor[ ]{ }
\vspace{-0.2in}
 \addtocontents{toc}{
     \leftskip3cm
    \scshape\small
    \parbox{5in}{\raggedleft Andrea Albert et al. on behalf of the SWGO Collaboration}
    \upshape\normalsize
    \string\par
    \raggedright
    \vskip -0.19in
    }
 
\noindent
\nocontentsline\chapterauthor[]{Andrea Albert$^{1}$}\orcidlink{0000-0003-0197-5646}
\nocontentsline\chapterauthor[]{Chad Brisbois$^{2}$}\orcidlink{0000-0002-5493-6344}
\nocontentsline\chapterauthor[]{Kristi Engel$^{1,2}$}\orcidlink{0000-0001-5737-1820}
\nocontentsline\chapterauthor[]{J. Patrick Harding$^{1,3}$}\orcidlink{0000-0001-9844-2648}
\nocontentsline\chapterauthor[]{on behalf of the SWGO Collaboration$^{4}$}
\\
 \begin{affils}
   \chapteraffil[1]{Physics Division, Los Alamos National Laboratory, Los Alamos, NM, 87545, USA}
   \chapteraffil[2]{University of Maryland, College Park, College Park, MD, 20742, USA}
   \chapteraffil[3]{Michigan State University, East Lansing, MI, 48824, USA}
   \chapteraffil[4]{\texttt{https://www.swgo.org/SWGOWiki/doku.php?id=collaboration}}
\end{affils}

\begin{figure}[htb]
\centering
	\includegraphics[width=\textwidth]{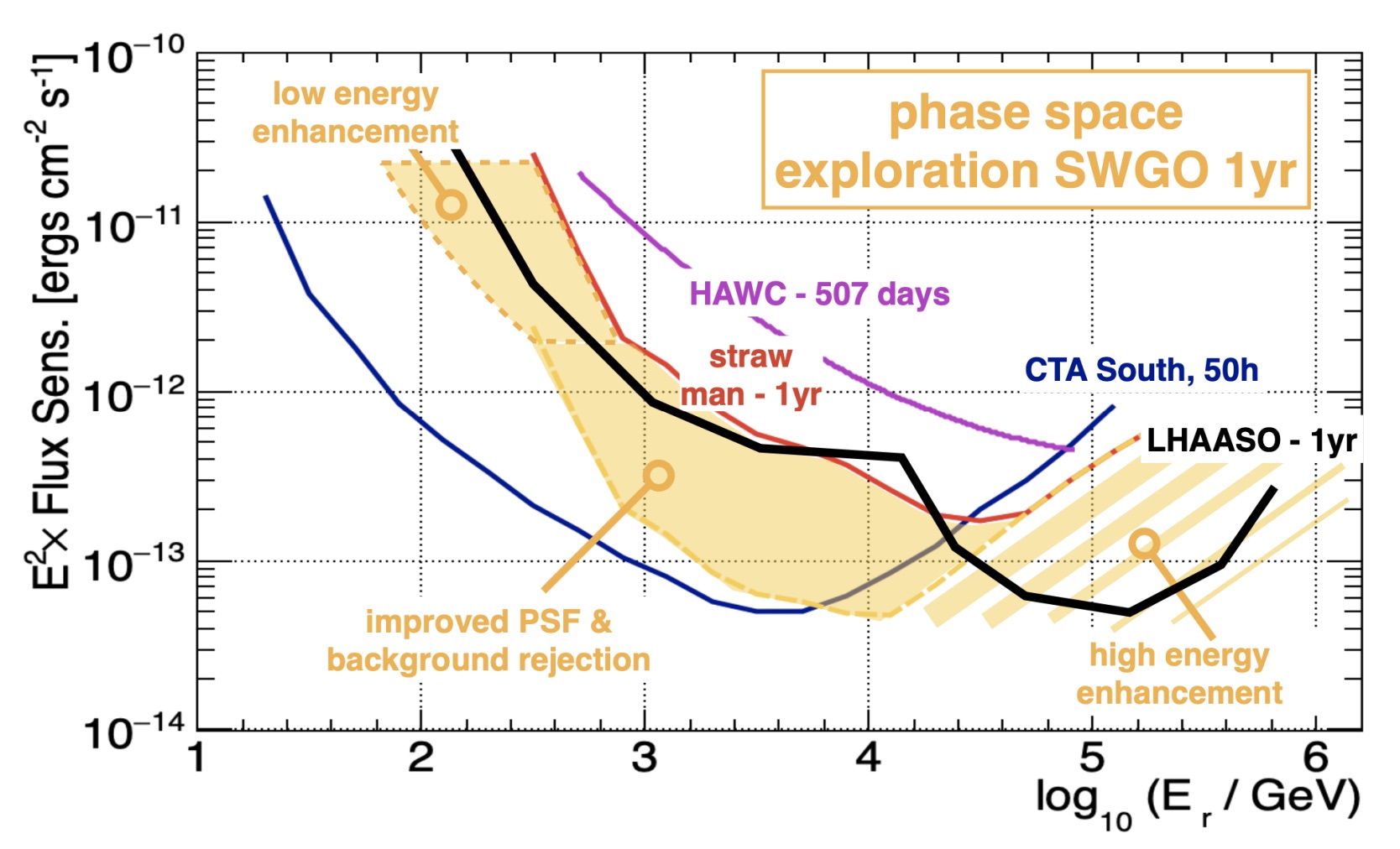}
	\caption{The differential sensitivity of the HAWC Observatory~\cite{Abeysekara_2019}, LHAASO~\cite{2019arXiv190502773B}, and the Southern portion of the Chrenkov Telescope Array (CTA)~\cite{CTAConsortium:2018tzg} with the phase space that will be explored in the design studies for SWGO. Figure from Ref.~\citenum{Hinton:2021rvp}.}	
	\label{fig:SWGO-sensi}
\end{figure}

The Southern Wide-field Gamma-ray Observatory (SWGO) is a proposed EAS experiment that would be located in South America. The design will build on the technology and successes of the High-Altitude Water Cherenkov (HAWC) Observatory~\cite{Abeysekara_2019}, which also inspired the Large High-Altitude Air-Shower Observatory (LHAASO)~\cite{2019arXiv190502773B}. Since 2015, HAWC has discovered new TeV sources and source classes, set new world-leading limits on dark matter decay and annihilation, and played a crucial role in multi-messenger observations~\cite{2015ApJ...800...78A,2017ApJ...841..100A,2017A&A...607A.115I,2017ApJ...842...85A,2017ApJ...843...39A,2017ApJ...843...40A,2017ApJ...843...88A,2017ApJ...843..116A,2017ApJ...848L..12A,2017Sci...358..911A,2018ApJ...853..154A,2018JCAP...02..049A,2018JCAP...06..043A,2018Sci...361.1378I, 2018PhRvD..98l3012A,2018Natur.562...82A,2019JCAP...08..023F, 2019arXiv190512518H,2020MNRAS.497.5318F}. Like HAWC, SWGO is planned to be a ground-based array of water Cherenkov detectors (WCDs) that detect particles in extensive air showers created by incident gamma rays in the upper atmosphere with a duty cycle of $\sim$100\%. It will observe gamma rays from $<$500~GeV to $>$100~TeV and have an instantaneous field of view of $\sim$2~sr. More details on the design of SWGO and the impact of the scientific goals of the Collaboration on that design can be found in Refs.~\citenum{Albert:2019afb,Hinton:2021rvp,Schoorlemmer:2019gee}, with the phase space that SWGO will occupy, showcasing ideal complementarity with existing and planned experiments, being shown in Figure~\ref{fig:SWGO-sensi}. A diverse science portfolio is possible with SWGO with such a design approach and heritage, including multimessenger studies. \\

\noindent\textbf{SWGO and Cosmic Rays: }

Cosmic-ray science goals with SWGO include measuring the cosmic-ray spectrum up to the so-called `knee' ($10^{15}$~eV). If the composition near the knee is dominated by protons, then $10^{17}$~eV is the end of the Galactic component to the cosmic-ray spectrum, with extragalactic sources dominating beyond the knee~\cite{Thoudam:2016syr}. If the composition is mostly dominated by heavier nuclei (primarily Carbon, Nitrogen and Oxygen), the interpretation of the knee is less clear. The KArlsruhe Shower Core and Array DEtector (KASCADE) results suggest the particle flux at the knee predominantly consists of protons ~\cite{2009APh....31...86A}, while another EAS experiment, ARGO-YBJ (Astrophysical Radiation Ground-based Observatory at YangBaJing), finds that the spectral index at particle energies below the knee is $-2.63 \pm 0.06$~\cite{Bartoli_2014}, and steepens to $-3.34 \pm 0.28$ at higher energies~\cite{2015_argoypj}, meaning protons may not be dominant at the knee in favor of extragalactic cosmic rays~\cite{PhysRevD.74.043005,2015_argoypj}. Additionally, direct detection cosmic-ray experiments find a hardening of the He spectrum at rigidities (momentum/charge) larger than 100~GV~\cite{article_AMSHE}. EAS arrays (such as the Pierre Auger Observatory or Telescope Array) measure cosmic rays with energies from $\sim$100~TeV to $\sim$~a few EeV~\cite{PierreAuger:2020qqz,2013ApJ...768L...1A}, while the direct detection satellites measure the spectrum up to $\sim$10~TeV~\cite{Mocchiutti:2014jja}. While conceived primarily as a high-energy gamma-ray observatory, the design of SWGO enables sensitive tracing of the components of the `knee' up to $10^{16}$~eV, measuring the maximum acceleration energies of Galactic sources, making it well suited to bridge the energy range between indirect and direct experiments.

First observed by Milagro, a large scale anisotropy has been observed in the distribution of cosmic rays, which has been since confirmed by several experiments~\cite{Abdo:2008aw,DiSciascio:2014jwa}. 
The incomplete field-of-view of ground based experiments necessitates combining data between detectors to observe the full $4\pi$~sr coverage required to examine the large scale anisotropy. 
The distribution of cosmic rays at 10~TeV was obtained by combining HAWC (Northern Hemisphere) and IceCube (Southern Hemisphere) data, showing the all-sky anisotropy for the first time~\cite{2019_HAWC_IC_cr}. SWGO would be able to provide coverage in the Southern sky, with a maximum multipole scale $>$0.1~PeV~\cite{Hinton:2021rvp}. Combining data between SWGO, HAWC, IceCube, and other experiments will improve our understanding of this anisotropy.\\

\noindent\textbf{SWGO and Cosmic Neutrinos: }
Currently, the cosmic neutrinos Icecube has detected thus far have energies between 100~TeV and 10~PeV~\cite{IceCube:2016uab}, making SWGO's gamma-ray energy range complementary to IceCube's neutrino energy range. Therefore, can search for common sources of neutrinos and gamma rays, which would indicate acceleration sites of hadrons in those regions (Galactic or extragalactic). Additionally, many dark matter decay models predict the creation of neutrinos, making dark-matter-dominated sources key targets for multimessenger studies~\cite{IceCube:2018tkk}. A cosmic neutrino event was seen coincident with a flaring blazar TXS~0506+056 at 3$\sigma$, resulting in an extensive multimessenger search~\cite{IceCube:2018dnn}. Similar events in the future would give opportunities for SWGO to participate in the followup efforts. However, no localized cosmic neutrino source has been significantly detected~\cite{IceCube:2016tpw}. \\

\noindent\textbf{SWGO and Gravitational Waves: }
The first gravitational waves from a binary-black-hole merger were detected in 2015~\cite{LIGOScientific:2016aoc}. Subsequent to the binary-neutron-star merger, GW170817, several experiments performed observations to examine the electromagnetic (and potentially neutrino) component to the merger~\cite{LIGOScientific:2017ync}. TeV gamma rays were observed by the High-Energy Spectroscopic System (H.E.S.S.) array~\cite{HESS:2017kmv}, but only after five hours because the release of the LIGO localization maps delayed data taking. Since H.E.S.S. is a pointed instrument, it needs localization information to know where to perform observations. An instrument like SWGO would be able to see such an event immediately, due to its wide-field-of-view.



\subsection{Future Ground-Based Extensive Air Shower Detectors for Cosmic Rays, Neutrinos, and Ultra-High-Energy Gamma Rays}
\label{sec-futureAirShower}

 \noindent
 \chapterauthor[1]{Markus Ahlers}\orcidlink{0000-0003-0709-5631}
 \\
 \begin{affils}
   \chapteraffil[1]{Niels Bohr International Academy, Blegdamsvej 17, 2100 Copenhagen, Denmark}
 \end{affils}


Cosmic ray interactions in the atmosphere create extensive air showers that can be observed by ground-based observatories using large areas of water-Cherenkov detectors, scintillator surface detector or underground muon detectors. The particle cascades in the atmosphere are visible by fluorescence detectors, air-Cherenkov detectors or air-radio detectors which are often augmented with a surface array for hybrid detection mechanisms.

The shower characteristics allow to infer cosmic ray mass composition and also to look for characteristic (but rare) events produced by neutrino and UHE gamma rays. Neutrinos can be identified via quasi-horizontal ($\theta\lesssim60^\circ$) extensive air showers with a high electromagnetic component. Earth-skimming tau neutrinos are visible as up-going showers from the decay of tau leptons produced in charged-current interactions in the Earth's crust. The signatures of UHE gamma rays are air showers with a larger atmospheric depth at the shower maximum and a steep lateral distribution function, along with a lower number of muons with respect to  showers initiated by nuclei.

The next generation of ground-based UHE CR observatories are envisioned for the 2030s. The community has started to collect ideas for the scientific scope and detector concept of a Global Cosmic Ray Observatory (GCOS)~\cite{GCOS}. The detector design will need to be optimized for the competing requirements for energy and mass resolution on one hand -- typically achieved by small and dense surface arrays -- and event statistics from large exposures on the other hand -- requiring larger arrays with sparser detectors. At the moment, the GCOS detector concept envisions a hybrid design including a total surface area of the order of 40,000 ${\rm km}^2$, which will allow to reach an exposures of the order of $2\times10^5$~${\rm km}^2{\rm yr}$ for cosmic rays after ten years of operation. 

The next-generation neutrino telescope IceCube-Gen2 also envisions an extended surface detector component above the main in-ice optical Cherenkov detector~\cite{IceCube-Gen2:2020qha,Gen2surface}. This surface detector would allow to observe cosmic ray air showers by their electromagnetic component and low-energy muons on the surface whereas high-energy muons are measured in the ice. A combination of elevated scintillation and radio detectors would enable high measurement accuracy of air showers. Together with the surface detector enhancement presently underway for IceTop~\cite{IceTopPlus} the IceCube-Gen2 surface array aims to cover an area of about 6~${\rm km}^2$.

Ground-based detectors can identify ``earth-skimming'' tau neutrinos, {\it i.e.}~tau neutrinos that travel through the Earth’s crust at a shallow angle. At energies $E > 1$~PeV tau neutrinos have a high probability to create tau leptons in charged-current interactions with Earth matter near the surface. These tau leptons can emerge from ground and initiate air shower as they decay in flight. Even though tau neutrino production by cosmic ray interactions is suppressed, neutrino flavor oscillations of astrophysical neutrinos guarantee a strong contribution of tau neutrinos upon arrival at Earth. 

Various future observatories have been proposed to observe Earth-skimming tau neutrinos by the decay of tau leptons above ground. Trinity is a detector concept that aims to detect air-Cherenkov emission by using a novel optical structure design that points at the horizon from an elevated vantage point~\cite{Otte:2018uxj}. BEACON~\cite{Wissel:2020sec} and TAROGE~\cite{TAROGE} plan to detect air-radio emission from elevated locations using compact arrays. The Ashra NTA is a proposed neutrino (and gamma ray) detector to be located on Mauna Loa, Hawaii, that observes air-Cherenkov and fluorescence light emission above the volcano with four detector stations~\cite{Sasaki:2014mwa}. TAMBO is another proposed detector to be located on one side of an Andean canyon to detect air showers emerging from the opposite side with an array of water-Cherenkov tanks~\cite{Romero-Wolf:2020pzh}. GRAND~\cite{GRAND:2018iaj} is a planned large array of sparse radio antennas to detect the radio emission from air showers not only triggered by high-energy neutrinos but also cosmic rays and gamma rays.\\

\subsubsection{The Beamforming Elevated Array for Cosmic Neutrinos}

\noindent
 \chapterauthor[1]{Austin Cummings}\orcidlink{0000-0000-0000-0000}
 \\
 \begin{affils}
    \chapteraffil[1]{Pennsylvania State University, State College, PA 16801, USA}
 \end{affils}
 
The Beamforming Elevated Array for Cosmic Neutrinos (BEACON) is a detector concept involving a mountain-top radio array for the detection of tau neutrinos with energies exceeding 100~PeV \cite{Wissel:2020fav}. The reference design for the BEACON detector includes 100 stations positioned at 3~km altitude, spaced 5~km apart to reduce the fraction of overlapping triggers. Each station covers $120^{\circ}$ in azimuth about the horizon and consists of 10 beamforming antennas with a frequency bandwidth of either 30~MHz-80~MHz or 200~MHz-1200~MHz. Beamforming with multiple antennas in a single station both provides significant increases in the SNR for a given event, thereby lowering the detection threshold, and allows for excellent angular resolution, making source classification easier. By viewing such large areas about the horizon with minimal antenna numbers, BEACON is capable of cost-effectively achieving large geometric apertures: the all-flavor sensitivity estimates of BEACON reach $\sim7 \times 10^{-10}$\,GeV cm$^{-2}$ s$^{-1}$ sr$^{-1}$ after 3 years of integration, assuming 1000 antenna stations (10 times larger than the reference design), which allows for detection of the diffuse neutrino flux even under pessimistic models for the UHECR mass composition and source evolution (pure iron composition and Fanaroff-Riley type II AGN evolution). BEACON stations will be deployed at several sites around the world, allowing for full-sky coverage and increasing the probability for potential multi-messenger follow-up measurements for transient astrophysical events. BEACON is currently in the demonstration phase, with a prototype array deployed at the Barcroft Station in the White Mountains of California.

\subsection{Current Balloon-Borne and Space-Based Extensive Air Shower Detectors for Cosmic Rays, Neutrinos, Ultra-High-Energy Gamma Rays}
\label{sec-curentBalloon}

\noindent
 \chapterauthor[1]{Austin Cummings}
 \chapterauthor[2]{Johannes Eser}\orcidlink{0000-0003-3849-2955}
 \\
 \begin{affils}
    \chapteraffil[1]{Pennsylvania State University, State College, PA 16801, USA}
    \chapteraffil[2]{University of Chicago, Chicago, IL 60637, USA}
 \end{affils}

One of the fundamental difficulties in charged particle astronomy is the inherent deflection due to Galactic and extragalactic magnetic fields. Only protons with energies exceeding \unit[$10^{19}$]{eV} are capable of pointing back to their sources without significant deflection (the required energy for heavier nuclei is even larger). At these energies, the cosmic ray flux is strongly suppressed, making charged particle astronomy with ground detectors nearly impossible. One potential solution to this problem is to make observations from near-space or space-based altitudes, using the entire Earth atmosphere as the active volume, thereby significantly increasing the exposure to cosmic events \cite{Linsley:1963km}. Additionally, space-based observation offers the advantage of full sky coverage with a single instrument, reducing the uncertainties inherent to ground-based detectors, which are capable of viewing only a single hemisphere. The relatively short orbital period ($\mathcal{O}(1~\mathrm{h})$) of a space-based instrument also ensures an optimal capability to follow up transient sources \cite{Venters:2019xwi}. In this manner, upcoming near-space and space-based instruments will make useful compliments to the existing and upcoming ground-based observatories for multi-messenger observations.

\subsubsection{The Antarctic Impulsive Transient Antenna}

\noindent
 \chapterauthor[1]{Austin Cummings}
 \chapterauthor[2]{Johannes Eser}\orcidlink{0000-0003-3849-2955}
 \\
 \begin{affils}
    \chapteraffil[1]{Pennsylvania State University, State College, PA 16801, USA}
    \chapteraffil[2]{University of Chicago, Chicago, IL 60637, USA}
 \end{affils}

The Antarctic Impulsive Transient Antenna (ANITA) is a high altitude balloon-borne detector with a payload that consists of quad-ridge horn antennas with a bandwidth of 200~MHz to 1200~MHz. The first version of the instrument (ANITA-I) flew in 2006-2007, while the last flight (ANITA-IV) was launched in December 2016, bringing the integrated flight time to around 98 days. The final version of ANITA, ANITA-IV had an energy threshold of \unit[$\sim 10^{18}$]{eV}. Using the emission from the in-ice and in-air Askaryan effects as well as geomagnetic emission, ANITA has detected both direct (above the Earth limb) and ice-reflected UHECR and set the most stringent limits to date on the cosmic neutrino flux for energies exceeding 100~EeV. In addition to the direct and reflected UHECR signals observed during the flights of ANITA-I \& III, two upwards going events with large emergence angles were also observed. These events have non-inverted pulse shapes consistent with direct events, but are not consistent with tau emergence probabilities calculated using Standard Model interactions of neutrinos \cite{ANITA:2018sgj}. In contrast, the flight of ANITA-IV measured 6 signals consistent with upward-going showers with small emergence angles, where the probability of having tau neutrino induced air showers are expected to be maximized. However, this observation is in conflict with limits set by existing ground-based experiments, and the effect of refracted UHECR signals from above the limb cannot be excluded \cite{ANITA:2020gmv}. The successor to the ANITA mission, called the Payload for Ultrahigh Energy Observations (PUEO) is currently in production (see section \ref{subsubsec:pueo}).

\subsubsection{The Extreme Universe Space Observatory on a Super Pressure Balloon}

\noindent
 \chapterauthor[1]{Austin Cummings}
 \chapterauthor[2]{Johannes Eser}\orcidlink{0000-0003-3849-2955}
 \\
 \begin{affils}
    \chapteraffil[1]{Pennsylvania State University, State College, PA 16801, USA}
    \chapteraffil[2]{University of Chicago, Chicago, IL 60637, USA}
 \end{affils}
 
 The Extreme Universe Space Observatory on a Super Pressure Balloon 1 (EUSO-SPB1) is the second generation balloon-borne instrument designed and built by the EUSO collaboration, and the first with the capability to measure UHECRs with energies above \unit[$3\times10^{18}$]{eV} via fluorescence emission. The UV light within the $12^{\circ}$ by $12^{\circ}$ field of view was focused by two 1~m diameter Fresnel lenses onto a focal surface populated with 2304 Multi-Anode PhotoMultiplier (MAPMT) pixels. Due to a shortened flight in 2017, the instrument has not recorded any showers, in agreement with estimated event rates calculated via extensive simulation studies, leaving the proof of principle of the detection technique still open while advancing the technical readiness level for the next step towards space observation \cite{Eser:2019ciy}.
 
 EUSO-SPB2 not only builds on the technologies utilized in EUSO-SPB1, allowing for enhanced capability for detection of EAS induced by UHECR via fluorescence emission, but also contains a second telescope optimized to detect optical Cherenkov emission from EAS sourced from either Earth-skimming neutrinos or above-the-limb cosmic rays, making EUSO-SPB2 the first true precursor to future space-based, multi-messenger instruments such as POEMMA. The launch of EUSO-SPB2 is scheduled for Spring 2023 from Wanaka, NZ with an mission duration of up to 100 days \cite{Eser:2021mbp}. During the flight, EUSO-SPB2 will measure, for the first time, EAS with energies E$>$\unit[10$^{18.2}$]{eV} from above using the fluorescence technique \cite{Filippatos:2021noz}. The Cherenkov telescope will raise the technology readiness level for SiPMs in space instruments while measuring different background conditions for the detection of upward going EAS initiated by Earth-skimming neutrinos. While the sensitivity of EUSO-SPB2 to the diffuse neutrino flux is not competitive with respect to existing ground-based experiments, it has some capability to measure neutrinos from astrophysical events following multi-messenger alerts. The technique for measuring neutrino induced EAS will be validated by pointing the instrument above the limb and detecting the optical Cherenkov emission of upwards going cosmic rays, where the expected event rate is larger than 100 events with energies above 1~PeV per hour of live time \cite{Cummings:2021lqa}. A successful flight of EUSO-SPB2 will have a significant impact on the realization of future space instruments as described in section \ref{sec-futureBalloon}. 

\subsubsection{The Mini Extreme Universe Space Observatory}

\noindent
 \chapterauthor[1]{Austin Cummings}
 \chapterauthor[2]{Johannes Eser}\orcidlink{0000-0003-3849-2955}
 \\
 \begin{affils}
    \chapteraffil[1]{Pennsylvania State University, State College, PA 16801, USA}
    \chapteraffil[2]{University of Chicago, Chicago, IL 60637, USA}
 \end{affils}
 
 The Mini Extreme Universe Space Observatory (MiniEUSO) is taking data since October 7, 2019 from inside the International Space Station (with a total of 51 sessions of data taking completed by February 2022). The instrument has a 25~cm diameter aperture with a wide FoV, utilizing Fresnel lenses to focus light onto the camera, which consists of 2304 MAPMT pixels. The primary goals of this mission are to qualify technology for space flight, measure various atmospheric events (such as TLE), and to search for: nuclearites, strange quark matter, meteors and meteoroids, UV emission from sea bio-luminescence, artificial satellites, and man-made space debris. To accomplish these tasks, a multi-level trigger with different time scales was developed, where each time scale was optimized for a different physical phenomena, and all triggers ran concurrently. One level of the trigger was designed to detect cosmic ray signals even though the estimated energy threshold for such an event exceeds \unit[$10^{21}$]{eV} and no such signal is anticipated. More details about the instrument and its results can be found in \cite{Bacholle:2020emk, JEM-EUSO:2021cux}.

\subsubsection{The Tracking Ultraviolet Set-up}

\chapterauthor[1]{Austin Cummings}
 \chapterauthor[2]{Johannes Eser}\orcidlink{0000-0003-3849-2955}
 \\
 \begin{affils}
    \chapteraffil[1]{Pennsylvania State University, State College, PA 16801, USA}
    \chapteraffil[2]{University of Chicago, Chicago, IL 60637, USA}
 \end{affils}

The Tracking Ultraviolet Set-up (TUS) was launched as part of the Lomonosov satellite in 2016, pioneering the measurement of UHECR from space. The instrument focuses the light from its \unit[2]{m$^2$} mirror onto a focal surface of 256 pixels, with an overall FoV of $\pm$ 4.5\textdegree. The trigger system had 4 levels: (i) a \unit[0.8]{$\mu$s} frame length to look for EAS tracks (ii) an integration time of \unit[25.6]{$\mu$s} and (iii) 0.4 ms to record  Transient Luminous Events (TLEs) and (iv) an integration time of \unit[6.6]{ms} for the optimized detection of meteors. Data taking could only be commenced in one mode at a time. By late 2017, the instrument recorded 80000 triggers, observed multiple TLEs and meteors but no signal of a cosmic ray air shower. This non-detection of UHECR is consistent for an estimated energy threshold above \unit[$10^{20}$]{eV}. A detailed discussion of the instrument and the first results can be found in \cite{Klimov:2017lwx, BARGHINI2021}.

\subsection{Future Balloon-Borne and Space-Based Extensive Air Shower Detectors for Cosmic Rays, Neutrinos, Ultra-High-Energy Gamma Rays}
\label{sec-futureBalloon}

\subsubsection{The Payload for Ultrahigh Energy Observations}\label{subsubsec:pueo}

\noindent
 \chapterauthor[1]{Remy L. Prechelt}\orcidlink{0000-0002-7191-7110}
 \chapterauthor[2]{Austin Cummings}
 \chapterauthor[3]{Johannes Eser}\orcidlink{0000-0003-3849-2955}
 \\
 \begin{affils}
   \chapteraffil[1]{Department of Physics \& Astronomy, University of Hawai'i M\=anoa, Honolulu, HI 96822}
   \chapteraffil[2]{Pennsylvania State University, State College, PA 16801, USA}
   \chapteraffil[3]{University of Chicago, Chicago, IL 60637, USA}
 \end{affils}
 
 The Payload for Ultrahigh Energy Observations (PUEO) 
 is a long-duration Antarctic balloon-borne experiment designed to detect UHECRs and UHE neutrinos via geomagnetic radiation generated by extensive air showers or Askaryan radiation generated by upward-going neutrino showers generated in ice~\cite{PUEO:2020bnn}. It is a direct successor to the ANITA experiment, which conducted a total of four flights between 2006 and 2016~\cite{ANITA:2008mzi,2010PhRvD..82b2004G,ANITA:2018vwl,ANITA:2019wyx}. The PUEO design features several improvements that will enable it to achieve world-leading sensitivity to UHE neutrinos above 1~EeV, improving upon ANITA by more than an order of magnitude below 30~EeV. In combination with the largest target volumes for neutrino interactions ($\sim 10^6$ km$^{3}$), the improved sensitivity will position PUEO to either make the first significant measurement of cosmic neutrinos with energies above 1~EeV or to set the most stringent limits on the fluxes of cosmogenic and astrophysical neutrinos at these energies. With a large instantaneous aperture, PUEO will also be well-suited for searching for UHE neutrinos from transient astrophysical sources.
 
 PUEO's instrument design builds significantly on the ANITA design, more than doubling the number of antennas, as well as featuring a lower-noise radio-frequency signal chain and an advanced trigger system that includes an interferometric phased-array trigger. These design improvements will significantly increase PUEO's sensitivity to all four of ANITA's detection channels -- Askaryan signals from upward-going UHE neutrinos interacting in the Antarctic ice and geomagnetic signals from above-the-horizon UHECRs, downward-going UHECR EASs reflecting off of the Antarctic ice, and EASs generated by the decay of Earth-emerging tau leptons generated from upward-going tau neutrinos. 
 


In contrast to ANITA, PUEO will consist of two instruments, separately dedicated to different detection channels. The Main Instrument consists of 108 dual-polarization quad-ridged horn antennas (compared to the 48 used in
the last flight of ANITA), including a ring of downward-canted antennas that will search for anomalous steeply-inclined upward-going cosmic-ray-like events such as those reported by ANITA from previous flights~\cite{ANITA:2018sgj,ANITA:2017qmn}. The Main Instrument is designed to detect signals in the 300~MHz to 1200~MHz frequency range, allowing for smaller antennas. The Low Frequency instrument will cover the 50~MHz to 300~MHz frequency range and is designed to detect EAS signals from UHECR and decaying tau leptons.


In addition to the changes in antenna design, the trigger subsystem will utilize real-time interferometric beamforming, further enhancing PUEO's sensitivity. This beamforming trigger computes highly directional beams on the sky by coherently summing waveforms with different time delays, improving PUEO's trigger performance ($\sim 50$\% at a signal-to-noise ratio of \({\sim}1\))~\cite{PUEO:2020bnn}. The beamforming trigger system leverages extensive heritage from the phased-array trigger of the Askaryan Radio Array~\cite{Allison:2018ynt} and employs the Xilinx Radio-Frequency System-on-Chip platform, which combines high-bandwidth digitizers, large field-programmable gate-arrays, and digital signal processing cores onto a single chip~\cite{PUEO:2020bnn}. With these improvements PUEO will benefit from a lower trigger threshold that will increase its acceptance to neutrino and cosmic-ray events.

All told, PUEO's design features multiple augmentations and leverages new technology in order to maximize its science reach. \\



\subsubsection{The Probe of Extreme Multi-Messenger Astrophysics}

\noindent
 \chapterauthor[1]{Austin Cummings}
 \chapterauthor[2]{Johannes Eser}\orcidlink{0000-0003-3849-2955}
 \\
 \begin{affils}
    \chapteraffil[1]{Pennsylvania State University, State College, PA 16801, USA}
    \chapteraffil[2]{University of Chicago, Chicago, IL 60637, USA}
 \end{affils}
 
The Probe Of Extreme Multi-Messenger Astrophysics (POEMMA) is a proposed dual satellite mission to observe both UHECR and VHE neutrinos \cite{POEMMA:2020ykm}. The focal plane of each POEMMA satellite is divided into two sections which target different science requirements: (i) the POEMMA Fluorescence Camera (PFC), which occupies roughly 80\% of the focal surface and is composed of 126720 1 $\mu$s frame length MAPMTs and (ii) the POEMMA Cherenkov Camera (PCC), which occupies the remainder of the focal surface and is composed of 5360 1 ns frame length SiPMs. The PFC is designed to measure the fluorescence emission from EAS induced by UHECRs in the Earth atmosphere while the PCC is designed to measure the optical Cherenkov emission from upwards-going EAS sourced from neutrino interactions in the Earth and from above-the-limb cosmic rays. POEMMA is designed to encompass two operational modes: (i)``POEMMA-Stereo", where both satellites are separated by a distance of $\sim300$~km and tilt towards one another near nadir to observe a common volume, lowering the energy threshold for detection of UHECR via fluorescence emission, as well as greatly improving the resolution of $X_{\mathrm{max}}$ ($<30\, \mathrm{g} \, \mathrm{cm}^{-2}$ above 100~EeV) and (ii) ``POEMMA-Limb", where the two satellites move closer together to a minimum of 30~km separation and tilt upwards to monitor the Earth limb following a potential multi-messenger alert, tracking sources as they move across the sky. In this mode, UHECR are still observed, but with higher energy thresholds, and reduced imaging capabilities. While the current configuration of POEMMA is not expected to be competitive in observing the diffuse neutrino flux with respect to existing ground-based observatories \cite{Reno:2019, Cummings:2020}, the full sky coverage, fast pointing direction, and excellent angular resolution ($1.5^{\circ}$) allow for enhanced ``Target-of-Opportunity" multi-messenger follow-up observations \cite{Venters:2019xwi}.

\subsection{Current and Future In-Ice and In-Ocean Particle Detectors}
\label{sec-iceOceanDetectors}

\subsubsection{The Askaryan Radio Array}

\noindent
 \chapterauthor[]{Kaeli A. Hughes}\orcidlink{0000-0002-4551-9581}
 \\
 \begin{affils}
   \chapteraffil[]{Department of Physics, Enrico Fermi Institute, Kavli Institute for Cosmological Physics, University of Chicago, Chicago, IL 60637}
 \end{affils}

\begin{figure*}[htp]
  \centering
  \includegraphics[width=0.48\textwidth]{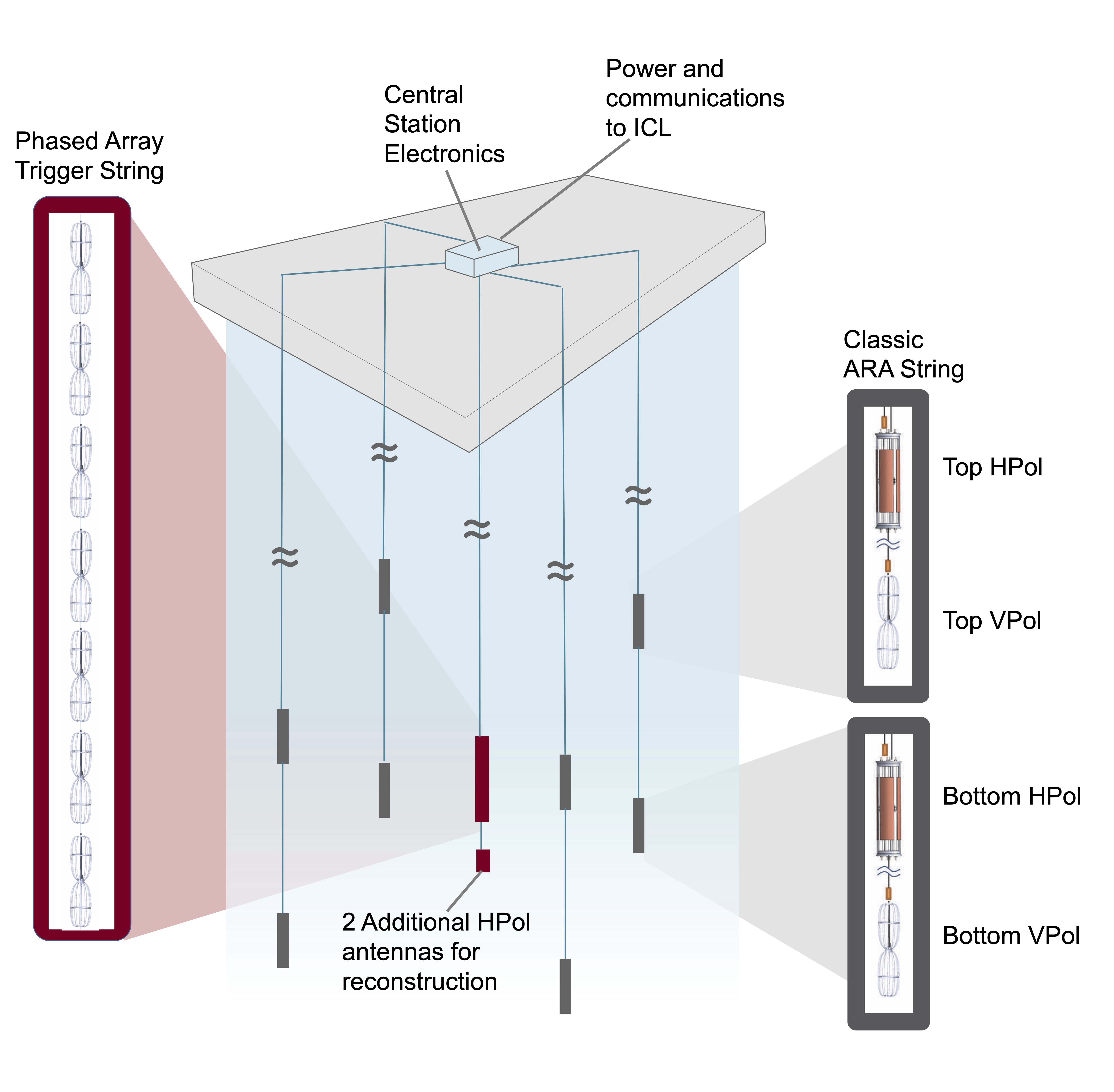}\label{fig:ARA-1}
  \includegraphics[width=0.48\textwidth]{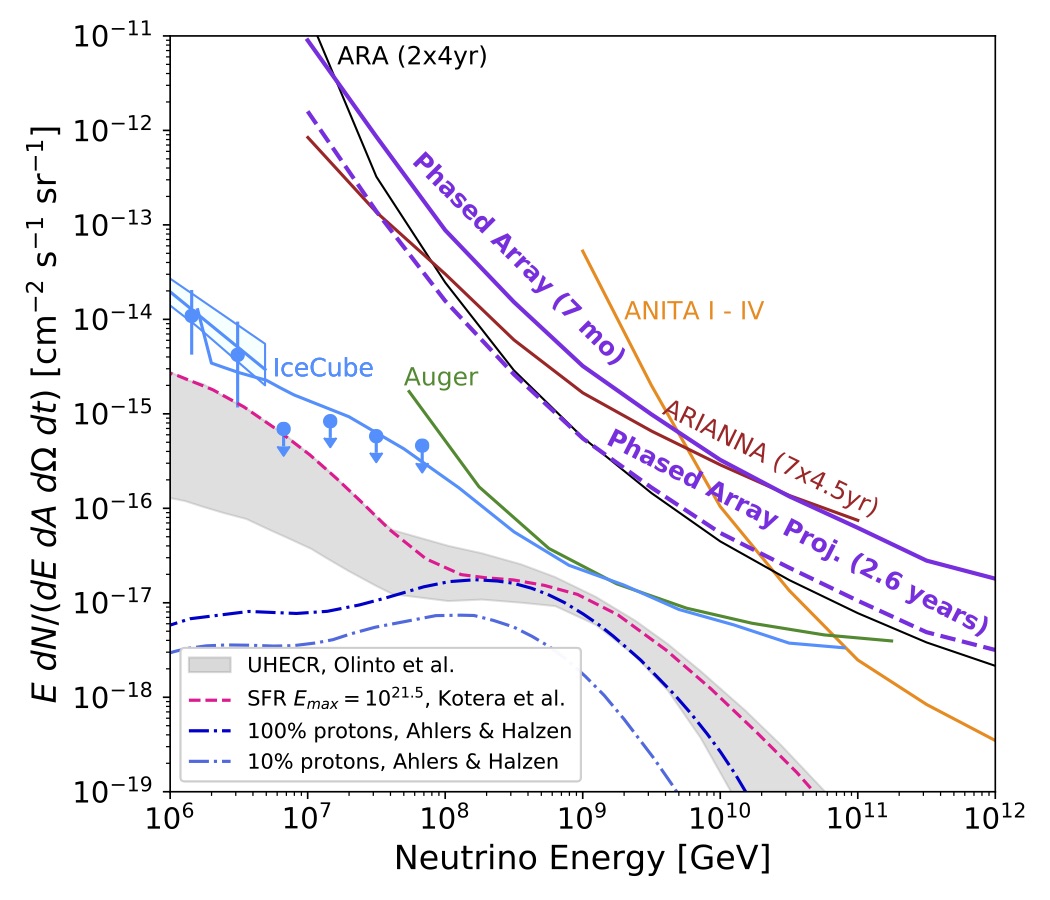}\label{fig:ARA-2}
  \caption{Left: A diagram of an ARA station, from \cite{PhasedArrayAnalysis}. Stations 1-4 only consist of the Classic ARA strings, shown in gray. Station 5 includes both the Classic ARA strings as well as the additional Phased Array triggering string, shown in red. Right: The best limits produced by the ARA Collaboration, shown in black (for the classic stations) and purple (for the new Phased Array instrument). Other experiments and models are shown \cite{ANITA_IV, ARIANNA, Auger101088, IceCube_limit1, IceCube_limit2, Olinto:2011ng, Kotera:2011cp, Ahlers:2012rz}. Adapted from \cite{PhasedArrayAnalysis}.}
\label{fig:ARA}
\end{figure*}

The Askaryan Radio Array (ARA) is a ground-based neutrino detector designed to detect radio Askaryan emission created by neutrinos interacting within the Antarctic ice \cite{AskaryanEffect}. This detection mechanism is most sensitive to neutrinos with energies above \SI{10}{PeV}. ARA was first deployed at the South Pole in 2011 and since then has built five independently-operating stations, each separated from its neighboring station by about 2 km \cite{ARAtestbed, Allison:2015eky}. An example station diagram is shown in Figure \ref{fig:ARA-1}.

A classic ARA station consists of a mixture of horizontally-polarized and vertically-polarized antennas buried to a maximum depth of 200 m in the Antarctic ice. These antennas are designed to target the frequency range of \SI{150}{MHz} to \SI{850}{MHz}, and the polarization information they record allows the incoming direction of neutrino signals to be reconstructed. This instrument triggers at a rate of around 5 Hz and has a 50\% trigger efficiency at a signal-to-noise ratio (SNR) of approximately 3.75 \cite{PhasedArrayInstrument}. A recent analysis of two of the five ARA stations resulted in ARA setting the best limit set by a radio detection experiment on the neutrino flux between \SI{100}{PeV} and \SI{30}{EeV} \cite{ARA23}, as shown in Figure \ref{fig:ARA-2}. Future analysis of all currently available ARA data will improve the livetime by approximately a factor of 5.

 ARA has also recently prototyped a phased array trigger, in which signals from neighboring antennas are summed in pre-determined directions called beams prior to the trigger, allowing impulsive signals to add coherently and effectively increasing the effective volume of the instrument. The 50\% trigger efficiency for the phased array trigger occurs at an SNR of approximately 2, a significant improvement compared to the classic ARA trigger. A recent analysis of data from this prototype phased array trigger, recently submitted for publication, shows that the phased array can improve the analysis efficiency as well, motivating this trigger design for future radio experiments such as RNO-G, PUEO, BEACON, and IceCube-Gen2 \cite{RNOG_2021, PUEO_white, Wissel:2020sec, IceCubeGen2_White}.

\subsubsection{The Radio Neutrino Observatory in Greenland}

\noindent
 \chapterauthor[1]{Kaeli A. Hughes}\orcidlink{0000-0002-4551-9581}
 \\
 \begin{affils}
   \chapteraffil[1]{Dept. of Physics, Enrico Fermi Institute, Kavli Institute for Cosmological Physics, University of Chicago, Chicago, IL 60637}
 \end{affils}
 
The Radio Neutrino Observatory in Greenland (RNO-G) is a new experiment under construction in Summit Station, Greenland \cite{RNOG_2021}. Its location in the Northern hemisphere makes RNO-G complementary to current and planned radio experiments at the South Pole \cite{ARA23, ARIANNA, IceCubeGen2_White}. In addition, there are potential sources in the Northern hemisphere visible to RNO-G that could have interesting multi-messenger implications, including blazars known to emit TeV gamma-rays \cite{1996ApJ...456L..83Q,1992Natur.358..477P}, the hotspot identified by the Telescope Array from anisotropies in the UHECR flux \cite{TelescopeArray:2014tsd}, and the blazar flaring in gamma-rays with a coincident neutrino detection from IceCube \cite{IceCube:2018cha,IceCube:2018dnn}.

 \begin{figure*}[htp]
  \centering
  \includegraphics[width=0.48\textwidth]{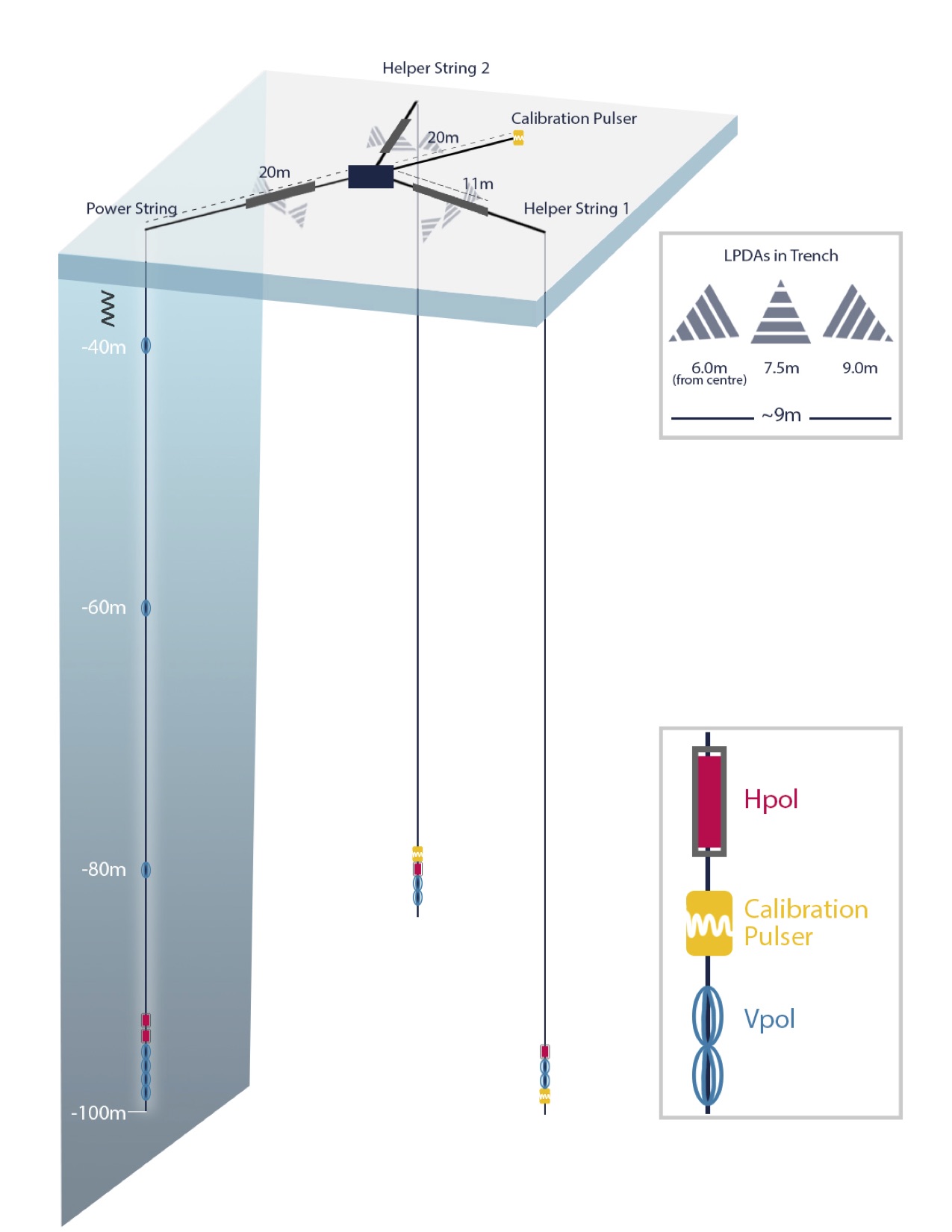}\label{fig:RNO_station}
  \includegraphics[width=0.48\textwidth]{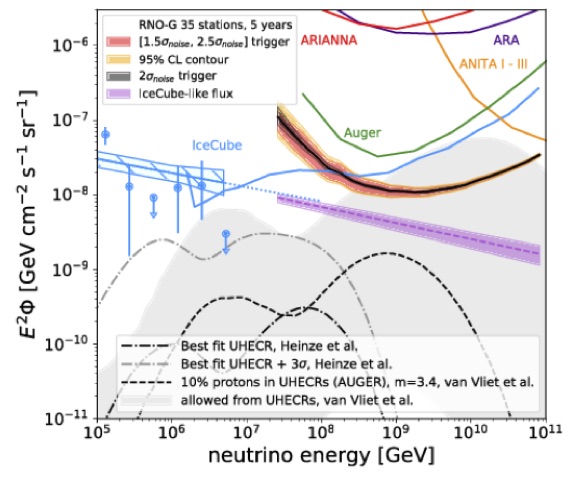}\label{fig:RNO_sensitivity}
  \caption{Left: A diagram of an RNO-G Station. The Power String, on the left, holds the antennas that make up the phased array trigger. Right: The five-year sensitivity of RNO-G to the all-flavor neutrino flux, from \cite{RNOG_2021}. The various bands represent the expected performance band of the phased array trigger, as well as the 95\% confidence level contours. Also shown are various expected flux models.}
\label{fig:RNO-G}
\end{figure*}

RNO-G currently has three stations deployed, with a planned 35 stations to be installed over the next few years. Like other experiments built to detect neutrinos above \SI{10}{PeV}, RNO-G is sensitive to the Askaryan emission created by neutrinos interacting in the ice. Each RNO-G station is independent and is built using a combination of surface antennas and deep antennas, achieving the maximum effective volume possible given the maximum drilling depth of \SI{100}{m}. An example of a station diagram is shown in Figure \ref{fig:RNO_station}.
 
RNO-G utilizes a phased array trigger design, first prototyped for in-ice use in the ARA experiment \cite{PhasedArrayInstrument}. Unlike the previously-deployed phased array prototype, the RNO-G stations are autonomous, getting power from a combination of solar panels and batteries and communicating via a wireless network. There are multiple operating modes for each station, allowing the power consumption to match the available power. The power consumption ranges between \SI{6}{W}-\SI{24}{W} for data-taking modes and down to \SI{70}{mW} for minimal winter operations during the polar night. 

Because of its scale, the development of RNO-G will show the feasibility of large-scale radio detectors for neutrinos above \SI{10}{PeV}. The expected sensitivity of the RNO-G experiment after five years, including the down time caused by polar night, is shown in Figure \ref{fig:RNO_sensitivity}.


\subsection{Optical Followup of Multimessenger Sources}
\label{sec-opticalFollowup}


\noindent
 \chapterauthor[1]{Robert Stein}\orcidlink{0000-0003-2434-0387}
 \\
 \begin{affils}
    \chapteraffil[]{California Institute of Technology, Pasadena, CA 91125, USA}
 \end{affils}





Optical telescopes are an integral component of the multi-messenger landscape. Wide-field optical telescopes discover the vast majority of transients such as core-collapse supernovae (SNe, see also section \ref{sec-CCSNeLongGRBs}) and Tidal Disruption Events (TDEs, see also section \ref{sec-TDE}) with predicted multi-messenger emission. Indeed, the electromagnetic signature of the first multi-messenger transient, SN1987A, was discovered by an optical telescope at Las Cumbras Obervatory (LCO).

In recent years, optical telescopes such as ASAS-SN \citep{2014AAS...22323603S,2017A&A...607A.115I},  DECam \cite{2019ApJ...883..125M}, MASTER \citep{lipunov_20}, Pan-STARRS \cite{2019A&A...626A.117P}, and ZTF \citep{2021NatAs...5..510S} perform dedicated follow-up programs to search for sources of TeV neutrinos detected by IceCube. 
The first probable TeV neutrino source, the blazar TXS 0506+056, exhibited a multi-wavelength flare coincident with the detection of a high-energy neutrino (see also section \ref{sec-astroAGN}). Optical observations of this flare were provided by ASAS-SN, Kanata/HONIR and Kiso/KWFC \citep{IceCube:2018dnn} as well as MASTER \citep{2020ApJ...896L..19L}. More recently, the TDE AT2019dsg and likely TDE AT2019fdr were identified as probable sources of TeV neutrinos as a direct result of observations by the optical telescope ZTF \citep{2021NatAs...5..510S, at2019fdr}.

The same optical telescopes, and many others, form the backbone of similar searches for kilonova counterparts to gravitational waves detected by LIGO/Virgo/KAGRA (see section \ref{sec-currentGW}). The first multi-messenger gravitational wave source, binary neutron star merger GW170817, was detected in coincidence with a gamma-ray burst. However, the localisation of this association was some $\sim$1100 sq deg, and it was not until the kilonova counterpart AT2017gfo was found by the LCO optical telescope Swope that broad multi-wavelength observations of the event could begin \citep{2017ApJ...848L..12A}. The identification of such a counterpart, including a measurement of the associated redshift, is essential to unlock key multi-messenger science such as studying heavy element formation \citep{2017ApJ...848L..27T} and measuring the Hubble constant \citep{2017Natur.551...85A}.





\chapter{Collaboration and Infrastructure}
\label{sec-Infrastructure}


\section{Forging Multimessenger Era Partnerships}
\label{sec-partnerships}

\noindent
 \chapterauthor[1]{Rita M. Sambruna}
 \chapterauthor[1]{Joshua E. Schlieder}
 \\
 \begin{affils}
    \chapteraffil[1]{NASA Goddard Space Flight Center, 8800 Greenbelt Rd., Greenbelt, MD 20771, USA}
     \end{affils}


The era of multi-messenger astrophysics (MMA) is here, bringing with it a renewed and more urgent need for the MMA (and time domain astrophysics, TDA) community to coordinate, collaborate, and communicate. Hosted virtually at NASA’s Goddard Space Flight Center (GSFC), the Multi-messenger Operational Science Support \& Astrophysical Information Center (MOSSAIC, \href{https://asd.gsfc.nasa.gov/mossaic}{https://asd.gsfc.nasa.gov/mossaic}) builds on current GSFC capabilities to provide a nexus for the ground- and space-based communities to come together and share information and planning. MOSSAIC’s services aim at fostering easier and more efficient paths for users to acquire, analyze, and interpret data from space-based observatories, and for planning future MMA/TDA missions. Future partnerships with other NASA Centers, academia, and industry will enable MOSSAIC to support the MMA/TDA community as they respond to the priority recommendation for this science from the Astrophysics 2020 Decadal Survey.


\subsection{Multimessenger Operational Science Support and Astrophysics Information Collaboration}
\label{sec-MOSSAIC}

\noindent
 \chapterauthor[1]{Rita M. Sambruna}
 \chapterauthor[1]{Joshua E. Schlieder}
 \\
 \begin{affils}
    \chapteraffil[1]{NASA Goddard Space Flight Center, 8800 Greenbelt Rd., Greenbelt, MD 20771, USA}
     \end{affils}


\textbf{\textit{Introduction}}
The advent of advanced ground-based observatories in a few years will expand the discovery horizon and drastically increase the number of transient and MMA sources needing prompt electromagnetic (EM) follow-up from the ground and in space. The needs of the MMA/TDA community will increase many-fold. This includes the need for coordination, collaboration, and communication between space and ground-based facilities, and between the astronomy and physics communities; the need for adequate infrastructure (data analysis and interpretation tools, modern and efficient alert systems, proposer and observer support, rapid data transmission links, etc.); and the need for common and frequent brainstorming together to anticipate future needs and develop solutions.  

\begin{figure}[ht]
    \includegraphics[width=\textwidth]{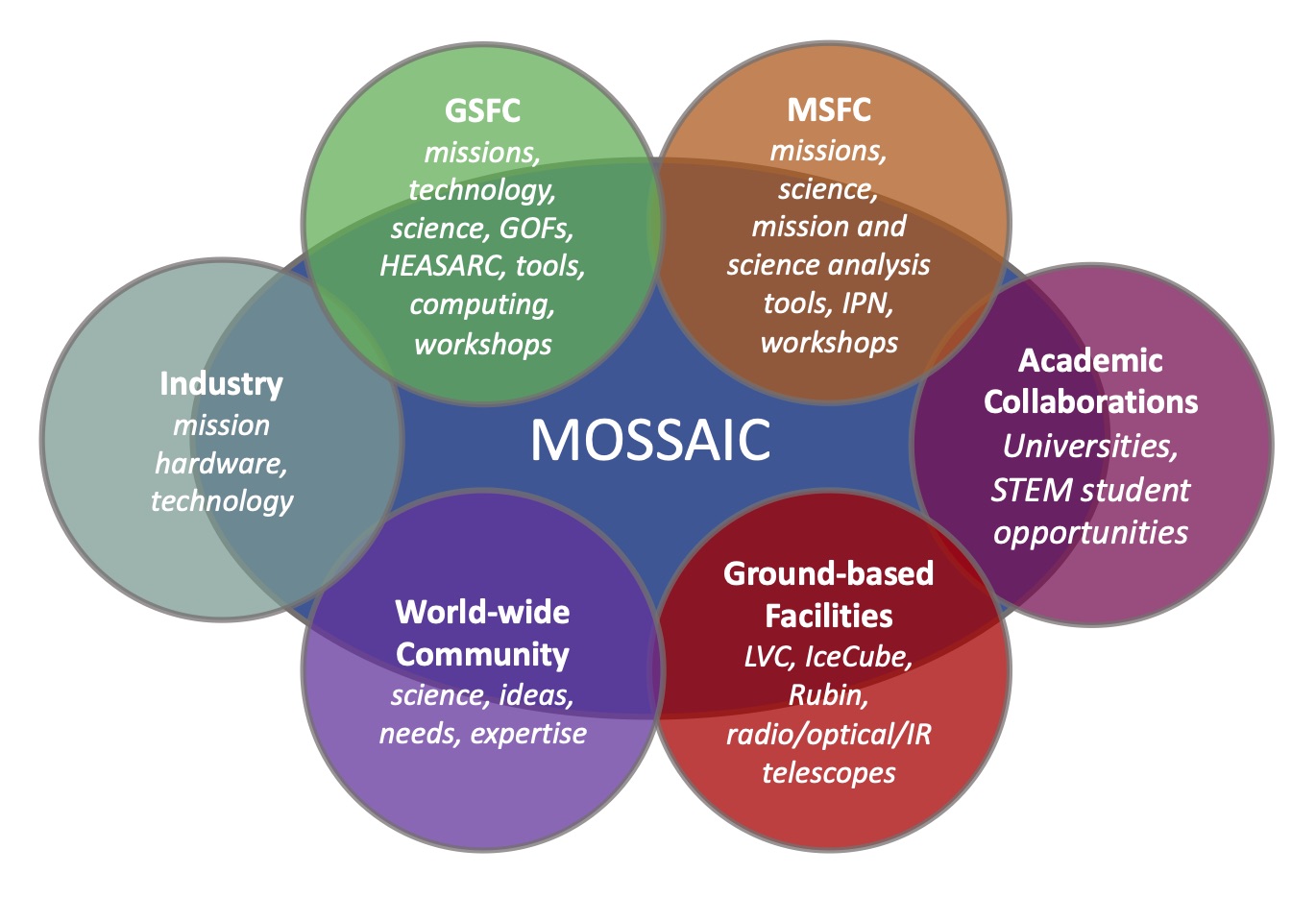}
    \caption{MOSSAIC is deeply connected to various MMA and TDA communities through research and services, including scientists worldwide, mission operating staff, and infrastructure developers. Connection is essential because MMA is a global enterprise and can't flourish without the concerted efforts of many parties. Facilitating collaboration, innovation, and exchange of ideas is one of the core values of MOSSAIC.}
    \label{fig:MOSSAIC}
\end{figure}

Hosted virtually at NASA's GSFC, MOSSAIC brings together current research, capabilities, and resources needed to support MMA/TDA scientists at NASA and around the world. MOSSAIC scientists, engineers, programmers, and managers are deeply engaged in research and infrastructure development for MMA/TDA science, and collaborate closely with the ground- and space-based, physics and astronomy communities. While the present MOSSAIC builds entirely on Goddard’s capabilities, we have a standing partnership with our colleagues at NASA's Marshall Space Flight Center (MSFC), where other MMA/TDA activities are underway, which will be incorporated in future augmentations of MOSSAIC’s capabilities. We also 
look forward to new partnerships with academia and industry (Figure~\ref{fig:MOSSAIC}).\\

\textbf{\textit{MOSSAIC's Functions}}
MOSSAIC’s services aim at providing the astrophysics community with: 1. A robust system for rapid alerts, and tools for data analysis, interpretation, and dissemination; 2. Mission development support and expertise, including formulating compelling and feasible science cases; 3. Space communication capabilities; and 4. Events to bring stakeholders together for planning and brainstorming. For more information, please visit \href{https://asd.gsfc.nasa.gov/mossaic}{https://asd.gsfc.nasa.gov/mossaic}. 

As the needs of the MMA/TDA community grow and evolve, so will MOSSAIC. Based on input from a variety of stakeholders, we will continue to expand the services and functions of MOSSAIC to better assist observers on the ground and in space. 

\textbf{\textit{Conclusions}}
MOSSAIC core values focus on communication, coordination, and collaboration. Another core value is service to the community, which builds on Goddard's tradition. Partnerships with other NASA Centers, academia, institutions, and industry are essential components of MOSSAIC. We invite you to join us in MOSSAIC and contribute to the discovery of the dynamic Universe. 
\\

\noindent {\small \textbf{\textit{Acknowledgements}}
The MOSSAIC concept received enthusiastic endorsement from many colleagues and institutions in the ground- and space-based MMA/TDA communities, who recognize the need for coordination and collaboration in this multifaceted discipline. We are grateful to the GSFC and Center leadership for their support of MOSSAIC.}

\subsection{Time-Domain Astronomy Coordination Hub (TACH) and the New Gamma-ray Coordinates Network (GCN)}
\label{sec-tachGCN}

\noindent
\chapterauthor[]{Judith Racusin}
\\
 \begin{affils}
   \chapteraffil[]{Astroparticle Physics Laboratory, NASA Goddard Space Flight Center, Greenbelt, MD}
 \end{affils}

The Gamma-ray Coordination Network (hereafter GCN Classic) has been the cornerstone of high-energy transient astrophyscis for the last 30 years, especially in the field of gamma-ray bursts (GRBs) and more recently serving alerts and coordinating the community for gravitational waves from neutron star and black hole mergers, and high-energy neutrinos.  The TACH project has been working to modernize GCN and build tools to aid in its expansion with the growth of multimessenger astrophysics.  We released the GCN Viewer (\url{https://heasarc.gsfc.nasa.gov/tachgcn}) in 2021 providing an interface to a searchable database of all GCN machine-generated Notices and human-written Circulars organized by astronomical events as well as by observatory and instrument.  In partnership with High-Energy Astrophysics Science Archive Research Center (HEASARC), the GCN Viewer will continue development over the next few years including cross-compatibility with other HEASARC archives and services.

TACH is building upon the legacy of GCN Classic to provide a modern cloud-based transient alert system known as the General Coordinates Network (GCN).  The new GCN utilizes the Apache Kafka protocol serving both GCN Classic formats (text, binary, VOevent) as well as a unified AVRO schema simplifying records across missions.  The new GCN will begin public operations in by Summer 2022, and both producers and consumers will be able to utilize any of the 3 systems (GCN Classic, GCN Classic over Kafka, or the new GCN) for the next few years until the GCN Classic system is retired.  The new GCN is built to be compatible with other kafka-based systems (e.g. SCiMMA, Rubin brokers), enabling coincidence searches and cross-system compatibility.

\section{Data Access and Archiving}
\label{sec-archiving}

\subsection{High-Energy Astrophysics Science Archive Research Center}
\label{sec-HEASARC}

\noindent
\chapterauthor[1]{Tess Jaffe}\orcidlink{0000-0003-2645-1339}
\chapterauthor[1]{Alan Smale}\orcidlink{0000-0001-9207-9796}
\\
 \begin{affils}
   \chapteraffil[1]{HEASARC Office, NASA Goddard Space Flight Center, Greenbelt, MD}
 \end{affils}
 
 The global astronomical community has recognized the importance of making data FAIR: findable, accessible, interoperable, and reusable.  As an early example, the High Energy Astrophysics Science Archive Research Center (HEASARC) was established in the 1990’s to change the way that x-ray and gamma-ray data were distributed and analyzed in the community.  HEASARC was created as a single user-friendly online facility through which x-ray and gamma-ray data from NASA missions and those of other agencies could be discovered and analyzed. It defined multimission standard data formats, similar standards for software to analyze the data, and provided generic libraries and tools for common tasks, plus domain expertise through active help desks. Today, HEASARC enables each new high energy mission to take advantage of the knowledge gained by previous missions rather than reinventing the wheel, and likewise researchers will find new missions familiar in many ways. The HEASARC is also a participant in the International Virtual Observatory Alliance (IVOA), which seeks to maximize the interoperability of all astronomy data worldwide. 

In addition to serving mission data on request, the HEASARC maintains the Gamma-ray Coordination Network (GCN) that has been helping the transient astronomy community automate the distribution of information about astronomical events in real time.  The GCN collaborates closely with the LIGO-Virgo gravitational wave group, with AMON (the Astrophysical Multi-messenger Observatory Network), and with IceCube to disseminate multi-messenger alerts for gravitational wave events, high energy neutrino detections, and other non-EM-spectrum-based alerts.  The number of alerts released ranges from 30-50k/day.  The GCN also provides follow-up notices giving new details about initial transient events, including detections or upper limits in correlative observations.  Work is already underway to prepare GCN for the Vera Rubin Observatory LSST era using new commercial technologies for event brokers. 
 
The revolutionary combination of multi-wavelength and multi-messenger data and the rapid community response to gravitational wave and gamma-ray burst events has illustrated dramatically how such coordination benefits all.  As a result the HEASARC has expanded from its original NASA remit to archive multi-messenger data such as the ground-based IceCube neutrino events and gravitational wave events, and includes the Legacy Archive for Microwave Background Data Analysis (LAMBDA), the go-to archive of CMB-related datasets whether space- or ground-based.  
 
In the next decade such archives around the world will more closely interoperate with each other through deeper VO interfaces and will continue to reduce barriers to multi-wavelength and multi-messenger astronomy.  The HEASARC GCN upgrade (see its new front-end viewer at: \href{https://heasarc.gsfc.nasa.gov/wsgi-scripts/tach/gcn_v2/tach.wsgi/}{https://heasarc.gsfc.nasa.gov/wsgi-scripts/tach/gcn\_v2/tach.wsgi/}) will allow users to subscribe to the alerts of interest from an even larger variety of multi-messenger sources, accelerating the advances in time domain astrophysics.  The Astro2020 Decadal Survey also recommended that archives funded by NASA and the NSF should increase their coordination in order to improve interoperability, and an effort to respond to this is under way.

\subsection{Space Science Data Center}
\label{sec-SSDC}

\chapterauthor[ ]{ }
\vspace{-0.2in}
 \addtocontents{toc}{
     \leftskip3cm
    \scshape\small
    \parbox{5in}{\raggedleft Gianluca Polenta et al. on behalf of the SSDC staff}
    \upshape\normalsize
    \string\par
    \raggedright
    \vskip -0.19in
    }
 
\noindent
\nocontentsline\chapterauthor[]{Gianluca Polenta$^{1}$}
\nocontentsline\chapterauthor[]{Stefano Ciprini$^{1,2}$}
\nocontentsline\chapterauthor[]{Valerio D'Elia$^{1}$}
\nocontentsline\chapterauthor[]{Dario Gasparrini$^{1,2}$}
\nocontentsline\chapterauthor[]{Marco Giardino$^{1}$}
\nocontentsline\chapterauthor[]{Cristina Leto$^{1}$}
\nocontentsline\chapterauthor[]{Fabrizio Lucarelli$^{1,3}$}
\nocontentsline\chapterauthor[]{Alessandro Maselli$^{1,3}$}
\nocontentsline\chapterauthor[]{Matteo Perri$^{1,3}$}
\nocontentsline\chapterauthor[]{Carlotta Pittori$^{1,3}$}
\nocontentsline\chapterauthor[]{Francesco Verrecchia$^{1,3}$}
\nocontentsline\chapterauthor[]{on behalf of the SSDC staff}
\\
 \begin{affils}
   \chapteraffil[1]{Space Science Data Center, Italian Space Agency, via del Politecnico snc, 00133, Roma, Italy}
   \chapteraffil[2]{INFN-Sezione di Roma Tor Vergata\^{\i}, 00133, Roma, Italy}
   \chapteraffil[3]{INAF-OAR, via Frascati 33, 00078 Monte Porzio Catone (RM), Italy}
\end{affils}



The Space Science Data Center\footnote{https://www.ssdc.asi.it} (SSDC) is a collaborative effort between the Italian Space Agency (ASI), National Institute for Astrophysics (INAF), and National Institute for Nuclear Physics (INFN) to provide a Research Infrastructure designed to facilitate collection, reduction, analysis, and distribution of data from supported science missions. The SSDC aims to develop a user-friendly, online, and public set of tools and services which realize the open science FAIR ({\it Findable, Accessable, Interoperable, and Reusable}) principles. 

These principles allow for non-expert users to easily navigate the large diversity of data which SSDC hosts: photon data from radio to $\gamma$-ray, as well as, cosmic ray and neutrino data. This is a crucial feature for practitioners of multi-wavelength and multi-messenger science to be effective. This point is best illustrated by the tools and services available on our web portal, including: the Sky Explorer\footnote{https://tools.ssdc.asi.it}, the Multi-Mission Interactive Archive\footnote{https://www.ssdc.asi.it/mma.html}, the SSDC Data Explorer, the SED Builder\footnote{https://tools.ssdc.asi.it/SED/}, the AGILE-LV3 tool, and the Fermi Online Data Analysis. Below we provide a brief description of these tools and services. A more detailed description can be found in ``The Future of Gamma-Ray Experiments in the MeV–EeV Range'' White Paper \cite{Engel:2022bgx}.

The Sky Explorer is the main gateway to access SSDC services, and allows users to easily investigate an astrophysical source via SSDC's web tools simply by specifying its name or coordinates. The SSDC Multi-Mission Interactive Archive (MMIA) allows users to easily access SSDC's high-energy astrophysics database, which contains extensive multi-wavelength data from several space missions (e.g. AGILE, 
Fermi, Swift, NuSTAR, Herschel), and interface with other SSDC web tools. The SSDC Data Explorer, for instance, enables users to easily visualize and analyze MMIA data. Similarly, the SED Builder builds and displays the spectral density distributions (SEDs) of astrophysical sources in the MMIA. Finally, for $\gamma$-ray data above 100 MeV, the AGILE-LV3 and the Fermi Online Data Analysis tools allow users to easily access data from AGILE and Fermi which may require substantial analysis time and to interface query results with other SSDC tools.

\subsection{Multiwavelength Classification Pipeline (MUWCLASS)}
\label{sec-MUWCLASS}

 \noindent
    \chapterauthor[1]{Hui Yang}
    \chapterauthor[2,3]{Jeremy Hare}
    \chapterauthor[1]{Oleg Kargaltsev}
 \\
 \begin{affils}
    \chapteraffil[1]{Department of Physics, The George Washington University, 725 21st St. NW, Washington, DC 20052, USA}
    \chapteraffil[2]{NASA Postdoctoral Program Fellow}
    \chapteraffil[3]{Astrophysics Science Division, NASA Goddard Space Flight Center, Mail Code 661, Greenbelt, MD 20771, USA}
 \end{affils}
 
 In the era of multimessenger, multiwavelength (MW), and multidomain  astronomy rapid classification  of a large number of sources becomes a particularly important task. The positional uncertainties  of gravitational wave sources, neutrino sources, or very high energy gamma ray sources can range from arcminutes to many square degrees. Depending on the location on the sky these regions may include millions of stars and galaxies and hundreds of X-ray and radio sources detected with new sensitive survey observatories such as eROSITA and SKA (or its prototypes). The classifications may need to be performed rapidly to identify  possible counterparts/progenitors of a high-energy event to enable a sensitive follow up to catch the fading EM emission from the event across the MW spectrum.
On-the-fly classification of all X-ray sources within the area of interest or continuous classification of all newly discovered X-ray sources will be an important component in the era of  multimessanger astronomy.

We have developed the multiwavelength machine learning pipeline for classification of unidentified X-ray sources (MUWCLASS) which uses information from both spectral and time domains.  The major component of our supervised machine learning pipeline is the training dataset (TD), which is a collection of 
several thousands X-ray sources with confident classifications. The current TD is categorized into 8 classes of X-ray emitters including active galactic nuclei (AGN), cataclysmic variables, high mass stars, high mass X-ray binaries, low mass stars, low mass X-ray binaries (this class includes non-accreting X-ray binaries), pulsars and isolated neutron stars (NS; this class also includes 11 magnetars), and young stellar objects which we constructed from multiple literature verified catalogs. We have built two versions of TD by cross-matching those literature verified sources with two X-ray catalogs, one from the Chandra Source Catalog Release 2.0 (CSCr2; \cite{2010ApJS..189...37E})  and the other from the 4XMM-DR11 catalog \cite{2020A&A...641A.136W} within their corresponding positional uncertainties. The CSC-based TD is now available online \cite{2021RNAAS...5..102Y}. The X-ray band fluxes and the hardness ratios are extracted as X-ray features
as well as two X-ray variability parameters, one for the inter-observation variability and the other for the intra-observation variability.
 The photometric properties at lower frequencies are extracted by cross-matching X-ray sources with the Gaia eDR3 \cite{2020yCat.1350....0G} in the optical, the Two Micron All-Sky Survey (2MASS; \cite{2003yCat.2246....0C}) in the near infrared (NIR), and the WISE All-Sky Data Release in the infrared (IR; \cite{2014yCat.2328....0C}) .    

At the heart of the MUWCLASS pipeline is a supervised ensemble decision-tree algorithm, Random Forest (RF), which is implemented via the scikit-learn python package. This algorithm offers a number of advantages over other ML algorithms (fast, does not require a distance metric, resistant to overfitting).
Before feeding our TD and unclassified source data into our RF classifier, we also apply a location-specific reddening/absorption correction to AGNs from TD (which come from surveys conducted away from the Galactic plan) while classifying sources in the Galactic plane. To handle the large imbalance of source types (e.g., there are substantially more X-ray detected AGNs than NSs), we use an implementation of the Synthetic Minority Over-sampling Technique to oversample our training data \cite{2011arXiv1106.1813C}. Measurement uncertainties are also taken into account by Monte Carlo (MC) sampling from feature probability density functions and averaging multiple MC sampling results to obtain confident classifications and measure their uncertainties. We have tested our pipeline which has an overall accuracy of about 86\%, up to 95\% for ``confident" classifications. The user-friendly automated MUWCLASS pipeline is now fully implemented in Python and will be made available to the astrophysical community via Github.

We are planning to generalize our pipeline to include radio properties and classify radio sources. It can also be used to classify optical, NIR and IR sources. In the future, we  will keep expanding  our TD  by making use of more sensitive modern surveys (e.g., PanSTARRS), radio surveys (from MeerKAT and VLASS) and optical/IR time-domain data from the Transiting Exoplanet Survey Satellite, the Zwicky Transient Facility as well. We will also include distance information (from Gaia) and account for the variable extinction and cross-matching confusion. We plan to adopt the pipeline to use eROSITA data as soon as the X-ray survey data are released. Additionally, as the sample of transient astrophysical X-ray sources (e.g., Tidal disruption events, gamma-ray bursts) grows we will include these sources in our TD. 

\section{Software}
\label{sec-software}

\subsection{Astrophysics Source Code Library}
\label{sec-ASCL}

 \noindent
 \chapterauthor[1]{Peter Teuben}\orcidlink{0000-0003-1774-3436}
 \chapterauthor[1,2]{Alice Allen}\orcidlink{0000-0003-3477-2845}
 \\
 \begin{affils}
   \chapteraffil[1]{Department of Astronomy, University of Maryland College Park, College Park, MD 20742, USA}
   \chapteraffil[2]{Editor, Astrophysics Source Code Library}
 \end{affils}

The development of software has undergone a dramatic transformation in
the past 60 years, and continues to do so. There are two aspects to
software in astronomy. On the one hand there is the software that
mirrors the development of the hardware, which was discussed in the
previous Section~\ref{sec-MUWCLASS}. Instruments need specific software for controlling the hardware, and often as well software that is used for instrument specific calibration. This type of software is often not
widely discussed, but generally makes its way in journals such as
SPIE/IEEE/ADASS. 

The second category is research software. This has traditionally
become more open---sometimes written by a collaboration of scientists
and professional programmers---and has sometimes even become less
domain-specific, enabling collaborations between instruments and
missions, and even across disciplines.

Additionally, instruments now deliver large amounts of data and groups
are collaborating on the analysis of this data, fundamentally changing the tools used. Who gets the credit for such software? How should it be credited? How can other scientists discover this software and re-use it?

We can view the corpus of research work in the framework of Papers,
Data, and Software, and we like to place these three on par with each
other when it comes to finding and citing them. With the onset of the
World Wide Web, astronomy has had several efforts in organizing this corpus and for Software, the sole survivor is the Astrophysics Source Code Library (ASCL; ascl.net). Since 1999, this repository of scientist-written software has become an asset to its community,
enabling researchers to find codes used in published works and discover new software that they might use. 
The ASCL now contains over 2,700 entries, which are also indexed by
the \href{http://ads.harvard.edu/}{NASA/SAO Astrophysics Data System}
(ADS) and Clarivate’s Web of Science Data Citation Index, and
\href{https://ascl.net/wordpress/?page_id=351}{are citable}; citations
to ASCL entries are tracked by ADS, Google Scholar, and Web of
Science.

So what are the current challenges?  We list a few:

\begin{itemize}
  
\item Availability: code is not made available, or on a website that
  proves to be ephemeral,

\item Documentation: code is poorly documented and not rigorously
  tested,

\item Findability: finding research software, most notably in other
  disciplines but applicable to ones own (reusability),

\item Funding: the price of code is often under-estimated,

\item Licensing: code is not or poorly licensed, restricting how
  others can use and amend the code.
  
\end{itemize}

Some of these challenges are historic--- the current perception leans towards the idea that building software is an art, not
a science. Compared to building an instrument or a house, building software arguably does not have rigorous methods. While hardware cannot be modified, the software often has to provide a solution ``in post.'' The lack of funding also places constraints on the researcher to provide a finished product in a reasonable time. Since there is no formal way to submit software as a polished product, the researcher often leaves this as the last item. A good counter example is the Journal for Open Source Software (JOSS), where authors are guided through an arguably complete list that brings the software on a high standard.




How should this look in 10 or 20 years? There is clear movement from many stakeholders towards making software more discoverable and citable. But, technology is hard to predict on these timescales.
Both software and hardware need rigorous procedures for testing and
verification, which arguably for research software is lagging that of
the hardware components. There needs to be a close collaboration between the publishing of papers, data, {\it and}  software, as well as more emphasis all around on their interplay!






\subsection{SCiMMA}
\label{sec-scimma}

\noindent
    \chapterauthor[1]{Adam Brazier}
\\
\begin{affils}
    \chapteraffil[1]{Cornell Center for Astrophysics and Planetary Science and Department of Astronomy, Cornell University, Ithaca, NY 14853, USA}
\end{affils}

\textbf{\textit{Introduction to SCiMMA:}}
The growing field of Multi-Messenger Astrophysics (MMA) has many needs, including reliable, low-latency communication of events to the MMA community and coordination of follow-ups; a platform enabling analyses by experts from the MMA community; cross-project and community exchange of MMA observation data and analyses; secured access allowing proprietary communications within long-term and also \textit{ad hoc} communications; and cross-archive searches to discover objects, build significance, and test theories~\cite{2019BAAS...51c.436C}. To help the community respond to these needs with robust cyberinfrastructure, the Scalable CyberInfrastructure for Multi-Messenger Astrophysics (SCiMMA) project was formed. The infrastructure to deliver these functions must scale to achieve the required performance and meet fluctuating demands while being affordable to deploy, operate, and maintain.  

The SCiMMA collaboration began as a conceptualization project funded by NSF OAC-1841590--- \textit{Collaborative Research: Community Planning for Scalable Cyberinfrastructure to Support Multi-Messenger Astrophysics}. This project concluded that an open-source effort mediated through a decentralized ``Institute'' would best sustainably achieve the identified goals, with a core team producing and maintaining services in response to community need through an open development process, with community involvement at all stages. Decentralizing the development team avoids the limits of excellence available at any one location, and also encourages community participation from outside that locus, but it requires a development process that directs and integrates the efforts of diverse Research Software Engineers (RSEs), who often have other demands on their time. The conceptualization project also identified federated identity and access management (IAM) as a key deliverable to allow the exchange of proprietary and public data.

Following the conceptualization project, a next phase of early design and prototyping was supported by NSF OAC-1934752, \textit{A Framework for Data Intensive Discovery in Multimessenger Astrophysics}. A design and development team was assembled at seven locations and the identified requirements were worked into a system design. The preferred platform for the core messaging service was identified as public cloud (e.g., Amazon Web Services (AWS), Google Cloud Platform, Microsoft's Azure, etc.); this carries the risk of vendor lock-in, which must be evaluated and mitigated when making architectural decisions. The SCiMMA security policy and operational controls team identified CILogon and COManage as the best infrastructure for the IAM service. The development process is a modified form of Agile scrum with two-week sprints, primarily using Slack for communications and GitHub for code management and continuous integration. An early test of the SCiMMA team's ability to integrate the needs and efforts of other services was with an engagement with the Supernova Early Warning System (SNEWS) team~\cite{2021arXiv210107779B}.

The two first prototype services from SCiMMA are the publish/subscribe messaging system, Hopskotch~\cite{hopskotch}, and the extensible and federated SCiMMA identity management system. Hopskotch is built on Apache Kafka~\cite{kafka} and hosted in AWS; the identity of the server software is irrelevant to most users as the SCiMMA architecture is designed to allow access to the Hopskotch service via a documented Python library, \texttt{hop-client}, distributed by SCiMMA through conda~\cite{anaconda} and PyPi~\cite{pip}; the client serves as the primary external API for users, hiding the details of the server architecture. With \texttt{hop-client}, users can identify themselves in the federated SCiMMA IAM system and subscribe and publish to communications channels (``topics'') as permitted by the authorization rules for that topic. The SCiMMA IAM service can be accessed programmatically or via the web interface, scimma-admin~\cite{scimma-admin}, allowing the creation and association of security groups and topics, testing scaling behaviour and managing resources, and implementing other policies as necessary.

\textbf{\textit{SCiMMA's Future Plans:}}
The next stage for SCiMMA is to bring the Hopskotch and IAM services into full production. Early testing of the Hopskotch prototype demonstrates that the requirements from the LIGO team~\cite{LIGOreqs} can be met and the necessary work and costs to deliver the production system in time to serve events originating with LIGO's O4 run, to begin end of 2022, are well understood. SCiMMA has applied for additional funding to provide services through the planned O4 and O5 runs~\cite{Aasi:2013wya}, Vera Rubin Observatory target-of-opportunity operations, and the IceCube Gen2 requirements-gathering process.

The immediate plans for additional SCiMMA development also include an archive of all data that have transited the Hopskotch system; this flexible archive will not apply schema restrictions at time of data ingress, because of the heterogeneous nature of MMA data and communications, but will allow specification of schema at time of query (this sort of service is often conceptualized as a ``data lake''); the data lake will also allow additional data to be ingested to increase the extent and value of data sets being queried. A JupyterHub analysis platform based on Astronomy Commons~\cite{astrocommonsgateways, astronomycommons} will be connected to the data lake, which will allow performant and flexible analysis of the real-time Hopskotch output as well as the archived data.


\subsection{FermiPy}
\label{sec-FermiPy}

 \noindent
 \chapterauthor[1,2,3]{Giacomo Principe}\orcidlink{0000-0003-0406-7387}
 \\
 \begin{affils}
   \chapteraffil[1]{Dipartimento di Fisica, Universit\'a di Trieste, I-34127 Trieste, Italy}
    \chapteraffil[2]{Istituto Nazionale di Fisica Nucleare, Sezione di Trieste, I-34127 Trieste, Italy}
    \chapteraffil[3]{INAF - Istituto di Radioastronomia, I-40129 Bologna, Italy}
 \end{affils}

The \textit{Fermi} LAT gamma-ray telescope~\cite{2009ApJ...697.1071A} detects photons by conversion into electron-positron pairs and has an operational energy range from 20\,MeV to 2\,TeV. LAT data, i.e., events classified as photon-like, are immediately publicly available at the NASA \textit{Fermi} Science Support Center\footnote{https://fermi.gsfc.nasa.gov/ssc/data/access/} (FSSC). The FSSC also offer a suite of public software tools---the \textit{Fermi} \textit{ScienceTools} (written in C++)---for the reduction and analysis of LAT data, as well as a python interface (pyLikelihood) which facilitates scripting analysis in python of LAT data.

Fermipy\footnote{https://fermipy.readthedocs.io/en/latest/}~\cite{2017ICRC...35..824W} is a python package that facilitates the analysis of \textit{Fermi}-LAT data. It is mainly based on the \textit{Fermi} Science Tools and it makes use of the pyLikelihood python interface. This tool depends on a few other open-source python libraries such as NumPy~\cite{2013A&A...558A..33A}, Scipy~\cite{2020NatMe..17..261V}, and Astropy~\cite{2018AJ....156..123A}, as well as some new functionalities imported from GammaPy~\cite{2015ICRC...34..789D}. In addition, an optional dependency needed for plotting and visualising the analysis results is given by Matplotlib~\cite{2007CSE.....9...90H}.

Fermipy is designed around a global analysis state object (GTAnalysis) which handles the data and model preparation, as well as provides some high-level analysis methods. The first step of the procedure is given by the creation of a configuration file that delineates analysis parameters, including  data selection, region-of-interest (ROI) geometry, and model specifications.
The high-level analysis methods are constituted by: model optimisation, search for possible additional faint sources, generation of TS maps (significance maps), re-localisation of the sources, and study of the extension, spectral and light-curve analyses.

\paragraph{Fermipy multi-wavelength and multi-messenger applications}
Fermipy is very suited for high level analyses of \textit{Fermi}-LAT gamma-ray data, in particular, thanks to its python framework, it makes possible to easily combine multi-wavelenght and multi-messenger results of celestial objects.

Related to the astrophysical topics discussed in Chapters~\ref{sec-astroBSM} and \ref{sec-astrophysics}, in this paragraph we highlight some examples of multi-wavelength and multi-messenger analyses performed with Fermipy. Starting with AGN studies, a remarkable example is represented by the extensive multi-wavelength campaign on M87 using ground- and space-based facilities performed during the first Event Horizon Telescope (EHT) observations~\cite{2021ApJ...911L..11E,2021AAS...23812503P}. In addition, there were many multi-wavelength studies of AGNe populations, which investigated the origin of the gamma-ray emission in young radio galaxies~\cite{2020A&A...635A.185P,2021MNRAS.507.4564P}, or bright blazars~\cite{2021ApJS..257...37P}. Moving to our Galaxy, Fermipy was used for studying different classes of Galactic sources, such as SNRs, PWNe and gamma-ray halos~\cite[][respectively]{2022arXiv220105567A,2020A&A...640A..76P,2021PhRvD.104j3002D}.

Fermipy is well suited for the search of gamma-ray transient emission on different time scales: from few seconds~\cite[such as the search of high-energy emission from FRBs,][]{2021arXiv210903548P}, to months ~\cite[e.g. the first catalog of \textit{Fermi}-LAT transient sources,][]{2021ApJS..256...13B}, or even several years~\cite[as in the case of the study FSRQs variability,][]{2019ApJ...877...39M}.

Directly related to multi-messenger phenomena, Fermipy has been recently used for studying the high-energy emission of the tidal disruption event (AT2019dsg) associated with a high-energy neutrino \citep{2021NatAs...5..510S}.

Finally, it was adopted also in the search of dark matter in our Universe, like the search of axion-like particles in extragalactic core-collapse supernovae~\cite{2020PhRvL.124w1101M}, or the search for dark-matter sub-halos in extended \textit{Fermi}-LAT Galactic sources~\cite{2020PhRvD.102j3010D}.

These ground-breaking multi-messenger and multi-wavelenghts science results herald the great science still to come from \textit{Fermi}-LAT, as well as the importance and flexibility of the Fermipy package, which may be utilised for different sources and missions.

\subsection{3ML}
\label{sec-3ML}

\chapterauthor[ ]{ }
\vspace{-0.2in}
 \addtocontents{toc}{
     \leftskip3cm
    \scshape\small
    \parbox{5in}{\raggedleft Henrike Fleischhack, J. Michael Burgess,  et al.}
    \upshape\normalsize
    \string\par
    \raggedright
    \vskip -0.19in
    }
 
\noindent
\nocontentsline\chapterauthor[]{Henrike Fleischhack$^{1,2,3}$}\orcidlink{ 0000-0002-0794-8780}
\nocontentsline\chapterauthor[]{J. Michael Burgess$^4$}\orcidlink{0000-0003-3345-9515}
\nocontentsline\chapterauthor[]{Nicola Omodei$^5$}
\nocontentsline\chapterauthor[]{Niccol\`{o} Di Lalla$^5$}
\nocontentsline\chapterauthor[]{Chad Brisbois$^{6}$}\orcidlink{0000-0002-5493-6344}
\\
 \begin{affils}
   \chapteraffil[1]{Catholic University of America, Washington DC}
   \chapteraffil[2]{NASA Goddard Space Flight Center, Greenbelt, MD}
   \chapteraffil[3]{Center for Research and Exploration in Space Science and Technology, NASA/GSFC, Greenbelt, MD}
   \chapteraffil[4]{Max-Planck Institut f\"ur Extraterrestrische Physik, Giessenbachstrasse 1, 85740 Garching, Germany}
   \chapteraffil[5]{W. W. Hansen Experimental Physics Laboratory, Kavli Institute for Particle Astrophysics and Cosmology, 
Department of Physics and SLAC National Accelerator Laboratory,
Stanford University, Stanford, CA 94305, USA}
 \chapteraffil[6]{University of Maryland, College Park, College Park, MD, 20742, USA}
\end{affils}

The Multi-Mission Maximum Likelihood framework (\texttt{3ML}, also: \texttt{threeML}, see Ref.~\citenum{2015arXiv150708343V}) is a python-based software package for astronomical joint-likelihood analyses of multi-wavelength data. By fitting models with measured data, a likelihood analysis aims to produce estimates for certain free parameters, $\theta$, of said ``model''---be it phenomenological or physical---describing the energy spectrum, shape, and/or time evolution of a gamma-ray source. The data with which this model is matched, $X$, can be of a variety of formats, 
analyzed together using the likelihood function ${\mathcal{L}(\theta\mid X)} = {P(X\mid\theta)}$, where ${P(X\mid\theta)}$ describes the probability of measuring $X$, given model parameters, $\theta$.

\texttt{ThreeML} handles data access, convolution of the models with the instrument responses, and calculation of the likelihood through the use of plugins, each built to handle data and instrument response files--- both proprietary formats (e.g., HAWC, \textit{Fermi}-LAT) and community standards (OGIP). Plugins can be fully implemented within the framework as wrappers around existing likelihood analysis tools 
or provided as standalone packages
. Plugins for a particular instrument are tailored as needed, e.g., providing the capability to set active bins or define the region of interest for an analysis. 
Plugins are also used to facilitate \texttt{ThreeML}'s options for minimizers and sampling algorithms, typically implemented for this use as wrappers around external libraries. This design allows a user to switch between options without modifying the rest of the analysis script.

A binned, forward-folding likelihood approach, where ${P(X\mid\theta)}$ above is represented by $\prod\limits_{i} \mathrm{Poisson}\left( x_i \mid N_i \left( \theta, \alpha \right)\right)$), is utilized in plugins for gamma-ray instruments such as HAWC and \textit{Fermi}-LAT. Here, data are binned in one or more dimensions (often energy, arrival direction, and/or time) such that $x_i$ is the number of measured photon candidates in bin $i$. The predicted counts, $N_i$, are a function of the model parameters, $\theta$, and nuisance parameters, $\alpha$. The nuisance parameters are those that are internal to a given plugin, e.g., background normalization. The predicted counts, $N_i$, are derived by folding the photon emission predicted by the model with the detector response, which is derived from simulation and includes angular resolution, energy resolution, and effective area. 
Separate plugin instances are used for analyses using multiple independent data sets, thereby calculating the likelihoods separately for each data set, multiplying their results to obtain to final likelihood value. The plugin approach makes \texttt{threeML} ideal for multi-wavelength analyses as they allow the user to easily add or remove data sets from the analysis. Additionally, since the majority of the code is agnostic to which plugins are used and how they are configured, a multi-instrument fit is no different than a single-instrument one except for setting up the plugins. 

In practice, calculation of the likelihood can be computationally expensive, due to the large number of factors considered when simulating inputs to the Instrument Response Functions and the number of bins. This means that the likelihood minimization may take significantly longer than performing a simple $\chi^2$-fit to a set of data points. To do this, one must derive a Spectral Energy Distribution (SED) or energy spectrum for each source from given data sets and then perform a $\chi^2$-fit to the data from that model. However, to do this $\chi^2$-fit, one must first deconvolve the instrumental data, a nontrivial task (see Ref.~\citenum{Starck_2002}). In addition to maximum likelihood estimation of model parameters (allowing for frequentist interpretation of results), \texttt{threeML} also supports Bayesian posterior distribution sampling. The posterior probability is given by Bayes' Theorem, $p(\theta|X) = \frac{P(x|\theta)}{p(x)}p(\theta)$, where $P$ is the likelihood as defined previously. The prior probability distribution of the free parameters, $p(\theta)$, and $p(x)$ is chosen so that the distribution is normalized appropriately.

\texttt{ThreeML} relies on the \texttt{astromodels} package (also written in Python) to model the underlying gamma-ray emission. In \texttt{astromodels}, a ``model'' consists of all sources used to describe a given region of interest. Sources consist of an emission spectrum, position, and  morphology. Many commonly used spectral and spatial functions, such as power laws and point sources, are already implemented. The user may also supply external templates for the morphology and spectrum, or define new functions as needed for an analysis. Analyzing sources exhibiting energy-dependent morphology or time-dependent spectra is also possible. The user can freely select which model parameters to free or fix when performing the fit, as appropriate to their task. Multiple parameters may be linked to each other and, for Bayesian analyses, have prior distributions associated with them. \texttt{threeML} also provides an interface to download publicly accessible data (such as the \textit{Fermi}-LAT's 4FGL catalog), generate models from that data, fit the free parameters in the model of that data, plot fitted spectral energy distributions with propagated uncertainties, and investigate the goodness-of-fit of the optimized model.

\texttt{ThreeML} and \texttt{astromodels} is freely available, and may be found on \texttt{github} (\url{https://github.com/threeML/}). Both packages are released via \texttt{conda} (channel ``\texttt{threeML}'') and \texttt{pip}. Documentation and worked examples can be found at  \url{https://threeml.readthedocs.io/} and \url{https://astromodels.readthedocs.io/}. New development on \texttt{threeML} is focused on further optimization and resource usage improvements, adding plugins for new and future missions/instruments, and ongoing efforts to enhance existing plugins.\\

\noindent {\small \textbf{\textit{Acknowledgements}} H.F. acknowledges support by NASA under award number 80GSFC21M0002. Any opinions, findings,
and conclusions or recommendations expressed in this material are those of the author(s) and do not necessarily reflect the views of the National Aeronautics and Space Administration.} 



\section{Outreach, Public Engagement, and Citizen Science}
\label{sec-outreach}

\noindent
    \chapterauthor[1,2]{Tiffany R. Lewis}\orcidlink{0000-0002-9854-1432}
\\
\begin{affils}
    \chapteraffil[1]{NASA Postdoctoral Program Fellow}
    \chapteraffil[2]{Astroparticle Physics Laboratory, NASA Goddard Space Flight Center, Greenbelt, MD}
\end{affils}

\paragraph*{Outreach in High-Energy \& Multimessenger Astrophysics}

Outreach is often used as a tool of recruiting - share your science in order to make more scientists. However, this limited view has broad implications for the inaccessibility of basic information to the average person. Educational or public outreach that is not primarily for recruiting is about providing a positive impression of science, scientists, and hopefully of a specific area of science that you discussed to society.  A democratic society that has broad popular respect and enthusiasm for science and scientists is more likely to support the infrastructure for new discoveries.

The first time a person interacts with a topic they are unlikely to come away with a deep understanding and a week later, they may not remember a single basic fact about the interaction, but they will remember how they felt during the discussion. Positive feelings about science and communicating with scientists form the basis of a soft-infrastructure for explaining the basics of multimessenger science for a non-expert audience that is part of the society pursuing it. 

At the core of K12 education in the U.S. is the production of an educated electorate, a population that is able to consider their values in relation to the problems and opportunities available to local, state, and national governance. In that context, education should provide a framework for organizing information in the world that may become relevant through decisions about scientific priorities, long term innovation supported by ongoing fundamental discoveries and an ability to use new information and innovations.  

One of the key components of presenting physics for an audience that needs to understand some basic information that the modern world is built on, but most of whom will not become physicists or astronomers, is to give precedence to modern physics topics (rather than kinematics and the debate around whether one needs calculus to parse it). Kinematics can be covered for those with an interest in pursuing science, engineering or math formally, but it is not reasonable or beneficial to assume that desire or background as a prerequisite for a conceptual overview of modern topics and how they are studied. For example, most adults know that cells have organelles but do not know that black holes are real or the difference between fission and fusion. 

One way that scientists can help with this is to develop age appropriate activities and demonstrations to help work understanding of high-energy astrophysics and multimessenger science into the public conscience over time. Work of this nature should be supported as an investment in the long term health of the field. A model for some of these activity recipes and their distribution may be found with the education materials developed by the Astronomical Society of the Pacific and their ongoing AAS Ambassador program, which puts resources and communication techniques in the hands of early career astronomers with a desire to engage with their local communities.

Community outreach is most effective when scientists are able to build rapport within the local area. It can help to hold predictable repeat events with similar structure and science representatives in attendance to foster a sense of community. It can also be helpful to develop branded programming, memberships, newsletters, etc to make sure that interested persons feel included and know how to access the public-facing resources that physicists can provide. Some examples of such programming include public lecture series, local facility tours, open house events, Astronomy on Tap, Meet a Scientist, and tabling events with themed activity stations. It is important to foster a sense of community and to be mindful of addressing barriers to diverse, equitable, inclusive, and accessible participation. Some ways to help thinking about preventing bias in event, information, and community access is to consult with local institutions (public schools, public libraries, museums or venues that might draw a similar crowd) and to connect with national programs for science education and outreach.

Outreach is a major driver of public engagement in individual science topics. Putting a face and a personality to the stories told about scientific discoveries can help to make them more real. Public support for scientific work is an important aspect of congressional support for scientific work.

\paragraph*{Public Engagement in High-Energy \& Multimessenger Astrophysics}

Public engagement can occur through traditional media, which is often not managed by scientists. Programs that cover a wide range of topics may have difficulty finding experts to consult on each one individually. It should be noted that an expert in one area is often not a sufficient substitute for an expert in another. While scientists should feel comfortable in admitting where their personal knowledge ends as a matter of respect for their colleagues, it ultimately falls to producers to do enough personnel research to ask the right person. Traditional media is so far reaching and the science they cover can last for decades in the public consciousness without the average consumer being able to verify it elsewhere, so there is a special imperative that the content be accurate. 

Public engagement online may occur though actively managed social media accounts, informational websites, online activities like scientist chats, topical walk throughs, and home guides for engineering crafts. Virtual engagement is often more immediately accessible to scientists who are interested in sharing a particular message with a target audience, but many of the principles we use in the organization of local events also apply online: foster a sense of community and to be mindful of addressing barriers to entry or participation. The building of an online community comes through repetition, intentional branding, and advertisement across platforms.

\paragraph*{Citizen Science in High-Energy \& Multimessenger Astrophysics}

Citizen science makes use of publicly available data or easily accessible facilities to perform simple tasks and interface with a scientific community by participating in a small part of a research project. Some notable examples in astronomy include linking amateur astronomers to optical followup networks for transient events, and providing vast libraries of images online with the goal of crowd sourcing pattern recognition as with Zooniverse.com.

For high-energy or multimessenger astrophysics these canned at-home projects might look like hosting a small detector as part of a local network, and sharing with the host what their detector is doing. Or otherwise to create simple, minimal time commitment, easily reproducible activities and make them available to the general public through a website, app, or public library equipment rental program. 

One concept study to use people's phones as detectors is discussed below, and additional effort should be devoted to improving upon the user experience and scientific utility of publicly accessible scientific efforts.

\subsection{CREDO}
\chapterauthor[ ]{ }
\vspace{-0.2in}
 \addtocontents{toc}{
     \leftskip3cm
    \scshape\small
    \parbox{5in}{\raggedleft Piotr Homola, et al. on behalf of the CREDO Collaboration}
    \upshape\normalsize
    \string\par
    \raggedright
    \vskip -0.19in
    }

\noindent
\nocontentsline\chapterauthor[]{Piotr Homola$^{1}$}
\nocontentsline\chapterauthor[]{Roger Clay$^{3}$}
\nocontentsline\chapterauthor[]{Dmitri Beznosko$^{2}$}
\nocontentsline\chapterauthor[]{Sławomir Stuglik$^{1}$}
\nocontentsline\chapterauthor[]{\L ukasz Bibrzycki$^{4}$}
\nocontentsline\chapterauthor[]{Marcin Piekarczyk$^{4}$}
\nocontentsline\chapterauthor[]{Olaf Bar$^{4}$}
\nocontentsline\chapterauthor[]{Michał Frontczak$^{4}$}
\nocontentsline\chapterauthor[]{Krzysztof Rzecki$^{5}$}
\nocontentsline\chapterauthor[]{Michał Nied\'{z}wiecki$^{8}$}
\nocontentsline\chapterauthor[]{Tomasz Sośnicki$^{5}$}
\nocontentsline\chapterauthor[]{Thomas Andersen$^6$}
\nocontentsline\chapterauthor[]{Tadeusz Wibig$^7$}
\nocontentsline\chapterauthor[]{on behalf of the CREDO Collaboration}
\\
 \begin{affils}
   \chapteraffil[1]{Institute of Nuclear Physics Polish Academy of Sciences, Radzikowskiego, Krak\'{o}w, Poland}
   \chapteraffil[2]{Clayton State University, Morrow, Georgia, USA}
   \chapteraffil[3]{University of Adelaide, Adelaide, S.A., Australia}
   \chapteraffil[4]{Pedagogical University of Kraków: Krakow, Małopolska, PL}
   \chapteraffil[5]{AGH University of Science and Technology in Krakow, Poland}
   \chapteraffil[6]{NSCIR - 046516 Meaford, Ontario N4L 1W7, Canada}
   \chapteraffil[7]{Faculty of Physics and Applied Informatics, University of Lodz, 90-236 Łódź, Pomorska 149/153, Poland}
   \chapteraffil[8]{Department of Computer Science, Cracow University of Technology, Warszawska, kraków, Poland}
\end{affils}

The Cosmic Ray Extremely Distributed Observatory (CREDO) Collaboration \cite{sym12111835} advocates studies of cosmic ray phenomena called Cosmic Ray Ensembles (CRE) which are currently not mainstream in the field but show promise of new, interesting, astrophysics. There is a theme of looking for correlations in the cosmic ray beam, spatially over large areas with innovative techniques, and temporally through the search for bursts in cosmic ray data sets, see e.g. Ref. \cite{Clay:2021ep}. 
To date, most of the data collected by CREDO comes from smartphones with the CREDO Detector app\footnote{\href{https://credo.science/credo-detector-mobile-app/}{https://credo.science/credo-detector-mobile-app/}}, operating on the Android system with already more than $10.5$ million detections, and with the Cosmic Ray App\footnote{\href{https://cosmicrayapp.com/}{https://cosmicrayapp.com/}} dedicated to iOS devices with more than $7$ million detections. 
Another important example of the infrastructure working within the  CREDO Collaboration, although not yet connected to the central system, is  the High Energy Astrophysics Muon System\footnote{\href{http://www.physics.adelaide.edu.au/astrophysics/muon/}{http://www.physics.adelaide.edu.au/astrophysics/muon/}} (HEAMS), an array of muon detectors operated by the University of Adelaide, Australia, consisting of several one square meter scintillator muon detectors in two locations distant by 40 km. These two example resources, and the corresponding studies, in particular the one demonstrating the feasibility of identification of muons with smartphones \cite{app11031185,Wibig:202192}, illustrate the main concept and potential of a global network of affordable radiation sensors. Correspondingly, established public interest, including in particular many young science enthusiasts and their teachers, promises a sustainable growth of the network and a continuous support for all the scientific projects to be carried out using the CREDO resources.
While CREDO aims at physical hosting of a multi-detector and multi-technique global sensor network, its full openness and free accessibility enables bridging and interoperability with practically all the detector systems receiving a cosmic signal of any type, as envisaged to be necessary for optimizing the observational strategies dedicated to CRE.
\section{Diversity, Equity, Inclusion, and Accessibility}
\label{sec-DEIA}

\noindent
    \chapterauthor[1,2,3]{Ronald S. Gamble}\orcidlink{0000-0002-8361-036X}
\\
\begin{affils}
    \chapteraffil[1]{Astrophysics Science Division, NASA Goddard Space Flight Center, Greenbelt, MD}
    \chapteraffil[2]{Department of Astronomy, University of Maryland, College Park, MD}
    \chapteraffil[3]{College Park/Center for Research and Exploration in Space Sciences \& Technology, University of Maryland, College Park, MD}
\end{affils}

Objectively, \textit{diversity} is defined as the quality or state of having many different forms, types, ideas, etc. It is rooted in the Latin language as \textit{Divertere}, which means \textit{to turn in different directions}~\cite{DiversityMW2021}
. Implying that usage of the word is inherently focused on the diversion or change of a chosen path. Taking into account that a diverted path, whether it be of a physical notion or a more psychological one, does not necessarily imply a negative outcome. Modern day usage of the word implies a more \textit{ethno-sociologically} constructed meaning. Today, diversity, equity, inclusion, and accessibility (DEIA) are viewed as fundamental elements of a well-rounded and sound workplace, school, and or team. Currently the field of Astronomy and Astrophysics is made up of a community of scientists,  engineers, technical staff, teachers, and science-enthusiasts. All having various levels of experiences and academic accolades, \textit{all} having one primary quality that is synergistic to the profession. We are all human. If we are to push the profession forward to new heights, then it is pertinent that we begin to build more diverse and inclusive collaborative teams. Historically, the profession is not kind to the people that bring the science to life. Often times disregarding social and ethical consequences in the pursuit of scientific discovery. The integrity of the field of astronomy and astrophysics has become very fragile with respect to increasing exposure to instances of racial and gender discrimination or exclusion.

An important aspect of building a diverse, inclusive, and effective collaborative team is rooted in the overall scientific goal and or objective. If one is to have a strong effective team of scientists collaborating on a body of work, then a certain degree of diversity in all its aspects should be considered. A term that is used a number of times within the social sciences is \textit{multimodal expertise}. Defined as utilizing the collective experiences of a team to reach a common goal, multimodal expertise is something that should be adopted more often when executing DEIA efforts in building these collaborative teams of [humans]. \textit{Neurodiversity} is often looked at as being a form of diversity of thought; supporting the notion that scientific advancements are done by collections of people working together towards an objective conclusion to an overarching hypothesis. Analogous to the Astro2020 decadal survey that puts heavy emphasis on multi-messenger astronomy~\cite{2021pdaa.book.....N}
, the importance of diversity, equity, inclusion, and accessibility in collaborative teams, introduced as incorporating multimodal expertise, are rooted in the advancement of the profession as well. With an ever increasing occurrence of large scientific collaborations, with lots of dynamical elements to the collaboration, placing emphasis on the institutional values that a collaboration is housed under becomes increasingly important~\cite{2016igamdfg.book.....P}
. Overall it can be seen that there exists significant facets to building teams of scientists and non-scientists, MSI and PWI alumni/students, binary and nonbinary genders, and the like.

\clearpage


\chapter*{Acknowledgements}
\addcontentsline{toc}{chapter}{\protect\textbf{Acknowledgements}}

J.P.W.V. is supported by the Deutsche Forschungsgemeinschaft(DFG) through the Heisenberg programme (Project No. 433075039). The MOSSAIC concept received enthusiastic endorsement from many colleagues and institutions in the ground- and space-based MMA/TDA communities, who recognize the need for coordination and collaboration in this multifaceted discipline. We are grateful to the GSFC and Center leadership for their support of MOSSAIC. H.F. acknowledges support by NASA under award number 80GSFC21M0002. Any opinions, findings,
and conclusions or recommendations expressed in this material are those of the author(s) and do not necessarily reflect the views of the National Aeronautics and Space Administration. 

\clearpage


\bibliographystyle{JHEP}
\bibliography{main,200FundamentalBSM-Kneller,361GammaBackground-Michela,352FastRadioBursts-Elijah,356PulsarHalos-Mattia,4518FutureGroundBasedGammaRayTele-JordanBrendaBrian,4517CurrentGroundBasedGammaRayTele-JordanBrendaBrian,240PrimordialBlackHoles-KristiPat.bib,361DGRB-Mora,311AGNJets-Haocheng,311AGNJets-Meli,624FermiPy-Giacomo,522MUWCLASS-Hare,351CCSNLGRB-FryerBurns,211H0-Coughlin,320TidalDisruption-Stein,CurrentOpticalFacilities-Stein,332Inspirals-Zachary,312UnjettedAGN-Inoue,354Magnetars-Wadiasingh,343NSBHMergers-Foucart,533SCiMMA-Brazier,342-Chirenti-NSNSMergers,451PulsarTiming-Joris,344BHBH-Chirenti,531ASCL-Teuben,362DiffuseNeutrino-Samalka,4512PUEO,4511CurrentBalloonSpaceEASDetectors-CummingsEser,4512PUEOPOEMMA-CummingsEser,331MassiveBlackHoleBinaries-Noble,364GWBackground-Meyers,4510AirShower-Ahlers,4513ARA,4513RNOG,536threeML-HenrikeMichael,4510BEACON-Cummings,260LIV_EngelHardingMartinezHuerta,353SNR-Filipovic,250DarkMatter-Slatyer,452currentGW,454SpaceBasedGWDetectors,Jordan-Brenda-Brian,4581SWGO-AlbertEngel,550DEIA-Gamble,260LIV-Castillo,455currentGamma-Colleen,456FutureSpaceBasedGamma-Kierans,CREDO,441AMON-Hugo,522SSDC-Gianluca_summarized}

\end{document}